
\magnification\magstep 1
\vsize=21 true cm
\hsize=16.7 true cm
\baselineskip=0.8 true cm
\parindent=1.1 true cm

\raggedbottom
\def\init{\tabskip 0pt\offinterlineskip}
\def\crr{\cr\noalign{\hrule}}

\def\intif{\int_0^\infty}

\font\bfmag=cmbx10 scaled\magstep3
\font\bfmagd=cmbx10 scaled\magstep2
\font\bfmagc=cmbx10 scaled\magstep1

\font\zmagd=cmr10 scaled\magstep2
\font\zmagc=cmr10 scaled\magstep1
\def\tr{{\rm Tr}}
\def\Lag{{\cal L}_{\pi\rho}}

\def\WMN{{W_{\mu\nu}}}
\def\Lgls{{\cal L}_{\pi \xi \rho A_1}}
\def\leviup{\epsilon^{\mu\nu\alpha\beta}}
\def\levido{\epsilon_{\mu\nu\alpha\beta}}
\def\L6{{\cal L}_{\omega}}

 2
 1

\def\dmuup{{\partial^\mu}}
\def\dmudo{{\partial_\mu}}
\def\dnuup{{\partial^\nu}}
\def\dnudo{{\partial_\nu}}

\def\di{{\partial_i}}
\def\dj{{\partial_j}}

{\rightline{}}
{\rightline{MC/TH 94/15}}

{\vskip 3cm}

\centerline{\bfmag Vers une th\'eorie unifi\'ee}

\medskip

\centerline{\bfmag des m\'esons et des baryons}

\bigskip

\bigskip

\centerline{(Towards a unified theory of mesons and baryons)}

\bigskip

\bigskip

\bigskip

\bigskip

\bigskip

\centerline{\zmagd Dimitri Kalafatis}

\bigskip

\bigskip

\bigskip

\medskip

\centerline{\zmagc Theoretical Physics Group}

\medskip

\centerline{\zmagc Department of Physics and Astronomy}

\medskip

\centerline{\zmagc The University of Manchester}

\medskip

\centerline{\zmagc Manchester M13 9PL}

\medskip

\centerline{\zmagc UNITED KINGDOM}

\bigskip

\bigskip

\bigskip

\bigskip

\centerline{28 juin 1994}

\eject

\centerline {\bfmagd Abstract}

\medskip

This review article is devoted to recent advances in the theoretical
description of nucleons as solitons of meson theories. Mesons are
the elementary fields of a highly nonlinear effective lagrangian of QCD and
baryons emerge as their topological soliton excitations. In chapter I we
construct
an effective lagrangian with the low mass mesons ($\pi,\rho,A_1$ and the scalar
meson $\epsilon$) generalising the Skyrme model. The vector meson fields are
introduced
as massive gauge fields in the linear sigma model. The parameters of the model
are
determined by fitting to low energy meson observables. We then look for
topological
soliton solutions of the model and investigate the nucleon-nucleon interaction
in the
product approximation. The results show that the scalar degrees of freedom give
rise to attractive intermediate range NN forces. This is in sharp contrast to
models incorporating vector degrees of freedom where only repulsive forces are
found.

In chapter II we evaluate the leading correction to the classical Skyrmion
mass, that is,
the Casimir energy. The main ultraviolet divergence is cancelled by well known
chiral counterterms. We show that the result is controlled by the low energy
behaviour
of the phase shifts which are large and positive due to the presence of two
normalisable
zero modes. As a consequence, the correction is found to be negative and rather
large,
of the order of 1 GeV for the original Skyrme model. We then show that the
situation
improves if one introduces higher chiral orders in the effective lagrangian.

The problem of the stability of topological solitons when vector fields enter
the chiral
lagrangian is the subject of chapter III. Isospin one vector mesons are usually
described as massive Yang-Mills particles in the chiral lagrangian. We
investigate here
some aspects of an alternative approach. It is found that within the theory we
propose for
the $\pi\rho$ system, the soliton is stable in very much the same way as with
the $\omega$-meson and that peculiar classical doublet solutions are ruled out.
The
formulation in terms of antisymmetric tensors is shown to be canonically
related
to a special case of our vector field description.

\eject

\centerline{\bfmag Introduction}

\bigskip

\bigskip

La Physique des Constituants \'El\'ementaires de la Mati\`ere
est le domaine dans lequel la Th\'eorie Quantique des Champs trouve ses
applications naturelles. Les progr\`es effectu\'es
dans la description du spectre des particules observ\'ees dans la
Nature, doivent beaucoup au principe de l' existence d' un
\ \ Lagrangien \ \ fondamental. Ces Lagrangiens sont habituellement
formul\'es en termes de multiplets de fermions \'el\'ementaires
interagissant par l' interm\'ediaire des bosons de jauge
associ\'es \`a des invariances d\'ecoulant de sym\'etries internes.
D' autre part il est bien connu que les interactions fondamentales
ob\'eissent \`a une hi\'erarchie compliqu\'ee, b\^atie sur l'
intensit\'e du couplage entre les champs de particules.
Pour les interactions  \'electromagn\'etiques et faibles il est
naturel de d\'ecrire les interactions dans le cadre de la th\'eorie des
perturbations. Celle-ci ne constitue une bonne
approximation \`a la solution exacte que si la constante de couplage est
faible, ce qui est le cas des interactions \'electrofaibles.

Cependant la th\'eorie des perturbations n' est pas d' une grande utilit\'e
pour les interactions fortes, responsables de la coh\'esion de la
mati\`ere. Dans la th\'eorie des interactions
fortes, le Lagrangien \lq\lq fondamental"
sur lequel beaucoup d' espoirs sont fond\'es, est celui de la
chromodynamique quantique (QCD). Les champs
fondamentaux de cette th\'eorie sont des multiplets \lq\lq color\'es" de
fermions (quarks) dont les interactions sont assur\'ees par des champs
de Yang-Mills de masse nulle (gluons).  Une propri\'et\'e essentielle
que la QCD assigne aux interactions fortes, la libert\'e
asymptotique, fait que les ph\'enom\`enes aux courtes distances
peuvent \^etre d\'ecrits par la th\'eorie des perturbations.  Pour
la description des propri\'et\'es \`a basse \'energie des hadrons (dont la
taille est de l' ordre du
fermi), tout
sch\'ema perturbatif est compl\`etement illusoire, car pour ces
distances,  la constante de couplage des interactions fortes est tr\`es grande.
Le fait que les quarks sont confin\'es [1] \`a
l' int\'erieur des hadrons n' est qu' une illustration du caract\`ere
non-perturbatif des interactions fortes. Comme il
n' existe pas \`a l' heure actuelle de m\'ethode non-perturbative fiable
et suffisamment simple, applicable au cas de la QCD, force est de
construire, pour la description des spectres des hadrons, des
mod\`eles o\`u les quarks sont confin\'es par un m\'ecanisme plus
ou moins \lq\lq ad-hoc" comme les mod\`eles de
quarks non-relativistes, de cordes, de sacs, di\'electriques etc...
Ces mod\`eles sont assez arbitraires car chacun d' eux simule
le confinement, alors que le m\'ecanisme de ce dernier reste inconnu.

Dans cet article nous adoptons une autre approche pour la
description du monde hadronique. Puisque les quarks et les gluons
sont confin\'es \`a l' int\'erieur des
hadrons il est, peut \^etre, plus appropri\'e d' \'eliminer ces
degr\'es de libert\'e et d' incorporer leurs effets dans une th\'eorie
effective des particules observ\'ees (les m\'esons et les baryons). Ce
point de vue est \'etay\'e par une conjecture de Witten [2], selon
laquelle \`a basse \'energie et \`a la limite
o\`u le nombre de couleurs $N_c$ devient grand, QCD est \'equivalente
\`a une th\'eorie effective de m\'esons interagissant faiblement. Les
baryons \'emergent de cette th\'eorie comme des solitons.
Cette id\'ee est attrayante car elle offre la possibilit\'e
de d\'ecrire les m\'esons et les baryons dans le cadre d' une seule th\'eorie.
En fait, un exemple d' une telle th\'eorie avait \'et\'e
propos\'e par Skyrme [3] il y a trente ans, bien avant
l' av\`enement de la QCD. Skyrme a construit
une th\'eorie o\`u le champ fondamental est le pion et o\`u les baryons sont
les solutions de type soliton topologique. Curieusement ses
travaux n' ont pas \'et\'e remarqu\'es en leur temps et ils n' ont \'et\'e
remis en vogue
que r\'ecemment.

\noindent L' id\'eal serait de d\'eriver \`a partir de la QCD une
th\'eorie effective de m\'esons. L' impossibilit\'e d' effectuer une
telle  d\'erivation \`a l' heure actuelle laisse une certaine libert\'e pour le
choix de la forme sp\'ecifique du Lagrangien effectif de la QCD.
Certes, ce dernier doit traduire les propri\'et\'es essentielles de
la QCD, comme la brisure spontan\'ee de la sym\'etrie
chirale etc..., mais ceci est encore tr\`es g\'en\'eral. Une
possibilit\'e est de se laisser
guider par la ph\'enom\'enologie bien connue des m\'esons afin de
r\'eduire le degr\'e d' arbitraire dans le choix du Lagrangien effectif,
 en respectant toujours les sym\'etries essentielles des interactions fortes.

L' objet de l' \'etude pr\'esent\'ee dans cet
article est d' examiner \`a quel point une th\'eorie effective de
m\'esons a des chances r\'eelles de fournir en m\^eme temps une description
quantitative et
physiquement coh\'erente des observables baryoniques via le concept des
solitons topologiques.

\noindent Ce concept est bas\'e sur l' existence d' un type de
 solutions pour des \'equations classiques non-lin\'eaires,
localis\'ees dans l' espace. La diff\'erence
entre ces solutions et les paquets d' onde usuels de la m\'ecanique quantique
est que les solitons, contrairement aux paquets d' onde, gardent
leur \'energie dans une r\'egion finie de l' espace. Cette r\'egion ne
s' \'etale pas au cours du temps et elle  \'evolue comme un objet en
entier, le soliton. Ces solutions ont \'et\'e trouv\'ees
d\`es  le si\`ecle dernier [4]-[5]\footnote\dag{Le nom soliton a \'et\'e pour
la
premi\`ere fois employ\'e par les auteurs de [6]}. Quelques unes d' entre
elles poss\`edent la propri\'et\'e remarquable d' \^etre obtenues par
une superposition alg\'ebrique de solutions lin\'eaires [7].
Un exemple  d' \'equation de ce type qui a \'et\'e utilis\'ee pour la
description
de particules \'etendues en $1+1$ dimensions, est l' \'equation  de Sine-Gordon
[8]. Ses solutions ont des propri\'et\'es topologiques non-triviales qui
r\'esultent des
conditions aux limites dans l' espace. Il appara\^\i t alors une
charge conserv\'ee que l' on peut identifier au nombre
baryonique. La conservation de cette charge repose sur la topologie de
la solution et caract\'erise le soliton [9].

\noindent Alors que les particules des th\'eories quantiques usuelles
sont les quanta \'el\'ementaires du champ, les solitons pr\'eservent leur
caract\`ere de particules m\^eme \`a la limite classique $\hbar \to 0$.
La quantification naturelle de ces solutions peut s' appuyer sur des
m\'ethodes semiclassiques [10]-[11]. De plus il a \'et\'e montr\'e
que dans certains cas la fonction d' onde du soliton d' une
th\'eorie de champs de bosons, peut avoir un comportement fermionique [12].
Cet aspect remarquable, montre qu' il n' est pas d\'eraisonnable d' esp\'erer
d\'ecrire
un spectre fermionique  (les baryons) \`a partir d' un Lagrangien
fondamental de bosons (les m\'esons).

La suite sera organis\'ee de la fa\c con suivante:
dans le chapitre I, nous proposons un mod\`ele g\'en\'eralisant le
mod\`ele original de Skyrme. Ce mod\`ele comprend outre le pion, les m\'esons
les
plus l\'egers. Il est bas\'e sur le
mod\`ele $\sigma$ lin\'eaire dans lequel les m\'esons vecteurs sont
introduits comme des champs de jauge non-ab\'eliens. La sym\'etrie de jauge est
bris\'ee
ensuite par des termes ph\'enom\'enologiques de masse. Le Lagrangien sera
construit de fa\c con \`a respecter les sym\'etries importantes des
interactions fortes. Il d\'epend, comme on le verra, de plusieurs
param\`etres qui, conform\'ement \`a l' esprit de notre approche, vont
\^etre fix\'es par la physique des m\'esons. Nous nous int\'eressons
ensuite aux solutions de type soliton de la th\'eorie, qui
vont \^etre identifi\'ees aux baryons. Pour savoir
dans quelle mesure cela  est raisonnable, il faut tester le mod\`ele sur
des observables bien connues
exp\'erimentalement de la physique des baryons.
Nous nous sommes essentiellement concentr\'e sur l'
interaction nucl\'eon-nucl\'eon que nous d\'erivons de l' interaction
soliton-soliton. A l' issue de cette \'etude nous saurons plus pr\'ecis\'ement
si notre g\'en\'eralisation du mod\`ele de Skyrme peut \^etre
consid\'er\'ee comme un bon candidat pour la th\'eorie unifi\'ee des
m\'esons et des baryons aux basses \'energies.

Comme nous l' avons mentionn\'e plus haut les solitons sont des objets
semi-classiques.
Il s' av\`ere important alors d' estimer les fluctuations quantiques \`a
leur masses. D' une mani\`ere g\'en\'erale nous pensons que les chances de
r\'eussite d' une th\'eorie effective peuvent \^etre mesur\'ees dans le
cadre de l' approximation semi-classique. Les premi\`eres corrections
quantiques
aux observables (par exemple la masse) du soliton sont particuli\`erement
importantes \`a cet \'egard. Le chapitre II est consacr\'e \`a l' \'etude de
l' \'energie d\^ue aux fluctuations du vide autour des solutions classiques. Le
mod\`ele de Skyrme y est analys\'e dans cet optique afin de d\'eterminer s' il
est coh\'erent avec le d\'eveloppement semi-classique. Si ce n' est pas le cas,
nous verrons dans quelle direction il faudrait chercher \`a le g\'en\'eraliser.

Nous avons \'et\'e aussi int\'eress\'es par le m\'ecanisme de stabilisation du
soliton par un m\'eson $\rho$. Ce probl\`eme n' a \'et\'e que partiellement
trait\'e par
le pass\'e, donnant naissance ainsi \`a des ambiguit\'es sur le r\^ole jou\'e
par ce
m\'eson dans la stabilit\'e du soliton. Dans le chapitre III nous
allons \'elucider ce probl\`eme en mettant l' accent sur l' importance
des lois de transformation chirales du m\'eson $\rho$. Nous proposons un
nouveau
lagrangien minimal pour le syst\`eme $\pi\rho$ et nous \'etudions la
stabilit\'e
du soliton correspondant.

Tirant le\c con de nos travaux nous allons esquisser une conclusion concernant
la tentative d' interpr\'eter les baryons comme les solitons topologiques d'
une
th\'eorie de m\'esons et proposer des directions de recherche pour am\'eliorer
notre connaissance actuelle du probl\`eme.

\eject

\rightline {}
\rightline {}
\rightline {}
{\rightline{}}
{\rightline{}}
\bigskip
\bigskip
\bigskip
\bigskip
\centerline {\bfmag Chapitre I}
\bigskip
\bigskip
\bigskip
\centerline {\bfmagd Un mod\`ele des interactions fortes}
\bigskip
\centerline {\bfmagd \`a basses \'energies}
\eject

Comme nous l' avons mentionn\'e dans l' introduction, Skyrme a propos\'e,
il y a trente ans, un mod\`ele unifiant les m\'esons et les baryons.
Dans ce mod\`ele le Lagrangien de d\'epart ne contient que des m\'esons
et les baryons appara\^\i ssent comme des solitons topologiques. Plus
r\'ecemment la recherche d' un sch\'ema d' approximation de la QCD
non-perturbative a fourni des arguments confirmant l' id\'ee de Skyrme.

\noindent D' abord t' Hooft [13] a montr\'e qu' il existe un param\`etre
de d\'eveloppement non-trivial dans les th\'eories de jauge
non-ab\'eliennes: le nombre
de couleurs. Quand ce param\`etre devient grand QCD se r\'eduit \`a une
th\'eorie effective de m\'esons faiblement coupl\'es entre eux. Dans les
th\'eories \`a couplage faible il existe parfois des solutions de type
monopole ayant une masse inversement proportionnelle \`a la constante
de couplage. Witten [14] a appliqu\'e cette id\'ee au cas de la QCD pour
\'emettre la conjecture que les baryons peuvent \^etre consid\'er\'es
comme des solitons \`a la limite o\`u $N_c$ devient infini.

Dans ce chapitre nous allons suivre la conjecture de
Witten et essayer de construire un Lagrangien effectif
bas\'e sur notre connaissance ph\'enom\'enologique de la physique des m\'esons.
Notre motivation principale est la construction d' un mod\`ele
th\'eorique simple et coh\'erent pour la description unifi\'ee de la
physique des hadrons aux basses \'energies.
Dans la section 1 nous pr\'esentons le mod\`ele de Skyrme et
r\'esumons son pouvoir pr\'edictif
dans le secteur des baryons tel qu' il appara\^\i t dans la
litt\'erature. La section 2 mettra en evidence la n\'ec\'essit\'e
de g\'en\'eraliser ce mod\`ele primitif pour rendre compte de la
ph\'enom\'enologie des hadrons. Nous allons proposer ensuite (section 3) un
mod\`ele bas\'e sur la physique des m\'esons les plus l\'egers:
$\pi-\rho-\omega-A_1-\epsilon$. Les param\`etres du mod\`ele vont \^etre
fix\'es sur les observables du secteur des m\'esons dans la section 4.
Les solutions classiques de type soliton repr\'esentant les baryons sont
recherch\'ees dans la section 5. L' interaction
baryon-baryon  est calcul\'ee dans la section 6. Le chapitre se terminera par
la section
7 dans laquelle nous allons pr\'esenter les r\'esultats et nos conclusions sur
le
mod\`ele propos\'e ici.

\bigskip

\bigskip

\leftline{\bfmagc 1. Les solitons du mod\`ele de Skyrme}

\bigskip

Le Lagrangien effectif le plus simple formul\'e en termes de pions et
r\'ealisant la brisure spontan\'ee de la sym\'etrie chirale est celui du
mod\`ele $\sigma$ non-lin\'eaire:

$$\eqalign{
{\cal{L}}_{\sigma NL}={{f_\pi^2}\over 4}\tr(\partial_\mu U\partial^\mu
U^\dagger)\cr}\eqno(1.1)
$$
\medskip

\noindent $f_\pi$ est la constante de
d\'esint\'egration du pion exp\'erimentalement bien d\'etermin\'ee
($f_\pi=0.093$ GeV). La matrice
$U=exp(\displaystyle{i{{\vec\tau.\vec\pi}\over{f_\pi}}})$ contient le
champ du pion $\vec\pi$ et est un \'el\'ement du groupe
$SU(2)$.  La cons\'equence imm\'ediate de l' unitarit\'e de la matrice
$U$ est la pr\'esence d' un courant conserv\'e,

$$\eqalign{
B^\mu={{\epsilon^{\mu\nu\alpha\beta}}\over{24\pi^2}}\tr(\partial_\nu U
U^\dagger\partial_\alpha U U^\dagger\partial_\beta U
U^\dagger)\cr}\eqno(1.2)
$$
\medskip

\noindent Il est clair sur cette formule m\^eme que
$B^\mu$ satisfait \`a une loi de conservation $\partial_\mu
B^\mu=0$ qui ne r\'esulte pas  d' une sym\'etrie du Lagrangien. Ce
courant est identifi\'e au courant baryonique.

\noindent Il a \'et\'e montr\'e dans la r\'ef. [14] que  si
le terme de Wess-Zumino est ajout\'e au
Lagrangien (1.1) les solitons de ce mod\`ele ont les
nombres quantiques des baryons.

\noindent En fait
pour le cas des deux saveurs, qui est celui que l' on
consid\'erera, le terme d' anomalie est nul si l' on consid\`ere
une th\'eorie de pions uniquement. Mais il a \'et\'e montr\'e dans la
r\'ef. [15] que le courant baryonique (1.2) m\^eme dans le
cas de $SU(2)$ peut \^etre retrouv\'e en consid\'erant la contribution
des boucles de fermions au courant $U(1)$.

\medskip

\noindent Cependant le mod\`ele (1.1) ne poss\`ede pas de solutions statiques
d' \'energie finie car il est instable par rapport aux dilatations [16].
Des solutions classiques stables existent uniquement en la
pr\'esence  de termes d' ordre sup\'erieur. Le premier candidat
sugg\'er\'e pour la description des baryons comme des solitons d' une
th\'eorie o\`u les m\'esons sont les champs \'el\'ementaires est le
mod\`ele de Skyrme, dont le Lagrangien est de la forme:

$$\eqalign{
{\cal{L}}_{SK}=&{\cal{L}}_{\sigma NL} +
{1\over {32e^2}}\tr([\partial_\mu U U^\dagger, \partial_\nu U
U^\dagger][\partial^\mu U U^\dagger, \partial^\nu U
U^\dagger])\cr}\eqno(1.3)
$$
\medskip

\noindent $e$ \'etant un param\`etre sans dimension. En d\'eveloppant le
champ $U$ en puissances du champ du pion

$$\eqalign{
U=&\exp{(i\vec\tau.\vec\pi(x))}\cr
=&1+i\vec\tau\vec\pi(x)-{{\vec\pi^2}\over 2}+ \ ... \cr}\eqno(1.4)
$$
\medskip

\noindent on obtient le secteur \`a charge topologique nulle.
Les diff\'erents termes obtenus en rempla\c cant l' expression (1.4) dans (1.3)
d\'ecrivent les interactions dans le
 secteur des pions. Le secteur des
 baryons correspond \`a des configurations non-triviales dans
lesquelles le champ $U$ pointe radialement dans l' espace de configuration
et dans l' espace d' isospin:

$$\eqalign{
U=&\exp{(i\vec\tau.\hat r F(r))}\cr
=&\cos F(r) +i(\vec\tau.\hat r )\sin F(r)\cr}\eqno(1.5)
$$
\medskip

\noindent  $\hat r$ \'etant le vecteur unitaire radial dans l' espace
des coordonn\'ees. Cette configuration est commun\'ement appel\'ee
h\'erisson (hedgehog) pour des raisons \'evidentes. Les solutions
classiques de charge baryonique $n$ , $n$ \'etant un nombre entier,
doivent verifier les conditions aux
limites $F(0)=n \pi$ et $F(\infty)=0$. On peut en fait montrer que
la charge topologique du soliton est:

$$\eqalign{
Q=\int d^3 x B_0 [F(r)]={1\over \pi}(F(0)-F(\infty))=n.\cr}\eqno(1.6)
$$
\medskip

\noindent La fonction \lq\lq chirale" $F(r)$ est trouv\'ee en extr\'emisant
l' \'energie statique du soliton pour une valeur donn\'ee de $n$. Cette
extr\'emisation consiste \`a r\'esoudre les \'equations du mouvement
classiques. L' objet qui en r\'esulte ne peut  cependant pas encore \^etre
identifi\'e aux baryons de la Nature (nucl\'eons, deltas etc..), car la
configuration
unitaire $U$ (eq. (1.5)) est d\'eg\'en\'er\'ee en  spin et en isospin.
Un proc\'ed\'e de quantification de ces degr\'es de libert\'e
consiste \`a effectuer une rotation de la solution $U$ dans l' espace de spin
et d' isospin et ensuite la projeter
sur les \'etats propres de spin et d' isospin.
Ce  proc\'ed\'e a \'et\'e utilis\'ee par les auteurs de [17], qui ont
ainsi calcul\'e les propri\'et\'es statiques du  secteur B=1 (nucl\'eon,
delta). Leurs r\'esultats montrent
 que les baryons du mod\`ele de Skyrme sont assez proches de ceux observ\'es
dans la nature, l' accord \'etant de $\approx$30\%.

Cependant, dans leur approche la constante de d\'esint\'egration du pion
$f_\pi$ a \'et\'e prise \'egale \`a $\approx 0.06$ $GeV$,
largement inf\'erieure \`a la
valeur experimentale. Si on veut pr\'eserver la description de la
physique des m\'esons au niveau du Lagrangien effectif on ne peut pas se
donner la libert\'e de modifier la valeur experimentale de $f_\pi$. On
comprend pourquoi la conclusion principale de plusieurs
auteurs\footnote\dag{qui
ult\'erieurement ont remis $f_\pi$ \`a sa valeur experimentale.} sur le secteur
$B=1$
du mod\`ele de Skyrme est finalement la suivante:

\medskip

La masse du baryon s' av\`ere \^etre trop large dans le mod\`ele de
Skyrme si les param\`etres du Lagrangien sont fix\'es sur la
physique des m\'esons.

\medskip

Une autre observable de la physique des hadrons \`a basses \'energies
tr\`es bien d\'etermin\'ee exp\'erimentalement et fournissant un test
s\'ev\`ere pour les mod\`eles th\'eoriques est l' interaction
nucl\'eon-nucl\'eon. Il est donc naturel de se demander, si le mod\`ele
de Skyrme reproduit les caract\'eristiques bien connues de cette interaction
. Pour ceci, il faut construire le secteur
$B=2$ du Lagrangien (1.3), et le projeter sur les canaux de spin et d'
isospin de l' op\'erateur potentiel $NN$.
Une m\'ethode simple et syst\'ematique a \'et\'e invent\'ee dans [18].
Ces auteurs ont trouv\'e (leurs
r\'esultats sont en accord avec ceux de la r\'ef. [19] qui utilisent une
m\'ethode s\'erieusement plus compliqu\'ee car elle fait appel \`a des
degr\'es de libert\'e de quarks) que
les canaux spin-spin et tenseur de l' interaction entre deux baryons du
Lagrangien (1.4) sont bien reproduits, surtout \`a
longue distance. Par contre, les forces dans le canal central de l' interaction
nucl\'eon-nucl\'eon sont r\'epulsives pour toute s\'eparation $R$ entre les
nucl\'eons,
comme il est montr\'e sur la figure (1.1).

Quand on sait que la liaison des noyaux est
 justement d\^ue \`a une (faible) attraction dans ce canal de l'
interaction, on peut faire la remarque suivante sur
le secteur $B=2$ du mod\`ele de Skyrme

\medskip

Les baryons du mod\`ele de Skyrme ne peuvent pas former des noyaux,
car \`a moyenne port\'ee leurs interactions sont r\'epulsives.

\medskip

Notons au passage que r\'ecemment, un calcul num\'erique [20] a montr\'e qu'
une
attraction \'emerge quand on r\'esoud exactement le syst\`eme $B=2$.
Cette attraction est cependant d' une port\'ee plus grande que celle qui
est n\'ecessaire \`a la coh\'esion des noyaux.

Ces r\'esultats sur la masse du nucl\'eon et l' interaction $NN$ peuvent
conduire \`a penser
que les mod\`eles o\`u les nucl\'eons sont des solitons topologiques ne
sont pas vraiment r\'ealistes. Toutefois, avant de remettre en cause
l' approche toute enti\`ere, on peut se demander si le Lagrangien (1.3)
d\'ecrit de fa\c con r\'ealiste la physique des m\'esons.  Il est \`a cet
\'egard essentiel de pr\'eciser que le terme de Skyrme n' est qu' un terme
parmi
l' infinit\'e de termes qui consituent le d\'eveloppement en
puissances de d\'eriv\'ees du champ du pion. A priori il n' y
pas de raison de s\'electionner ce terme particulier d' ordre quatre, et
non pas l' autre terme d' ordre quatre
(sym\'etrique en d\'eriv\'ees du champ du pion), ou m\^eme des termes d'
ordre six, etc... .

\noindent D' un autre c\^ot\'e, il existe une pl\'eiade de r\'esultats
exp\'erimentaux montrant l' existence de r\'esonances
dans les ondes $S$, $P$ etc. de l' amplitude de diffusion $\pi\pi$.
Par exemple, on sait que l' onde $S$ de la diffusion $\pi\pi$ est
domin\'ee par une r\'esonance autour de $1$ GeV (le m\'eson $\epsilon$).
 Les experiences de diffusion $\pi\pi$ ont rev\'el\'e aussi
l' existence de r\'esonances dans les ondes de spin-1,  bien en dessous de
$1$ GeV: les m\'esons $\rho$ et $\omega$ (de masses $0.77$ GeV et $0.78$
GeV respectivement). Ces m\'esons constituent des p\^oles de l'
amplitude de diffusion et ils  ne peuvent pas \^etre contenus dans des
Lagrangiens locaux comme ceux du type (1.3) d\'efinis en
termes du champ du pion uniquement. Comme \`a l' heure actuelle, il n' existe
pas d' approche syst\'ematique capable d' engendrer dynamiquement des
r\'esonances en th\'eorie des perturbations, un bon point de d\'epart
pour rendre compte du r\^ole de ces r\'esonances dans la physique des
hadrons \`a basse \'energie est de les introduire directement dans le
Lagrangien
effectif.

L' avantage des Lagrangiens effectifs
en termes de m\'esons (pion, m\'eson scalaire, m\'esons vecteurs etc...)
est de fournir une description unifi\'ee des m\'esons et des
baryons comme il a \'et\'e sugg\'er\'e dans [21].

\bigskip

\leftline{\bfmagc 2. G\'en\'eralisations du mod\`ele de Skyrme}

\bigskip

Ces observations sur la n\'ec\'essit\'e de g\'en\'eraliser le
mod\`ele de Skyrme ont conduit
de nombreux auteurs \`a construire des Lagrangiens \`a partir des m\'esons
au-del\`a
du pion, et examiner leurs effets sur les propri\'et\'es
statiques des baryons. On a vite r\'ealis\'e que le terme de Skyrme, aussi
bien qu' un des termes d' ordre six dans le d\'eveloppement chiral,
peuvent \^etre interpr\'et\'es comme les limites locales d' un
Lagrangien contenant les m\'esons $\rho$ [22] et $\omega$ [23] explicitement.
Par exemple le terme d' ordre six qui
n' est rien d' autre que le courant baryonique au carr\'e:

$$\eqalign{
{\cal{L}}_{6}=-bB^\mu B_\mu \ ,\cr}\eqno(2.1)
$$
\medskip

\noindent est la limite du Lagrangien de la r\'ef. [23] quand le
m\'eson $\omega$ devient tr\`es massif (le param\`etre $b$ peut \^etre
reli\'e au couplage $\omega\pi\pi\pi$ comme on verra par la suite).
Une \'etude des effets des m\'esons vecteurs ($\omega,\rho,A_1$)
dans le secteur baryonique a \'et\'e \'efectu\'ee dans [24-25].
Il y a \'et\'e trouv\'e que les propri\'et\'es
statiques des baryons sont sensiblement am\'elior\'ees par rapport au
mod\`ele de Skyrme original (\`a part la masse du baryon qui reste trop
large).

Pour ce qui est des degr\'es de libert\'e scalaires, ils  ont  d' abord
\'et\'es introduits dans le Lagrangien effectif \`a travers un  terme
d' ordre quatre dans le d\'eveloppement chiral, sym\'etrique dans les
d\'eriv\'ees du champ $U$:

$$\eqalign{
{\cal{L}}_{S}=\gamma \big[\tr(\partial_{\mu} U\partial^{\mu}
U^\dagger)\big]^2\cr}\eqno(2.2)
$$
\medskip

La constante $\gamma$ est directement li\'ee aux longueurs de diffusion
$a_0$ et $a_2$ de l' interaction $\pi\pi$. En ce qui concerne les baryons, il
n' est pas difficile de voir sur
cette relation que la contribution \`a la masse du soliton provenant de
${\cal{L}}_{S}$ est n\'egative. Alors que cette observation montre d' abord que
le
seul m\'eson ayant la propri\'et\'e de faire baisser la masse du soliton
au niveau m\^eme de l' approximation locale est bien le m\'eson scalaire,
le signe n\'egatif de cette contribution peut entra\^\i ner une
d\'estabilisation du soliton.  Dans la r\'ef. [26] la valeur critique de
la constante $\gamma$ (au-del\`a
de laquelle les solutions sont instables) est
d\'etermin\'ee aussi bien que les propri\'et\'es statiques des baryons
dans un r\'egime de $\gamma$ inoffensif pour la stabilit\'e. Il y a
\'et\'e trouv\'e que les pr\'edictions pour la masse du baryon sont
am\'elior\'ees
par la seule inclusion de ce terme. Par ailleurs, une
extension \lq\lq non-locale" de ce Lagrangien, de fa\c con \`a inclure les
effets du m\'eson scalaire quand ce dernier est de masse finie a \'et\'e
\'etudi\'ee dans [27]\footnote\dag{o\`u le m\'eson $\omega$ est aussi
pr\'esent} dans le cadre de l' interaction
nucl\'eon-nucl\'eon. Des forces attractives appara\^\i ssent dans le
canal central de cette interaction, mais elles sont de longue port\'ee.
A moyenne port\'ee la r\'epulsion, bien que diminu\'ee par rapport \`a
celle du  mod\`ele de Skyrme, persiste. En fait l' inclusion
du champ scalaire dans cette approximation n' a que peu d' effets sur le champ
du $\omega$ qui est le principal responsable de la r\'epulsion. Dans la
section 3, on \'etudiera  une fa\c con d' inclure le champ scalaire qui
aura aussi des effets non-triviaux sur le champ du $\omega$.

\noindent Il faut aussi pr\'eciser ici que la pr\'esence du m\'eson
scalaire n' est pas seulement dict\'ee par la
ph\'enom\'enologie m\'esonique. Elle peut aussi traduire, au niveau du
Lagrangien effectif, l'
anomalie d' \'echelle de la QCD [28]. Dans ce contexte,  l'
interaction nucl\'eon-nucl\'eon acquiert
des contributions attractives dans son canal central, mais l\`a encore
cette attraction est de longue port\'ee [29].

Nous allons maintenant construire un Lagrangien effectif contenant
les m\'esons les plus l\'egers ($\pi, \ \rho, \ A_1, \omega$ et le m\'eson
scalaire
$\epsilon$), d' une fa\c con qui respecte la sym\'etrie chirale, les
anomalies du secteur m\'esonique et de fa\c con g\'en\'erale
les propri\'et\'es de basse \'energie de ces m\'esons.

\bigskip
\leftline{\bfmagc 3. Un mod\`ele unifiant les m\'esons et les baryons}
\bigskip

Nous allons pr\'esenter dans cette section un mod\`ele [30] qui
constitue une extension du Lagrangien effectif propos\'e par les auteurs de
la r\'ef. [24], de fa\c con \`a inclure des degr\'es de lib\'ert\'e scalaires.
En gardant les motivations des auteurs de cet article, on examinera
les effets de l' inclusion de la r\'esonance scalaire du canal $S$
de la diffusion $\pi\pi$, sur le secteur des
m\'esons et surtout sur la physique des baryons.

\noindent Le Lagrangien sera divis\'e en deux parties, celle qui correspond au
secteur $SU(2)\otimes SU(2)$ d\'ecrivant la physique des
interactions des m\'esons $\pi,\epsilon,\rho,A_1$ \`a laquelle on ajoutera la
partie d\'ecrivant les int\'eractions du pion dans le
secteur $U(1)$ avec le m\'eson $\omega$. Dans cette deuxi\`eme partie
nous allons inclure
les couplages anormaux du pion aux m\'esons vecteurs par l'
interm\'ediaire du terme d'
anomalie de Wess-Zumino. Ecrivons d' abord ce Lagrangien,
on verra ensuite comment y arriver:

$$\eqalign{
{\cal{L}}={\cal{L}}_{\pi\xi\rho A_1}+{\cal{L}}_{\pi\omega}\cr}\eqno(3.1)
$$
\medskip

\noindent avec $\Lgls$ donn\'e par:

$$\eqalign{
\Lgls &={1\over 2}\dmudo \xi \dmuup \xi+{\xi ^2\over
4}\tr( D_{\mu }UD^{\mu }U^\dagger )-\lambda (\xi ^2-\Gamma ^2)^2-{1\over 8}
\tr(X_
{\mu \nu }^2+Y_{\mu \nu}^2)\cr
&+{m_{\rho }^2\over 4}\tr (X_{\mu
}^2+Y_{\mu }^2)+{{\Delta m^2}\over 8}\tr (X_{\mu }^2+Y_{\mu }^2-2U^\dagger
X_{\mu }UY^{\mu })\cr}\eqno(3.2)
$$
\medskip

\noindent et ${\cal{L}}_{\pi\omega}$ par l' expression:

$$\eqalign{
{\cal L}_{\pi\omega}&=-{1\over 4}\omega_{\mu\nu}^2+{m_{\omega}^2\over
2}\omega_{\mu}^2+ \beta_{\omega}\omega_{\mu}B^{\mu}\cr
&-i{\beta_{\omega}g\over{8\pi^2}}\leviup
\dmudo\omega_{\nu}\tr(X_{\alpha}L_{\beta}-Y_{\alpha}R_{\beta}
+ig(UY_{\alpha}U^\dagger X_{\beta}-Y_{\alpha}X_{\beta}))\cr}\eqno(3.3)
$$
\medskip

\noindent Dans les expressions (3.2) et (3.3) les champs  $\xi$ and
$\omega_\mu$
repr\'esentent le champ scalaire et celui du $\omega$ respectivement.
Nous avons utilis\'e les d\'efinitions suivantes:

$$\eqalign{
U&=e^{{i\over f_\pi}\vec \tau.\vec \pi}\cr
D_{\mu }U&=\dmudo U+ig(X_{\mu }U-UY_{\mu })\cr
X_{\mu \nu }&=\dmudo X_{\nu}-\dnudo X_{\mu}+ig[X_{\mu},X_{\nu}]\cr
Y_{\mu \nu }&=\dmudo Y_{\nu}-\dnudo Y_{\mu}+ig[Y_{\mu},Y_{\nu}]\cr
B^{\mu}&={\leviup\over {24\pi^2}}\tr(L_{\nu}L_{\alpha}L_{\beta})\cr
L_{\mu}&=\dmudo UU^\dagger \cr
R_{\mu}&=\dmudo U^\dagger U\cr}\eqno(3.4)
$$
\medskip

\noindent o\`u $\vec\pi$ est le champ du pion, et les champs
$X_\mu=\vec\tau.\vec X_\mu$
et $Y_\mu=\vec\tau.\vec Y_\mu$ sont des champs d' isospin unit\'e de
chiralit\'e gauche et droite respectivement.
On verra dans la section suivante comment ces champs gauche et droit
sont reli\'es au champs des m\'esons $\rho$ et $A_1$.  Regardons maintenant
de plus pr\`es ce Lagrangien.

a) Le Lagrangien $\Lgls$

\noindent La premi\`ere partie du Lagrangien de l' \'equation (3.2) est
bas\'ee sur le mod\`ele $\sigma$ lin\'eaire. Ce mod\`ele d\'ecrit
les interactions entre le \lq\lq pion" $\vec\phi$ et son partenaire
chiral, le champ $\sigma$. Il est commode de param\'etriser ces deux
champs de fa\c con compacte par un champ matriciel
${\cal{V}}=\sigma(x)+i\vec\tau.\vec\pi(x)$.
Alors la densit\'e Lagrangienne
du mod\`ele $\sigma$ lin\'eaire s' \'ecrit:

$$\eqalign{
{\cal{L}}={1\over 4}\tr(\dmudo {\cal V}\dmuup {\cal V}^\dagger)
-\lambda
\left(
{1\over 2}\tr ({\cal V}{\cal V}^\dagger)-\Gamma^2
\right)^2\cr}\eqno(3.5)
$$
\medskip

\noindent Il est clair sur cette expression que la forme
bilin\'eaire $\tr({1\over 2}{\cal{V}}{\cal{V}}^\dagger)$ do\^\i t atteindre la
valeur $\Gamma^2$ pour minimiser l' \'energie potentielle. La  constante
$\Gamma$ est reli\'ee \`a la constante de d\'esint\'egration du
pion $f_\pi$.

\noindent R\'eecrivons maintenant le champ du quaternion
${\cal{V}}$ de fa\c con a en extraire son contenu  isoscalaire $\xi$:

$$\eqalign{
{\cal V}=\xi(x)U(x)=\xi(x)\exp
\left(
{{i\over f_\pi}\vec\tau .\vec\pi(x)}
\right)\cr}\eqno(3.6)
$$
\medskip

\noindent o\`u $\xi(x)$ est d\'efini par
$\xi^2(x)=\sigma^2(x)+\vec\pi^2(x)$, cette d\'efinition manifestant son
caract\`ere scalaire par une rotation chirale. L' \'equation (3.5) est mise
maintenant sous la forme:

$$\eqalign{
{\cal{L}}={1\over 2}\dmudo \xi \dmuup \xi +{\xi^2\over 4}\tr(\dmudo U\dmuup
U^\dagger)-\lambda(\xi ^2-\Gamma ^2)^2\cr}\eqno(3.7)
$$
\medskip

\noindent Le mod\`ele $\sigma$ non-lin\'eaire
peut \^etre retrouv\'e lorsque le param\`etre $\lambda$ tend
vers l' infini. En effet, quand $\lambda\to\infty$, $\xi$ doit tendre
vers la valeur $\Gamma$ pour que l' \'energie potentielle dans
l' \'equation (3.7) soit non-infinie.

\noindent Il faut remarquer que cette \'ecriture du mod\`ele $\sigma$
lin\'eaire simplifie consid\'erablement le formalisme, car les champs
$U$ et $\xi$ ont des propri\'et\'es de transformation simples pour une
rotation chirale.

En effet, la loi de transormation des champs
$U$ et $\xi$ par une rotation $SU(2)_L\otimes SU(2)_R$ est:

$$\eqalign{
U& \ \to A(x) \ U \ B^\dagger (x)\cr
\xi& \ \to \xi\cr}\eqno(3.8)
$$
\medskip

\noindent avec les matrices $A(x)\in SU(2)_L$ et $B(x)\in SU(2)_R$.
Il suffit de rendre maintenant le Lagrangien (3.7) invariant sous une
transformation chirale locale. La substitution des gradients
$\partial_\mu U$ par les gradients covariants $D_\mu U$ d\'efinis
plus haut, assure l' invariance locale de l' action. Au Lagrangien obtenu
il faut maintenant ajouter les termes cin\'etiques des champs de
jauge gauche et droit, et le r\'esultat sera invariant sous les
transformations (3.8) pour $U$ et $\xi$ et

$$\eqalign{
X_\mu& \ \to AX_\mu A^\dagger-{i\over g}A\partial_\mu A^\dagger\cr
Y_\mu& \ \to BY_\mu B^\dagger-{i\over g}B\partial_\mu B^\dagger\cr}\eqno(3.9)
$$
\medskip

\noindent pour les
champs $X_\mu$ et $Y_\mu$.
La sym\'etrie locale $SU(2)_L\otimes SU(2)_R$ est ensuite
bris\'ee de fa\c con minimale en ajoutant des termes de masse
pour les champs gauche et droit.
Remarquons ici que le m\'eson scalaire introduit par le
biais du mod\`ele sigma lin\'eaire, donne lieu \`a une contribution positive
\`a l' \'energie comme l' \'equation (3.7) le montre.
Un autre avantage de jauger le groupe $SU(2)_L\otimes SU(2)_R$
est d' avoir une description
sym\'etrique des m\'esons $\rho$ et $A_1$, car ils sont d\'ecrits en tant que
partenaires chiraux.

b) Le Lagrangien ${\cal{L}}_{\pi\omega}$

\noindent L' \'equation (3.3) d\'ecrit le secteur $U(1)$ des interactions
m\'esoniques aux basses \'energies. Le premier terme dans cette \'equation
d\'ecrit le couplage du $\omega$ aux trois pions dans le secteur
des m\'esons, et son couplage au courant baryonique. Ensuite vient  le
terme d' anomalie que l' on doit inclure, pour tenir compte des anomalies du
s\'ecteur $U(1)$ [31]. Mis \`a part ces consid\'erations dans
le secteur des m\'esons, le couplage du $\omega$ au courant baryonique
est essentiel pour la stabilit\'e des solutions classiques [32] dans notre
cas o\`u le champ chiral se couple \`a des champs de
Yang-Mills massifs. Si on adopte ce point de vue, la pr\'esence du $\omega$
dans le Lagrangien effectif est donc d' une importance capitale pour la
physique des baryons (stabilit\'e) aussi bien que pour celle des
m\'esons (anomalies).

\noindent En fait, ces anomalies sont d\^ues
 \`a  la pr\'esence d'
une sym\'etrie discr\`ete que le mod\`ele
sigma (non)-lin\'eaire poss\`ede, et qui n' est pas
observ\'ee dans la nature. Cette sym\'etrie \lq\lq redondante" interdirait des
processus o\`u le nombre des bosons n' est pas conserv\'e [14]. Le terme de
Wess-Zumino effectif, assure que l' on tient
compte de ces processus dans le secteur m\'esonique du Lagrangien
effectif, car il brise cette sym\'etrie. La d\'erivation du terme d'
anomalie quand les champs de jauge sont pr\'esents,
peut \^etre trouv\'ee dans la litt\'erature [33].

On n' a pas consid\'er\'e le couplage direct des degr\'es de libert\'e
scalaires au m\'eson $\omega$. Ces couplages ont \'et\'e \'etudi\'es
dans [34], pour la photoproduction de pions. Les r\'esultats
montrent que les couplages du type $\xi^2\omega^2_\mu$ engendrent un
\'etat li\'e dans l' onde S, en d\'esaccord avec la
ph\'enom\'enologie.

On va s' int\'eresser maintenant aux propri\'et\'es de transformation du
Lagrangien
(3.1). Celles-ci peuvent \^etre
utilis\'ees pour calculer les courants de Noether associ\'es \`a des
rotations de $SU(2)_L$ et $SU(2)_R$ aussi bien que celles de $U(1)$.

\noindent Effectuons des transformations
locales sur les champs du Lagrangien (3.1)

$$\eqalign{
U  & \to U+iQ_L U\cr
X_\mu & \to X_\mu+ i[Q_L,X_\mu]-{{\partial_\mu Q_L}\over g} \ \ \ \
Q_L={{\vec\epsilon_L(x).\vec\tau}\over 2}\cr
Y_\mu & \to  Y_\mu\cr}\eqno(3.10a)
$$
\medskip

\noindent pour $SU(2)_L$ et

$$\eqalign{
U \ & \to U-i UQ_R\cr
Y_\mu & \to Y_\mu+ i[Q_R,Y_\mu]-{{\partial_\mu Q_R}\over g} \ \ \ \
Q_L={{\vec\epsilon_R(x).\vec\tau}\over 2}\cr
X_\mu & \to  X_\mu\cr}\eqno(3.10b)
$$
\medskip

\noindent pour $SU(2)_R$. $\vec\epsilon_L(x)$ et $\vec\epsilon_R(x)$ sont des
fonctions arbitraires. La variation du Lagrangien satisfait alors les
relations suivantes (\`a une int\'egration par parties pr\`es):

$$\eqalign{
\delta_{SU(2)_L}{\cal{L}}=&-(\partial_\mu \vec
j^\mu_L)\vec\epsilon_L+(-\vec
j^\mu_L-{{\partial{\cal{L}}}\over{\partial\vec
X_\mu}}{1\over{2g}})\partial_\mu\vec\epsilon_L\cr
\delta_{SU(2)_R}{\cal{L}}=&-(\partial_\mu \vec
j^\mu_R)\vec\epsilon_R+(-\vec
j^\mu_R-{{\partial{\cal{L}}}\over{\partial\vec
Y_\mu}}{1\over{2g}})\partial_\mu\vec\epsilon_R\cr}\eqno(3.11)
$$
\medskip

\noindent avec les courants gauches et
droits $\vec j^\mu_L, \vec j^\mu_R$. En calculant les membres de gauche de ces
\'equations avec
les lois de transformation (3.10) et en combinant les expressions de ces
courants on arrive \`a l' expression des
courants vecteurs et axiaux:

$$\eqalign{
\vec V_{\mu}=&\vec j_\mu^L+\vec j_\mu^R=i\tr\bigg\{\vec\tau\big[{1\over
4}([X^{\nu},X_{\mu\nu}]+[Y^{\nu},Y_{\mu\nu}])+{\xi^2\over
4}(D_{\mu}UU^\dagger +D_{\mu}U^\dagger U)\cr
+&{{\beta_{\omega}}\over
{16\pi^2}}\epsilon_{\mu\rho\nu\alpha}[\omega^{\rho}\dnuup(L^{\alpha}-
R^{\alpha})+ig\partial^{\rho}\omega^{\nu}(-X^{\alpha}+Y^{\alpha}+U^\dagger
 X^{\alpha}U-UY^{\alpha}U^\dagger )]\big]\bigg\},\cr}\eqno(3.12a)
$$

$$\eqalign{
\vec A_{\mu}=&\vec j_\mu^L-\vec j_\mu^R=i\tr\bigg\{\vec\tau\big[{1\over
4}([X^{\nu},X_{\mu\nu}]-[Y^{\nu},Y_{\mu\nu}])+{\xi^2\over
4}(D_{\mu}UU^\dagger -D_{\mu}U^\dagger U)\cr
+&{{\beta_{\omega}}\over
{16\pi^2}}\epsilon_{\mu\rho\nu\alpha}[\omega^{\rho}\dnuup(L^{\alpha}+
R^{\alpha})-ig\partial^{\rho}\omega^{\nu}(X^{\alpha}+Y^{\alpha}+U^\dagger
 X^{\alpha}U+UY^{\alpha}U^\dagger )]\big]\bigg\}\cr}\eqno(3.12b)
$$
\medskip

\noindent On peut montrer que ces courants, toujours
en vertu de l' \'equation (3.11),  satisfont aux \'equations suivantes:

$$\eqalign{
\partial_\mu\vec V^{\mu}=& \ 0\cr
\partial_\mu\vec A^{\mu}=&-{i\over 8}\tr [\vec\tau{g^2\beta_{\omega}\over
\pi^2}\levido\dmuup\omega^{\nu}(X^{\alpha}Y^{\beta}+Y^{\alpha}X^{\beta})]\cr}\eqno(3.13)
$$
\medskip

\noindent Ces relations, t\'emoignent de la conservation du courant
vectoriel et de la non-conservation du courant axial, cette derni\`ere
\'etant d\^ue \`a l' anomalie. Elles sont bien connues  et v\'erifi\'ees
par l' exp\'erience. Pour retrouver PCAC, il faut ajouter
un terme de masse du pion comme il a \'et\'e fait dans [24]. Il est utile
d' observer aussi  que dans le  terme qui couple le $\omega$ aux
champs $\pi$, $X$ et $Y$ (\'equation (3.3)) le contreterme de Bardeen
(proportionnel \`a $g^2$)
assure  que l' anomalie subsiste uniquement dans le
courant axial (\'equation (3.13)). Cette soustraction a aussi pour effet de
briser explicitement la sym\'etrie chirale.

Le courant U(1) d\'eduit du Lagrangien (3.1) est identifi\'e au courant
baryonique. Son expression est:

$$\eqalign{
 {\cal J}^{\mu}_{I=0}={B^{\mu}\over 2}-{ig\over
{16\pi^2}}\leviup\tr[\dnudo (X_{\alpha}L_{\beta}-Y_{\alpha}R_{\beta}
+ig(UY_{\alpha}U^\dagger X_{\beta}-Y_{\alpha}X_{\beta}))]\cr}\eqno(3.14)
$$
\medskip

\noindent Ce courant est normalis\'e \`a la moiti\'e du nombre
baryonique et satisfait \`a la relation $\dmudo {\cal
J}^{\mu}_{I=0}=0$, car comme on a vu dans la section 1 $B_\mu$ est
conserv\'e et la contribution du terme de Wess-Zumino \`a (3.14) est
une divergence totale.

\noindent Notre Lagrangien de m\'esons
poss\`ede une structure conforme aux propri\'et\'es g\'en\'erales
des interactions fortes, nous pouvons maintenant examiner sa limite locale,
c' est \`a dire les termes du d\'eveloppement chiral en puissances de
d\'eriv\'ees du champ du pion qui en r\'esultent lorsque les autres m\'esons
($\rho,\omega,A_1,\epsilon$) deviennent tr\`es massifs. Nous avons d\'ej\`a
remarqu\'e que quand le
param\`etre $\lambda$ (qui est proportionnel au carr\'e de la  la masse du
m\'eson scalaire) devient tr\`es grand on
retrouve le mod\`ele $\sigma$ non-lin\'eaire.
Pour ce qui est de la contribution des m\'esons vecteurs,
on peut montrer (Appendice C) en utilisant les equations d'
Euler-Lagrange pour les champs
$X_\mu$,$Y_\mu$ et $\omega_\mu$, qu' en l' absence du terme de
Wess-Zumino et quand les masses $m_\omega,m_\rho,m_{A_1}$
deviennent tr\`es grandes le Lagrangien se r\'eduit \`a:

$$\eqalign{
{\cal{L}} \ \to {\cal{L}}_{2+4+6}={{f_\pi^2}\over 4}\tr(\partial_\mu
U\partial^\mu U^\dagger)
+{1\over {64g^2}}\tr([\partial_\mu U U^\dagger, \partial_\nu U
U^\dagger]^2)-{{\beta_\omega^2}\over{2m_\omega^2}}B_\mu B^\mu\cr}\eqno(3.15)
$$
\medskip

\noindent On obtient donc les premiers termes du d\'eveloppement
chiral avec l' absence du terme d' ordre quatre sym\'etrique en
d\'eriv\'ees du champ du pion. Cette propri\'et\'e est la bienvenue car
elle emp\^eche la d\'estabilisation du soliton comme il a \'et\'e vu
plus haut. Elle est une cons\'equence de l'
introduction du champ scalaire par le biais du mod\`ele sigma lin\'eaire:
seuls les termes ayant une contribution
positive \`a l' \'energie sont pr\'esents.

On a donc r\'epondu \`a la demande des sections 1 et 2 en construisant
un Lagrangien qui g\'en\'eralise le mod\`ele de
Skyrme. Pour savoir si ce mod\`ele a des chances d' \^etre un bon
mod\`ele effectif, on doit d' abord \'examiner les pr\'edictions de ce
Lagrangien pour les observables dans le secteur m\'esonique.

\bigskip
\leftline{\bfmagc 4. Le secteur des m\'esons}
\bigskip

Nous allons nous int\'eresser maintenant aux masses, couplages et
propri\'et\'es des m\'esons dans notre mod\`ele.
Les diff\'erentes observables de ce secteur sont obtenues en
d\'eveloppant le champ $U$ en puissances du champ du pion:

$$\eqalign{
U=&\exp{({i\over {f_\pi}}\vec\tau.\vec\pi(x))}\cr
=&1+{i\over {f_\pi}}\vec\tau.\vec\pi(x)-{1\over{f_\pi^2}}{{\vec\pi^2(x)}\over
2}+ \ ... \cr}\eqno(4.1)
$$
\medskip

\noindent Le champ du m\'eson $\epsilon$ est reli\'e au champ $\xi$ par
la relation $\epsilon=\Gamma-\xi$. En rempla\c cant cette expression
dans l' \'equation (3.2) on trouve la relation pour la masse du scalaire:
$m_\epsilon^2=8\lambda\Gamma^2$.

Consid\'erons maintenant la contribution dans le secteur des m\'esons
des termes qui ne d\'ependent pas du champ scalaire dans l' \'equation (3.1),
et rempla\c cons $U$ par son d\'eveloppement en champ du pion. Il est
facile de voir que si l' on utilise la d\'efinition naive du champ axial
$A^\mu=(X_\mu-Y_\mu)$ on obtient des couplages lin\'eaires avec le
champ du pion :
$(Cte)\partial_\mu\vec\pi.\vec A^\mu$. Pour les faire
dispara\^\i tre, on  doit diagonaliser le
Lagrangien \`a l' ordre le
plus bas. Le r\'esultat de cette diagonalisation aussi bien que de l'
identification de la masse physique du m\'eson $A_1$ est exprim\'e par
les relations suivantes:

$$\eqalign{
\vec{\rho_{\mu}}&={1 \over{\sqrt2}}(\vec {X_{\mu }}+\vec {Y_{\mu }})\cr \vec
{A_{\mu}}&={1 \over{\sqrt2}}(\vec {X_{\mu }}-\vec {Y_{\mu  }}+{2gf_{\pi}\over
{m_{\rho}^2+\Delta m^2}}\dmudo \vec \pi)\cr
m_{A_1}&={{m_{\rho}^2+\Delta m^2}\over
{{(m_{\rho}^2+\Delta m^2-2g^2f_\pi^2)}^{1\over 2}}}\cr}\eqno(4.2)
$$
\medskip

\noindent La pr\'esence des champs de jauge conduit \`a la relation
suivante entre la constante $\Gamma$ et $f_\pi$:
$\displaystyle{{\Gamma^2}=f_{\pi}^2{m_{A_1}^2\over  {m_{\rho }^2+\Delta
m^2}}}$.

En s\'electionnant maintenant les
termes cubiques et quartiques dans les champs, on obtient les
couplages d' interaction forte des m\'esons. Il est aussi possible
d' obtenir les interactions {\'electromagn\'etiques} de ces m\'esons.
Pour cela, il suffit de jauger le groupe $U(1)$-\'electromagnetique
dans l' \'equation (3.1).

\noindent Ces couplages peuvent \^etre utilis\'es pour le calcul d'
observables m\'esoniques au
premier ordre des perturbations. Nous donnons ici l' expression de la largeur
$\Gamma_{A_1\to\epsilon\pi}$ calcul\'ee avec le Lagrangien (3.1), d'
autres expressions pouvant \^etre trouv\'ees dans [24]:

$$\eqalign{
\Gamma_{A_1 \to\pi\epsilon}={8\Gamma^2g^2\over 192\pi m_{A_1}^5f^2_\pi}
\left[1-{2g^2f^2_\pi\over m^2_\rho+\Delta m^2}\right]^2
\left[(m_{A_1}^2+m_\pi^2-m_\epsilon^2)^2-4m_{A_1}^2m_\pi^2\right]^{3\over
2}\cr}\eqno(4.3)
$$
\medskip

\noindent Les observables m\'esoniques, comme l' illustre la relation (4.3)
d\'ependent
des  param\`etres pr\'esents dans le Lagrangien
(3.1). On fixe ces param\`etres en ajustant certaines
de ces observables, ce qui donne:
\medskip

\centerline{$m_\rho=0.769$ GeV}
\centerline{$m_\omega=0.782$ GeV}
\centerline{$g=3.78$}
\centerline{$\Delta m^2=-0.462 m_\rho^2$}
\centerline{$f_\pi=0.093$ GeV}
\centerline{$\beta_\omega=9.3$}

\medskip

\noindent $m_\rho, m_\omega$ ont \'et\'e prises \'egales aux masses
physiques de ces m\'esons. Les valeurs de $g$ et $\Delta m^2$
reproduisent la valeur experimentale de la masse du m\'eson $A_1$
($m_{A_1}=1.194$ GeV) et la largeur $\Gamma_{\rho\to\pi\pi}$.
Celle de $\beta_\omega$ reproduit la largeur \'electromagn\'etique de
d\'esint\'egration du $\omega$ en un pion et un photon. En ce qui
concerne la masse du m\'eson $\epsilon$ (proportionnelle \`a $\sqrt{\lambda}$),
remarquons que ce dernier a une largeur assez grande. Nous avons pr\'ef\'er\'e
pour
cette raison laisser libre $m_\epsilon$, \`a varier entre $0.5$ GeV et $1.0$
GeV.

Une fois les param\`etres fix\'es sur les observables
m\'esoniques, il faut remarquer que le Lagrangien (3.1) peut pr\'edire
d' autres observables
dans le secteur des m\'esons. Ceci a \'et\'e illustr\'e dans la
r\'ef\'erence [24] o\`u le mod\`ele $\sigma$
non-lin\'eaire \'etait \`a l' \'etude (la limite $\lambda\to\infty$ de
notre mod\`ele). Dans le cadre de notre mod\`ele, le choix de la valeur
de $1$ GeV pour la masse du scalaire donne pour la largeur partielle du
$A_1$ (\'equation (4.3)) une valeur de $0.1$ \%, qui est en accord avec
la limite experimentale ($\le 1.0$ \%).

\noindent Le couplage du m\'eson $A_1$ au champ scalaire $\epsilon$ modifie la
constante de couplage  $\epsilon\pi\pi$:
 $\delta_\epsilon={\Gamma\over 2}(1-{{4g^2f^2_{\pi}}\over{m_{\rho}^2+\Delta
m^2}})$. En utilisant cette relation, on peut \'eventuellement donner
une pr\'ediction pour les longueurs de diffusion $\pi\pi$ qui
d\'ependent de $\delta_\epsilon$. Avec $m_\epsilon=1$ GeV on obtient
$a_{0}=0.235$ fm
et $a_{2}=-0.056$ fm, \`a comparer aux valeurs experimentales $a_{0}=0.36\pm
0.07$ fm
et $a_{2}=-0.039\pm 0.017$ fm [35].

En r\'esum\'e, nous avons determin\'e les param\`etres du
mod\`ele $\sigma$ lin\'eaire, de fa\c con \`a  ce qu' il puisse
d\'ecrire correctement la physique des m\'esons  $\pi,\rho,\omega, A_1,
\epsilon$ et
de leurs interactions. Une fois ces
param\`etres fix\'es sur un nombre d' observables \'egal au nombre de
ces param\`etres, il est clair que ce Lagrangien poss\`ede encore un pouvoir de
pr\'ediction important dans le secteur des m\'esons [24].

\noindent Nous allons \'examiner maintenant \`a quel point une th\'eorie
r\'ealiste de m\'esons peut \^etre
utilis\'ee pour d\'ecrire aussi les baryons. Ceux-ci se trouvent dans le
secteur non-trivial du Lagrangien effectif, celui des solitons
topologiques.

\bigskip

\leftline{\bfmagc 5. Le secteur baryonique}

\bigskip

Construisons d' abord les solutions \`a une unit\'e de charge
baryonique. Ces solutions des \'equations du mouvement, sont statiques
et ont une \'energie finie.

\noindent Les composantes $X_0,Y_0,\omega_i=0$
sont nulles \`a la limite statique. On consid\`ere les champs classiques
(h\'erisson) suivants:

$$\eqalign{
U=&\exp{(i\vec\tau.\hat r F(r))},\cr
X_i=&\alpha(r)\tau_i+(\beta(r)-\alpha(r))(\vec\tau \hat r)\hat
r_i+\gamma(r){(\vec\tau \times \hat r)}_i,\cr
Y_i=&-\alpha(r)\tau_i-(\beta(r)-\alpha(r))(\vec\tau \hat r)\hat
r_i+\gamma(r){(\vec\tau \times \hat r)}_i,\cr
\omega_{0}=&\omega(r),\cr
\xi=&\xi(r)\cr}\eqno(5.1)
$$
\medskip

\noindent o\`u les profils $F,\alpha,\gamma,\beta,\omega,\xi$ sont des
fonctions radiales. L' \'energie de la configuration de type
soliton est \'egale \`a $-\int {\cal{L}} d^3 \vec r$. Son expression
est donn\'ee par:

$$\eqalignno{
E&=4\pi\intif r^2dr\bigg\{4({\gamma\over r}-g(\alpha^2+\gamma^2))^2+2(
\dot\gamma+{\gamma\over r}-2g\alpha\beta)^2+2(
\dot\alpha+{{\alpha-\beta}\over r}+2g\beta\gamma)^2\cr
&+{\xi^2\over 2}[(\dot F+2g\beta)^2+2({\sin F\over
r}+2g(\alpha\cos F-\gamma\sin F))^2]+{\dot\xi^2\over
2}+\lambda(\xi^2-\Gamma^2)^2&(5.2)\cr
&+m_{\rho}^2[\beta^2+2(\alpha^2+\gamma^2)]
+\Delta m^2[2(\alpha\cos F-\gamma\sin F)^2+\beta^2]
-{1\over 2}(\dot\omega^2+m_{\omega}^2\omega^2)\cr
&+\beta_{\omega}\omega{\sin ^2 F\over {2\pi^2 r^2}}\dot F
+\beta_{\omega}g{\dot\omega\over {2\pi^2}}[-2{{\alpha\sin^2 F}\over
r}-{{\gamma\sin 2F}\over r}+4g\alpha\gamma\sin^2 F-g\sin
2F(\alpha^2-\gamma^2)]\bigg\}\cr}
$$
\medskip

\noindent Nous allons rendre cette fonctionnelle stationnaire par
rapport aux variations arbitraires des champs
$F,\alpha,\beta,\gamma,\xi,\omega$ en r\'esolvant
les \'equations d' Euler-Lagrange associ\'ees. Ces \'equations
sont des \'equations non-lin\'eaires coupl\'ees.
On trouvera leurs expressions assez
longues et compliqu\'ees dans l' Appendice A. Les configurations de
nombre baryonique unit\'e et d' \'energie finie satisfont aux conditions aux
limites $F(0)=\pi$, $F(\infty)=0$ pour le champ chiral et $\dot\xi(0)=0$,
$\xi(\infty)=\Gamma$ pour le champ scalaire. Celles pour les champs de
jauge r\'esultent des \'equations du mouvement et sont donn\'ees par:
$\dot\alpha(0)=\dot\beta(0)=\dot\omega(0)=0,\ \gamma(0)=0,
\ \alpha(\infty)=\beta(\infty)=\gamma(\infty)=\omega(\infty)=0$. Les solutions
classiques pour les profils $F$ et $\epsilon$ sont pr\'esent\'ees sur la
figure (1.2).

\noindent Les solutions des \'equations du mouvement diff\`erent tant
qualitativement que quantitativement de celles du
mod\`ele non-lin\'eaire. Un des aspects essentiels du couplage
du scalaire au champ chiral est la diminution de la fonction de profil
$F$ \`a l' int\'erieur du soliton comme il est clair sur la figure (1.2). Cet
effet du champ
scalaire a \'et\'e aussi observ\'e dans la r\'ef. [29]. Mais dans
notre mod\`ele, la pr\'esence du champ scalaire influence aussi la solution
classique des champs $\alpha,\beta,\gamma,\omega$, comme il est
manifeste sur la figure (1.3).

\noindent En effet, comme nous pouvons le constater sur cette figure, le
couplage du champ scalaire au champs de jauge  via le terme
$\displaystyle{{{\xi^2}\over 4}\tr(D_\mu U D^\mu U^\dagger)}$ supprime
fortement ces champs au niveau classique (d' un facteur
5 quand la masse du scalaire est \'egale \`a $0.5$ GeV).
Plus \'etonnant encore est
l' effet sur le champ classique du $\omega$. Ce dernier voit sa port\'ee
r\'eduite dans le cas lin\'eaire. Ce ph\'enom\`ene,  est observ\'e
uniquement quand les m\'esons $\rho$ et $A_1$ sont pr\'esents comme
nous l'  ont indiqu\'e des analyses num\'eriques du cas o\`u la
constante $g$ tend vers z\'ero. Ceci montre
l' importance d' inclure les m\'esons de masse finie dans le Lagrangien
effectif.

\noindent Quels sont maintenant les effets sur les propri\'et\'es statiques du
soliton? Pour le savoir, il faut calculer des observables telles que la masse,
la constante de couplage axiale, etc... Nous avons calcul\'e
des quantit\'es qui d\'ependent seulement des solutions classiques des
\'equations du mouvement, c' est \`a dire la masse du soliton, le rayon
isoscalaire du baryon et la constante de couplage axiale. La masse du
soliton est donn\'ee par l' expression (5.2). Pour
ce qui est du rayon isoscalaire, en rempla\c cant les expressions (5.1)
dans celle de la densit\'e baryonique ($2{\cal J}^{\mu}_{I=0}$), on
trouve:

$$\eqalign{
<r^2>_{I=0}&={2\over \pi}\intif r^2dr\bigg\{-\dot F\
sin^2F\hfil\cr
&+4g\sin F[\alpha(1-2g\gamma r)\sin F+\gamma \cos F+gr\cos F
(\alpha^2-\gamma^2)]\bigg\}\cr}\eqno(5.3)
$$
\medskip

\noindent La constante de couplage axiale $g_A$ du nucl\'eon peut \^etre
calcul\'ee par projection directe sur l' \'el\'ement de matrice de la
troisi\`eme composante du courant axial entre  deux \'etats
de nucl\'eon comme il a \'et\'e fait dans [17]. En appliquant le
th\'eor\`eme de la divergence \`a l' \'equation (3.14), on trouve l'
expression suivante:

$$\eqalign{
g_A=-{4\over 3}\pi f_{\pi}^2\lim_{r\to\infty}(r^3\dot
F)-{{4g^2\beta_{\omega}}\over
{3\pi}}\intif r^3(\alpha^2-\gamma^2) \dot\omega dr\cr}\eqno(5.4)
$$
\medskip
\noindent Comme il est illustr\'e sur la figure (1.4) la pr\'esence du
champ scalaire conduit \`a une diminution notable de la masse
classique du soliton, mais aussi de $<r^2>^{1\over 2}_{I=0}$
et de $g_A$ en comparaison avec le mod\`ele non-lin\'eaire.

\bigskip

\leftline{\bfmagc 6. L' interaction nucl\'eon-nucl\'eon}

\bigskip

Pour calculer l' interaction entre deux solitons il est n\'ecessaire d'
approximer la forme
des champs $U,X,Y,\omega,\xi$ dans le secteur $B=2$,
car la forme exacte de ceux-ci dans le cadre de
notre mod\`ele est tr\`es compliqu\'ee \`a d\'eterminer. Une
approximation simple connue sous le nom d' approximation du produit,
offre une description raisonnable de l' interaction entre deux solitons
statiques, c. a. d. deux solutions du secteur $B=1$. Cette approximation
consiste \`a supposer que la configuration des champs, quand les deux
solitons sont s\'epar\'es d' une distance $\vec R$, est un produit de la
forme:

$$\eqalign{
U_0^{B=2}(\vec r,\vec R)=U_0^{B=1}(\vec r-{{\vec R}\over 2})U_0^{B=1}(\vec
r+{{\vec R)}\over 2}\cr}\eqno(6.1)
$$
\medskip

\noindent o\`u $U_0^{B=1}$ est la solution $B=1$ trouv\'ee dans la
section pr\'ec\'edente. Les m\'erites, aussi bien que les d\'efauts de cette
approximation sont largement discut\'es dans la litt\'erature [19].
Pour notre part, nous allons insister sur deux aspects importants que seul l'
ansatz du produit contient de fa\c con naturelle. D' abord, \'etant
donn\'e que l' on veut calculer un potentiel local, la
d\'efinition de la s\'eparation entre les deux solitons do\^\i t
\^etre d\'epourvue d' ambiguit\'es, ce qui est le cas de l' approximation
(6.1). Ensuite, le nombre baryonique associ\'e
\`a la configuration du membre de droite de
(6.1)  est automatiquement \'egal \`a 2. Dans un calcul exact on
ne peut tenir compte de ces
conditions essentielles sans aggraver la
complexit\'e du probl\`eme [36].

\noindent On ne s' attend pas \`a ce que l'
approximation du produit puisse donner une bonne description de
l' interaction aux courtes
distances. Cependant, le centre de notre int\^eret est la moyenne port\'ee
($R\ge
1$ fm) du
potentiel nucl\'eon-nucl\'eon, r\'egion o\`u l'
approximation du produit devrait \^etre ad\'equate.

La configuration (6.1) est d\'eg\'en\'er\'ee en spin et en isospin, car
elle est purement classique. Il en r\'esulte que l' interaction entre les deux
objets du
membre de droite de cette \'equation n' est pas identifiable \`a
l' interaction baryon-baryon. Pour construire l' interaction
nucl\'eon-nucle\'on il faut lever cette d\'eg\'en\'erescence en
effectuant des rotations dans l' espace $SU(2)$ des deux solitons
s\'epar\'ement:
$U_{0}(\vec r+{\vec R\over 2})\to AU_{0}(\vec r+{\vec R\over 2})A^\dagger$,
$U_{0}(\vec r-{\vec R\over 2})\to BU_{0}(\vec r-{\vec R\over 2})B^\dagger$
avec les matrices $A=a_0+i\vec\tau.\vec a$ et  $B=b_0+i\vec\tau.\vec b$.

\noindent On suppose maintenant que les champs de jauge gauche et droit
se transforment comme:

$$\eqalign{
X_i(\vec r,\vec R,C)&=A[X_i(\vec r_1)+U_{0}(\vec r_1)CX_i(\vec
r_2)C^\dagger U_{0}^\dagger (\vec r_1)]A^\dagger\cr
Y_i(\vec r,\vec R,C)&=B[Y_i(\vec r_2)+U_{0}^\dagger (\vec r_2)C^\dagger
Y_i(\vec
r_1)CU_{0}(\vec r_2)]B^\dagger\cr}\eqno(6.2)
$$
\medskip

\noindent pour ces m\^emes rotations $A$ et $B$. On a utilis\'e les
notations $\vec r_1=\vec r+\vec {R\over 2}$ and $\vec r_2=\vec r-\vec {R\over
2}$. Il est \'evident que seule la rotation relative $C=A^\dagger B=c_0
+i\vec\tau\vec c$ importe dans le syst\`eme \`a deux solitons et l'
interaction baryon-baryon comme on le verra
par la suite s' \'ecrit comme une fonction de ce produit $C=A^\dagger
B$. La transformation (6.2) des champs de jauge $X_i$ and $Y_i$ est conforme
\`a leur loi de transformation \`a la  limite locale (voir section 3).
En effet, \`a cette limite les champs de jauge tendent vers les gradients du
champ du pion: $X_{i}\to{{ig\Gamma^2}\over m_A^2}\di UU^\dagger ={i\over
2g}L_{i}$
et $Y_{i}\to{{ig\Gamma^2}\over m_A^2}\di U^ \dagger U={i\over 2g}R_{i}$,
et il est facile de montrer que les courants $L_i$ et $R_i$ se transforment
comme la loi (6.2) par une rotation de $SU(2)\otimes SU(2)$.

\noindent Pour ce qui est du champ scalaire il est clair qu' il ne peut
d\'ependre que des variables $\vec r$ et $\vec R$. Une forme
compatible avec l' approximation du produit pour le champ $U$ et l'
\'equation (3.6) est:

$$\eqalign{
\xi(\vec r,\vec R)={{\xi(\vec r_1)\xi(\vec r_2)}\over \Gamma}\cr}
\eqno(6.3)
$$
\medskip

\noindent Ces formes (6.2) et (6.3) pour les champs de jauge et le champ
scalaire dans le secteur $B=2$ bien que coh\'erentes, ne satisfont pas aux
\'equations du mouvement dans ce secteur. Mais pour ce qui est du champ
du $\omega$ on n' a pas cette libert\'e, car ce champ ne se propage pas,
il est simplement contraint. La composante temporelle de ce champ
ob\'eit \`a l' \'equation suivante:

$$\eqalign{
(\di\di-m^2_{\omega})\omega=S\cr}
\eqno(6.4)
$$
\medskip

\noindent o\`u $S$ est la fonction source $S=2\beta_\omega{\cal
J}^0_{I=0}$. ${\cal J}^0_{I=0}$ est le courant baryonique d\'efini en
(3.14). Il est clair que les
configurations des champs $U$, $X_i$ et $Y_i$ une fois choisies celle du
$\omega$ est
automatiquement donn\'ee par l' expression (6.4).

\noindent L' \'equation (6.4) peut \^etre re\'ecrite comme $\omega(\vec r)=\int
G(\vec
r-\vec r\,')S(\vec r\,')d^3\vec r\,'$, avec $G(\vec
r)={-1\over {4\pi}}{{\exp(-m_{\omega}\vert\vec r\vert)}\over{\vert\vec
r\vert}}$. En ins\'erant les configurations des champs $U$,$X_{i}$
et $Y_{i}$ (6.1-6.2) dans cette expression on obtient celle du champ
$\omega$ dans le secteur $B=2$:

$$
\omega(\vec r,\vec R,c_\mu)=\omega(\vec r_1)+\omega(\vec r_2)+\omega_{T}(\vec
r,
\vec R,c_\mu)
\eqno(6.5)
$$
\medskip

Les deux premiers termes du membre de droite de cette \'equation ont
d\'ej\`a \'et\'e consid\'er\'es dans [27] o\`u le terme mixte $\omega_T$
avait \'et\'e n\'eglig\'e parce qu' il est de
courte port\'ee. En fait le calcul exact des termes contenant $\omega_T$
dans l' interaction est s\'erieusement compliqu\'e du point de vue
num\'erique, la raison est la dimensionalit\'e \'elev\'ee ($R^5$) des
int\'egrales qu' il faut \'evaluer. N\'eanmoins nous avons  eff\'ectu\'e
les calculs num\'eriques de ces termes pour certains points autour de $1$
fm., pour s' apercevoir que pour le cas de notre mod\`ele ces termes
sont petits. Nous avons par cons\'equent n\'eglig\'e $\omega_T$ dans l'
expression (6.5).
Pour \^etre complets nous donnons son expression:

$$
\omega_{T}(\vec r,\vec
R,c_\mu)=\beta_{\omega}\displaystyle\int d^3 r\,' G(\vec r-\vec
r\,')\bigg\{T_{B}(\vec r\,',\vec R,c_\mu)+T_{A}(\vec r\,',\vec  R,c_\mu)\bigg\}
$$
\medskip

\noindent avec \hfill$\displaystyle{
 T_{B}(\vec r,\vec R,C)={\epsilon_{ijk}\over
8\pi^2}\tr(R^1_iCL^2_jL^2_kC^\dagger-R^1_iR^1_jCL^2_kC^\dagger)}$
\hfill(6.6)
\medskip
\noindent et

$$\eqalignno{
T_{A}(\vec r,\vec R,c_\mu)={{ig\epsilon_{ijk}}\over
8\pi^2}\di&\bigg\{\tr\big[I^1_jCL^2_kC^\dagger-R^1_kCX^2_jC^\dagger-R^1_kCI^2_j
C^\dagger
+Y^1_jCL^2_kC^\dagger\cr
&+ig(I^1_kCI^2_jC^\dagger+Y^1_jCX^2_kC^\dagger-CY^2_jC^\dagger
X^1_k\cr
&-U^\dagger_1CU^\dagger_2C^\dagger Y^1_jCU_2C^\dagger
U_1CX^2_kC^\dagger
-X^1_kCU^\dagger_2C^\dagger
Y^1_jCU_2C^\dagger\cr
&+X^1_kY^1_j-U^\dagger_1CY^2_jC^\dagger
U_1CX^2_kC^\dagger+X^2_kY^2_j)\big]\bigg\}&(6.7)\cr}
$$
\medskip

Nous avons fait usage des
notations compactes $I^1_i=U^\dagger(\vec r_1)X_i(\vec r_1)U(\vec
r_1)$, \ \ \
$I^2_i=U(\vec r_2)Y_i(\vec r_2)U^\dagger(\vec r_2)$, \ \ \  $R^1_i=\di
U^\dagger(\vec r_1) U(\vec r_1)$, \ \ \ $L^2_i=\di U(\vec r_2) U^\dagger(\vec
r_2)$, $U_1=U(\vec r_1),\ \ \ \ U_2=U(\vec r_2)$,
$X^2_i=X_i(\vec r_2), \ Y^1_i=Y_i(\vec r_1)$,
$X^1_i=X_i(\vec r_1), \ Y^2_i=Y_i(\vec r_2)$ .

\noindent L' op\'erateur potentiel soliton-soliton est donn\'e par l'
expression:

$$\eqalign{
V(\vec R,c_\mu)=-\int d^3\vec r\bigg\{{\cal L}_{B=2}(\vec r_1,\vec
r_2,c_\mu)-2{\cal
L}_{B=1}(\vec r)\bigg\}\cr}\eqno(6.8)
$$
\medskip

\noindent Pour notre mod\`ele, il faut
remplacer les configurations du secteur $B=2$ pr\'esent\'ees plus haut
dans cette \'equation. Le r\'esultat peut \^etre \'ecrit comme la somme
de deux termes:

$$\eqalign{
V(\vec R,c_\mu)=\int d^3\vec r\bigg\{U_{\xi\rho A_1}(\vec r,\vec R,c_\mu)
+U_{\omega}(\vec r,\vec R,c_\mu)\bigg\}\cr}\eqno(6.9)
$$
\medskip

\noindent Dans cette expression la premi\`ere partie est la contribution
associ\'ee \`a $\Lgls$ d\'ecrivant les effects d\^us aux champs $\vec \rho,
\vec A_1,\xi$ et la seconde correspond \`a la contribution des couplages
du m\'eson $\omega$:

$$\eqalign{
&U_{\xi\rho A_1}(\vec r,\vec R,c_\mu)={1\over
8}\tr{\bf\bigg\{}4F^1_{ij}\big[C\di X^2_jC^\dagger -C\dj
X^2_iC^\dagger-R^1_iCX^2_jC^\dagger+R^1_jCX^2_iC^\dagger\cr
&\quad+CX^2_jC^\dagger R^1_i-CX^2_iC^\dagger
R^1_j+ig\big(C[X^2_i,X^2_j]C^\dagger+CX^2_iC^\dagger I^1_j-CX^2_jC^\dagger
I^1_i\cr
&\quad+I^1_iCX^2_jC^\dagger-I^1_jCX^2_iC^\dagger\big)\big]+2R^1_iCX^2_jC^\dagger\big[
R^1_iCX^2_jC^\dagger-R^1_jCX^2_iC^\dagger\cr
&\quad+2ig\big(I^1_jCX^2_iC^\dagger-I^1_iCX^2_jC^\dagger-[I^1_i,I^1_j]\big)\big]+
4R^1_iCX^2_j\big[-\di X^2_jC^\dagger\cr
&\quad+\dj X^2_iC^\dagger-X^2_jC^\dagger R^1_i+X^2_iC^\dagger
R^1_j+ig\big(X^2_j
C^\dagger
I^1_i-X^2_iC^\dagger I^1_j-[X^2_i,X^2_j]C^\dagger\big)\big]\cr
&\quad+4C\di X^2_j\big[X^2_jC^\dagger R^1_i-X^2_iC^\dagger R^1_j+ig\big(X^2_i
C^\dagger
I^1_j-X^2_jC^\dagger I^1_i+[I^1_i,I^1_j]\big)\big]\cr
&\quad+2CX^2_jC^\dagger R^1_i\big[CX^2_jC^\dagger R^1_i-CX^2_iC^\dagger R^1_j+
2ig\big(
CX^2_iC^\dagger I^1_j-CX^2_jC^\dagger I^1_i+[I^1_i,I^1_j]\big)\big]\cr
&\quad+4igX^2_jC^\dagger R^1_i\big[I^1_iCX^2_j-I^1_jCX^2_i+C[X^2_i,X^2_j]\big]
+4ig\di X^2_jC^\dagger\big[I^1_iCX^2_j-I^1_jCX^2_i\big]\cr
&\quad-4g^2\bigg[-CX^2_iC^\dagger\big(I^1_jI^1_jI^1_i+I^1_iI^1_iI^1_j-2I^1_j
I^1_iI^1_j\big)
-C^\dagger I^1_iC\big(X^2_jX^2_jX^2_i+X^2_iX^2_iX^2_j\cr
&\quad-2X^2_jX^2_iX^2_j\big)-I^1_iI^1_iCX^2_jX^2_jC^\dagger-I^1_iI^1_jCX^2_j
X^2_iC^\dagger
+2I^1_iI^1_jCX^2_iX^2_jC^\dagger\cr
&\quad+{1\over 2}\big[CX^2_iC^\dagger I^1_j(CX^2_iC^\dagger
I^1_j-CX^2_jC^\dagger
I^1_i)+C^\dagger I^1_iCX^2_j(C^\dagger I^1_iCX^2_j-C^\dagger
I^1_jCX^2_i)\big]\bigg]{\bf\bigg\}}\cr
&\quad+\bigg<F^1_{ij}\to F^2_{ij},C\to C^\dagger,I^1_i\to I^2_i,X^2_i\to
Y^1_i,R^1_i\to L^2_i\bigg>\cr}\eqno(6.10)
$$
\eject
$$\eqalign{
&+{{{\xi_1}^2{\xi_2}^2}\over
{2\Gamma^2}}\tr\bigg\{R^1_iC[L^2_i+ig(X^2_i-I^2_i)]C^\dagger+C^\dagger
[Y^1_i-I^1_i]C
[igL^2_i+g^2(I^2_i-X^2_i)]\bigg\}\cr
\noalign{\medskip}
\noalign{\medskip}
&+({m^2_{\rho}\over 2}+{\Delta m^2\over 4})\tr[I^2_iC^\dagger
Y^1_iC+I^1_iCX^2_iC^\dagger ]-{\Delta m^2\over
4}\tr[I^1_iCI^2_iC^\dagger+Y^1_iCX^2_iC^\dagger]\cr
\noalign{\medskip}
\noalign{\medskip}
&-{1\over \Gamma ^2}\xi_1\xi_2\di \xi_1\di \xi_2-\lambda(\Gamma
^2-{\xi_1}^2)(\Gamma^2-{\xi_2}^2)
\left(2-{(\Gamma ^2-\xi_1^2)(\Gamma^2-\xi_2^2)\over \Gamma^4}
\right)\cr
}\eqno(6.10, suite)
$$

\bigskip

$$\eqalign{
U_{\omega}(\vec r,\vec R,c_\mu)&=-{\beta_{\omega}\over
2}\bigg\{\omega_1(B_0^2+A^2)+\omega_2(B_0^1+A^1)\cr
&+(\omega_1+\omega_2)[T_{B}(\vec r,\vec R,c_\mu)+T_{A}(\vec r,\vec R,c_\mu)]\cr
&+\omega_{T}(\vec r,\vec R,c_\mu)
\left[
B_0^1+A^1+B_0^2+A^2+T_{B}(\vec r,\vec
R,c_\mu)+T_{A}(\vec r,\vec R,c_\mu)
\right]
\bigg\}\cr}\eqno(6.11)
$$
\medskip

\noindent avec les notations

$$\eqalign{
F^1_{ij}&=U^\dagger(\vec r_1)\di X_j(\vec r_1)U(\vec r_1),
F^2_{ij}=U(\vec r_2)\di Y_j(\vec r_2)U^\dagger(\vec r_2)
,B_0^1={-\epsilon_{ijk}\over{24\pi^2}}\tr(L^1_iL^1_jL^1_k),\cr
B_0^2&={-\epsilon_{ijk}\over{24\pi^2}}\tr(L^2_iL^2_jL^2_k),
A^1=A(\vec r_1),A^2=A(\vec r_2),
\xi_1=\xi(\vec r_1),\xi_2=\xi(\vec r_2),\cr
\omega_1&=\omega(\vec r_1),\omega_2= \omega(\vec r_2)\cr}\eqno(6.12)
$$
\medskip

\noindent Apr\`es avoir integr\'e sur la variable $\vec r$, il nous reste
une fonction de $\vec R$ et des variables $c_\mu$. Il n' est pas
difficile de montrer que cette fonction est en fait un polyn\^ome
d' ordre pair en ces $c_\mu$:

$$\eqalign{
V(\vec R,c_\mu)=\upsilon_0(\vec R)+\upsilon_2(\vec R,c_\mu)+\upsilon_4
(\vec R,c_\mu)+
\upsilon_6(\vec R,c_\mu)+ \ ...\cr}\eqno(6.13)
$$
\medskip

\noindent avec la forme g\'en\'erale pour les
 $\upsilon_{2p} \ \ (p=1,2...)$ :

$$\eqalign{
\upsilon_2(\vec R,c_\mu)&=\alpha_1+\alpha_2 c_{0}^2+\alpha_3 c_{3}^2\cr
\upsilon_4(\vec R,c_\mu)&=\beta_1+\beta_2 c_{0}^2+\beta_3
c_{3}^2+\beta_4 c_{0}^4+\beta_5 c_{0}^2c_{3}^2+\beta_6 c_{3}^4\cr
\upsilon_6(\vec R,c_\mu)&= \ ... \cr}\eqno(6.14)
$$
\medskip

\noindent Pour arriver \`a ces expressions nous avons pris le vecteur
$\vec R$ parall\`ele \`a l' axe $z$ de l' espace de
configuration. Maintenant, les fonctions $\alpha$ et $\beta$ peuvent
\^etre calcul\'ees en prenant diff\'erentes projections de la matrice
$C$ dans le membre de droite des \'equations (6.6), (6.7) et (6.10),
(6.11). Pour donner un exemple, la fonction $\alpha_1$ est \'egale au
membre de droite de l' \'equation (6.9), dans laquelle il faut retenir
seulement les termes quadratiques en $c_\mu$ et prendre $C=i\tau_1$.
L' expression exacte des $\alpha_i, \beta_i$ est donn\'ee dans l' appendice
B. Les termes d' ordre sup\'erieur \`a quatre dans le polyn\^ome de
l' \'equation (6.13) proviennent du terme de Wess-Zumino (ce
terme contribue aussi \`a $\upsilon_0$, $\upsilon_2$
et $\upsilon_4$). Nous avons n\'eglig\'e ces termes dans notre
\'etude car ils contribuent essentiellement \`a des op\'erateurs en
repr\'esentation de spin \'elev\'e. L' approximation de n\'egliger
$\upsilon_6$ peut \^etre partiellement
justifi\'ee  par les r\'esultats de la r\'ef. [24] o\`u il a
\'et\'e trouv\'e que le terme de Wess-Zumino a globalement un petit effet sur
les observables du secteur $B=1$.

Pour extraire maintenant les canaux physiques (de spin et isospin
d\'efinis) de l' interaction, on utilise la m\'ethode simple de
projection qui a \'et\'e introduite dans la r\'ef. [18]. La
d\'ecomposition naturelle de l' op\'erateur potentiel non-relativiste
agissant dans un \'espace \`a deux nucl\'eons est:

$$\eqalign{
V(\vec R)&=V_{C}^+(\vert \vec R\vert)+(\vec \tau_1.\vec \tau_2)V_{C}^-(\vert
\vec
R\vert)+(\vec\sigma_1.\vec\sigma_2)\big[V_{SS}^+(\vert \vec R\vert)+(\vec
\tau_1.\vec
\tau_2)V_{SS}^-(\vert \vec R\vert)\big]\cr
&+\big[3(\vec\sigma_1.\vec R)(\vec\sigma_2.\vec R)/\vec
R^2-(\vec\sigma_1.\vec\sigma_2)\big]\big[V_{T}^+(\vert \vec R\vert)+(\vec
\tau_1.\vec
\tau_2)V_{T}^-(\vert \vec
R\vert)\big]\cr}
\eqno(6.15)
$$
\medskip

\noindent Pour calculer les six composantes $V_C^+,V_C^-,...$ dans notre
mod\`ele, il faut six \'equations. On les obtient en \'egalant six
\'el\'ements de matrice de $V(\vec R)$ (6.15) avec ceux de l'
op\'erateur $V(\vec R,c_\mu)$ (6.9), ces derniers \'etant \'exprim\'es
en termes des fonctions d' onde de spin et d' isospin du nucl\'eon. La
solution de ce syst\`eme d' \'equations est:

$$\eqalign{ V_{C}^+&= \int da_{\mu} \int db_{\mu}\delta(\sum_{\mu=1}^4
a_{\mu}^2 -1) \delta (\sum_{\mu=1}^4 b_{\mu}^2 -1)V(\vec R,c_\mu)f_{C}^+
(a_{\mu}, b_{\mu})\cr
&=\upsilon_0+\alpha_1+\beta_1+{1\over
4}(\alpha_2+\beta_2+\alpha_3+\beta_3)+{1\over
8}(\beta_4+\beta_6)+{1\over 24}\beta_5\cr
V_{T}^-&=\int da_{\mu} \int db_{\mu}\delta(\sum_{\mu=1}^4 a_{\mu}^2 -1) \delta
(\sum_{\mu=1}^4 b_{\mu}^2 -1)V(\vec R,c_\mu)f_{T}^- (a_{\mu},
b_{\mu})\cr
&={1\over 54}(\alpha_3+\beta_3)+{1\over 432}\beta_5+{1\over
72}\beta_6\cr
V_{SS}^-&=\int da_{\mu} \int db_{\mu}\delta(\sum_{\mu=1}^4 a_{\mu}^2 -1) \delta
(\sum_{\mu=1}^4 b_{\mu}^2 -1)V(\vec R,c_\mu)f_{SS}^- (a_{\mu},
b_{\mu})\cr
&={1\over 36}(\alpha_2+\beta_2)-{1\over108}(\alpha_3+
\beta_3)+{1\over 48}\beta_4+{1\over 432}\beta_5-{1\over 144}
\beta_6\cr
V_{T}^+&=V_{SS}^+=V_{C}^-=0\cr }
\eqno(6.16)$$
\medskip

\noindent o\`u les $a_\mu$ sont les variables du nucl\'eon $1$ et $b_\mu$
celles du nucl\'eon $2$. Dans ces formules les fonctions
$f_C^+,f_C^-,...$ sont des densit\'es dans l' espace $SU(2)\otimes
SU(2)$. Leurs expressions sont donn\'ees par:

$$\eqalign{
f_{C}^+ (a_{\mu},b_{\mu})&= {1\over {2\pi^4}}(a_{1}^2+a_{2}^2)
(b^2_1+b^2_2+b^2_3+b^2_0)\cr
f_{T}^- (a_{\mu},b_{\mu})&= {1\over {6\pi^4}}
\bigg\{
(a_{1}^2+a_{2}^2)
(b_{1}^2+b_{2}^2-b_{0}^2-b_{3}^2)\cr
&+(a_{1}a_{0}-i(a_{2}a_{0}+a_{1}a_{3})-
a_{2}a_{3}) (b_{1}b_{0}+i(b_{2}b_{0}+b_{1}b_{3})-b_{2}b_{3})
\bigg\}\cr
f_{SS}^- (a_{\mu},b_{\mu})&= f_{T}^- (a_{\mu},b_{\mu})\cr
&+{1\over {2\pi^4}}
(a_{1}a_{0}-i(a_{2}a_{0}+a_{1}a_{3})-a_{2}a_{3}) (b_{1}b_{0}+i(b_{2}b_{0}+
b_{1}b_{3})-b_{2}b_{3})}
\eqno(6.17)$$
\medskip

\noindent Pour les calculs num\'eriques cependant les expressions en $\alpha_i$
et
$\beta_i$ sont d' une plus grande utilit\'e. C' est aussi
celles que nous avons utilis\'ees pour calculer les diff\'erents canaux de
l' interaction nucl\'eon-nucl\'eon dans notre mod\`ele.

\noindent Il est \`a noter ici que les fonctions $\alpha_i,
\beta_i$ peuvent \^etre utilis\'ees pour
calculer non seulement les interactions entre nucl\'eons mais aussi
celles entre des baryons de spin plus \'elev\'e. Ceci a \'et\'e fait
dans [18]. Dans notre mod\`ele le calcul de ces interactions
n\'ecessiterait l' inclusion des termes comme $\upsilon_6$ dans (6.13).

On va s' int\'eresser maintenant aux r\'esultats, pour estimer
si notre mod\`ele bas\'e sur une d\'escription
coh\'erente des m\'esons peut reproduire les aspects
caract\'eristiques de l' interaction nucl\'eon-nucl\'eon
telle qu' ils sont observ\'es exp\'erimentalement.

\bigskip

\leftline{\bfmagc 7. Discussion des r\'esultats}

\bigskip

Nous avons calcul\'e num\'eriquement les diff\'erentes composantes $V^+_C$,
$V^-_T$,
$V^-_{SS}$ avec les fonctions de profil $F$, $\alpha$, $\beta$, $\gamma$,
$\omega$, $\epsilon$ obtenues dans le secteur $B=1$ (section
5).

\noindent Tout d' abord, nous avons trouv\'e que la contribution du terme d'
anomalie de
Wess-Zumino au potentiel est tr\`es petite. Ceci est en accord avec l'
observation
des auteurs [24] qui ont trouv\'e (dans le cadre du mod\`ele $\sigma$
non-lin\'eaire) que ce terme bien qu' important pour la bonne
description de la physique des m\'esons est quantitativement moins
important dans le secteur des baryons.

\noindent La contribution essentielle \`a $U_\omega$ est d\^ue au
couplage du $\omega$ aux trois pions. On peut montrer (aussi bien
analytiquement que num\'eriquement) que si le terme de Wess-Zumino est
n\'eglig\'e, la contribution de $U_\omega$ au potentiel $V^+_C$
est r\'epulsive. Par ailleurs on a vu lors de l'
\'etude du secteur $B=1$ que le champ du $\omega$ dans le mod\`ele
lin\'eaire est d' une port\'ee inf\'erieure \`a celle qu' il a si le
param\`etre $\lambda$ est tr\`es large (mod\`ele non-lin\'eaire). Cet
effet est \`a l' origine de la suppression de la port\'ee du
terme $U_\omega$ dans le canal central de l' interaction (table 1).
La r\'epulsion d\^ue au champ du m\'eson $\omega$ est donc de
courte port\'ee dans notre mod\`ele. Cet effet est d\^u \`a l' inclusion
simultan\'ee de tous les m\'esons.

\noindent En ce qui concerne la contribution du terme $U_{\xi\rho A_1}$
\`a $V^+_C$, on peut montrer que le premier terme de l' \'equation
(6.10) est positif alors que les deux derniers (absents du
mod\`ele de Skyrme) sont n\'egatifs. Il est \`a noter
que ces deux derniers termes disparaissent dans le cas
non-lin\'eaire ($\xi\to\Gamma$). En comparant les profils
lin\'eaires \`a ceux du cas non-lin\'eaire pour
les fonctions $F,\alpha,\beta,\gamma$ (figure (1.3)), on comprend pourquoi
la partie r\'epulsive de $U_{\xi\rho A_1}$ est s\'erieusement diminu\'ee
dans notre mod\`ele (table 1).

\centerline{{\bf Table 1}}

Contribution des m\'esons vecteurs au canal central de l' interaction
nucl\'eon-nucl\'eon, quand ceux-ci sont introduits en jaugeant le mod\`ele
sigma non-lin\'eaire
($\lambda=\infty$) ou le mod\`ele sigma lin\'eaire ($\lambda=0.8$). $R$
est en fermi et le potentiel est en MeV.

\def\init{\tabskip 0pt\offinterlineskip}
\def\crr{\cr\noalign{\hrule}}

$$\vbox{\init\halign to 13 cm{
\strut#&\vrule#\tabskip=1em plus 2em&
\hfil$#$\hfil&
\vrule$\,$\vrule#&
\hfil$#$\hfil&
\vrule#&
\hfil$#$\hfil&
\vrule#&
\hfil$#$\hfil&
\vrule#&
\hfil$#$\hfil&
\vrule#\tabskip 0pt\crr
&&R&&\rho ,A_1(\lambda =\infty )&&\rho ,A_1(\lambda =0.8 )&&\omega
(\lambda=\infty )
&&\omega (\lambda =0.8)&\crr

&&0.0&&596.0&&122.0&&405.2&&872.5&\crr
&&0.5&&400.5&&108.0&&247.5&&405.6&\crr
&&1.0&&110.6&&22.0&&63.7&&54.9&\crr
&&1.5&&18.6&&2.3&&10.7&&5.7&\crr
&&2.0&&3.4&&0.4&&1.5&&0.6&\crr
}}$$

\noindent Nous montrons sur la figure (1.5) les r\'esultats pour $V_C^+$
obtenus avec une
masse du scalaire de $0.62$ GeV. Pour comparaison, nous y avons
dessin\'e les r\'esultats du mod\`ele non-lin\'eaire, (le cas o\`u le champ
du scalaire est \'elimin\'e) et les valeurs du potentiel de Paris [37].

Notre \'etude a montr\'e que quand la masse du
m\'eson isoscalaire augmente l' attraction est pouss\'ee au-del\`a de
$1.5$ fm. On peut penser que le fait d' avoir besoin d' une masse $m_\epsilon$
assez petite pour que l' attraction soit plac\'ee au bon endroit,
constitue  un probl\`eme car la ph\'enom\'enologie veut que la
r\'esonance $S$ de la diffusion $\pi\pi$ se situe plut\^ot vers $1$ GeV.
A ce sujet, rappellons-nous du cas des mod\`eles d' \'echange de bosons, l\`a
aussi, il est n\'ecessaire d' introduire un champ scalaire assez l\'eger
($m\sim
0.5$ GeV) pour reproduire
correctement l' interaction nucl\'eon-nucl\'eon. Dans le cadre de cette
description l' introduction de ce m\'eson scalaire fictif
peut \^etre \'evit\'ee en
consid\'erant  explicitement l' \'echange de deux pions [38].
Malheureusement, il n' existe pas \`a l' heure actuelle une m\'ethode
qui permettrait de tenir compte de ces effets dans notre probl\`eme. D'
un autre c\^ot\'e, comme nous le verrons plus loin, il n' est pas impossible
que dans notre mod\`ele on obtienne les forces attractives avec une masse
du m\'eson scalaire, plus proche de sa valeur exp\'erimentale.

On a donc montr\'e qu' une g\'en\'eralisation du mod\`ele de Skyrme
d\'ecrivant la physique du pion et des m\'esons scalaires et  vecteurs
est capable de reproduire l' essentiel de la
physique des interactions nucl\'eon-nucl\'eon, \`a savoir une
r\'epulsion \`a courte port\'ee et une attraction \`a moyenne
port\'ee dans le canal central de cette interaction. De l' attraction mais \`a
longue port\'ee
a \'et\'e obtenue par d' autres auteurs, toujours par le biais de l' inclusion
des degr\'es de libert\'e scalaires mais dans des contextes diff\'erents
(anomalie d'
\'echelle de la QCD [29], corrections \`a une boucle de pions [39]).
Notre mod\`ele sugg\`ere que pour que cette attraction soit vraiment de
moyenne port\'ee (en accord avec la ph\'enom\'enologie) il faut inclure
non seulement le m\'eson scalaire mais aussi les m\'esons vecteurs
$\rho,\omega,A_1$ dans le Lagrangien effectif. A cet \'egard, observons
sur la figure (1.5) que m\^eme en l' absence du m\'eson scalaire la r\'epulsion
est
fortement diminu\'ee par rapport au mod\`ele de Skyrme.

Pour illustrer la n\'ec\'essit\'e d' inclure simultan\'ement
les m\'esons $\epsilon,\rho,\omega,A_1$ nous avons effectu\'e un calcul
de l' interaction NN dans le cas ou seuls le pion et le
scalaire sont pr\'esents dans le Lagrangien effectif. Nous
avons consid\'er\'e le mod\`ele

$$\eqalign{
{\cal{L}}_{\pi\xi}=&{1\over 2}\dmudo \xi \dmuup \xi+{{\xi^2}\over
4}\tr(\partial_\mu U\partial^\mu U^\dagger)
-\lambda (\xi ^2-\Gamma ^2)^2\cr
&+{1\over {32e^2}}\tr([\partial_\mu U U^\dagger, \partial_\nu U
U^\dagger]^2)-{{b}\over{2f_\pi^2}}B_\mu
B^\mu\cr}\eqno(7.1)
$$
\medskip

Le mod\`ele (7.1) est tr\`es proche du mod\`ele de la r\'ef. [29]
mais ici le terme d' ordre six est inclus car il n' est pas justifi\'e de le
n\'egliger. A
la limite o\`u $\lambda\to\infty$ on retrouve le Lagrangien (3.15) en
termes de pions uniquement. Sur la figure (1.6) on voit que l'
attraction n' a plus la bonne port\'ee quand les masses des m\'esons
$\rho,\omega,A_1$ deviennent infinies.

Il est donc clair qu' un Lagrangien effectif de pions
et de scalaires seulement ne peut fournir que de l' attraction \`a longue
port\'ee, confirmant ainsi notre conclusion principale.

Faisons quelques remarques maintenant sur la validit\'e de l'
approximation du produit adopt\'ee au cours de notre
\'etude. Dans notre travail aussi bien que dans ceux des
r\'ef\'erences [29] et [39], le potentiel $NN$ est calcul\'e en approximant
le syst\`eme $B=2$ par la configuration simple de l' \'equation (6.1).
R\'ecemment les auteurs de [36] ont trouv\'e que la
r\'epulsion du mod\`ele de Skyrme est sensiblement diminu\'ee si l'
ansatz de Manton-Singer est utilis\'e dans un calcul semi-exact.
D' autre part les  auteurs de [20] trouvent m\^eme
de l' attraction dans le mod\`ele de Skyrme par des m\'ethodes
num\'eriques assez compliqu\'ees mais cette attraction n' a pas
la bonne port\'ee. Ces calculs num\'eriques utilisant des approximations
allant au-del\`a
de celle du produit ont l 'air de sugg\'erer que cette
derni\`ere n' est pas ad\'equate pour le calcul de l' interaction. En fait
l' avantage de l' approximation du produit est qu' elle est simple et
suffisament transparente pour extraire les aspects ph\'enom\'enologiques connus
de l' interaction
nucl\'eon-nucl\'eon. Comme on l' a vu auparavant, les signes des
diff\'erentes contributions au potentiel sont connus et coh\'erents
avec les ingr\'edients physiques du mod\`ele (3.1). Nous pensons que les
diff\'erentes approximations peuvent affecter la magnitude de ces
contributions mais pas leur signe. De plus, dans tous les calculs
pr\'ecit\'es, on calcule un potentiel local et seule l' approximation du
produit
offre une d\'efinition non ambigue de l' interdistance entre deux solitons.
Si les r\'esultats de [20] et [36] sont corrects, ils donnent \`a
penser que l' approximation du produit sous estime l' attraction que le
Lagrangien (3.1) produit. Autrement dit si notre mod\`ele est trait\'e
num\'eriquement avec les m\'ethodes des r\'efs. [20] ou [36], il est tr\`es
probable qu' il
fournisse trop d' attraction avec $m_\epsilon=0.62$ GeV. Dans ce cas, en
augmentant la valeur de $m_\epsilon$ jusqu' \`a la valeur
experimentale ($1$ GeV) on pourrait retrouver l' attraction $NN$ empirique. Il
serait
en effet tr\`es int\'eressant d' effectuer le traitement num\'erique de
[20] dans le cadre de notre mod\`ele qui, comme nous l' avons vu dans
les sections pr\'ec\'edentes, est plus r\'ealiste que le
mod\`ele de Skyrme.

\noindent L' interaction dans le canal spin-spin aussi bien que celle du
canal tenseur ont \'et\'e aussi calcul\'ees dans le mod\`ele lin\'eaire et
non-lin\'eaire (figures 1.7 et 1.8). La pr\'esence du champ scalaire
supprime fortement ces deux canaux. Nos
r\'esultats sont en accord qualitatif avec ceux de la r\'ef. [29].

Une derni\`ere remarque importante doit \^etre faite. Un
degr\'e de libert\'e qui ne doit pas manquer dans les  mod\`eles des
hadrons \`a basse \'energie est celui du m\'eson $\omega$. Il
est bien connu que sa contribution \`a l' interaction $NN$ est fortement
repulsive aussi bien dans les mod\`eles d' \'echange de m\'esons que dans
le cadre des mod\`eles que nous consid\'erons ici. Son
absence dans les travaux [20],[29],[36],[39] est
injustifi\'ee. Dans notre travail il est bien pr\'esent, et malgr\'e
cela l' attraction persiste. Ce ph\'enom\`ene est le r\'esultat d'
un m\'ecanisme non-trivial d\^u \`a l' inclusion simultan\'ee
de tous les m\'esons.

En r\'esumant les calculs de ce chapitre on peut dire que le
mod\`ele que nous avons \'etudi\'e ici et qui g\'en\'eralise le mod\`ele
de Skyrme de fa\c con \`a d\'ecrire correctement la physique des m\'esons,
est non seulement
capable de pr\'edire mieux les propri\'et\'es statiques des baryons,
mais aussi de produire une interaction nucl\'eon-nucl\'eon en accord
quantitatif avec la ph\'enom\'enologie.

\noindent Il est essentiel de garder \`a l' esprit qu' \`a  la limite  o\`u
le nombre des couleurs devient grand, QCD est \'equivalente
\`a une th\'eorie effective d' un nombre infini de
m\'esons\footnote\dag{ceci pour conserver la propri\'et\'e de
libert\'e asymptotique des fonctions \`a deux points}. Par cons\'equent
il n' y a aucune raison de consid\'erer le pion seulement.

\noindent La prise en compte des m\'esons les plus l\'egers dans le
Lagrangien effectif est en fait plus qu' une fa\c con de tester le
d\'eveloppement semi-classique des observables baryoniques comme il a
\'et\'e sugg\'er\'e par Witten [21]. Le Lagrangien (3.1) que nous
 avons propos\'e dans ce chapitre offre surtout  un
cadre th\'eorique simple et coh\'erent o\`u
les m\'esons (les champs \'el\'ementaires) et les baryons (leurs
excitations de type soliton topologique) sont d\'ecrits simultan\'ement.

\eject

\bigskip

\noindent{\bfmagc Appendice A}

\bigskip

On donne ici l' expression des \'equations diff\'erentielles
non-lin\'eaires coupl\'ees, dont la solution \'extr\'emise la
fonctionnelle d' \'energie dans le  secteur B=1 (section 5 du texte). Pour
simplifier quelque peu les formules, il est utile de d\'efinir les
diff\'erents \lq\lq moments" $u$, $v$ et $t$  qui sont associ\'es  aux champs
$F$, $\alpha$ et  $\gamma$ respectivement:

$$\eqalign{
u&=\xi^2r^2(\dot F+2g\beta)\cr
v&=\dot\alpha r+\alpha-\beta(1-2g\gamma r)\cr
t&=\dot\gamma r+\gamma-2g\alpha\beta r\cr}\eqno(A.1)
$$
\medskip

\noindent Alors les \'equations du mouvement sont de la forme:

$$\eqalignno{
&\dot u=\xi^2\big[\sin 2 F[(1-2g\gamma r)^2-4g^2\alpha^2r^2]+4g\alpha r
\cos 2F(1-2g\gamma r)\big]\cr
&-2\Delta m^2r^2[2\alpha\gamma\cos 2F+(\alpha^2-\gamma^2)
\sin 2F]+\beta_{\omega}{\dot\omega\over {2\pi^2}}\big[
-2g[\gamma r+gr^2(\alpha^2-\gamma^2)]\cos 2F\cr
&-2g\alpha r(1-2g\gamma r)\sin 2F-\sin^2F\big]&(A.2)\cr}
$$
\medskip

$$\eqalign{
&\dot v=4g\alpha r[g(\alpha^2+\gamma^2)-{\gamma\over r}]-2g\beta
t+g\xi^2\cos F[\sin F+2gr(\alpha\cos F-\gamma\sin F)]\cr
&+m_{\rho}^2\alpha r-\Delta m^2r\cos F
(\gamma\sin F-\alpha\cos F)-\beta_{\omega}g{\dot\omega\over
{4\pi^2}}[\sin^2 F(1-2g\gamma r)+g\alpha r\sin 2 F]\cr}
\eqno(A.3)
$$
\medskip

$$\eqalignno{
&\dot t=-2(1-2g\gamma r)[g(\alpha^2+\gamma^2)-{\gamma\over r}]+2g\beta
v-g\xi^2\sin F[\sin F+2gr(\alpha\cos F-\gamma\sin F)]\cr
&+m_{\rho}^2\gamma r+\Delta m^2r\sin F
(\gamma\sin F-\alpha\cos F)-\beta_{\omega}g{\dot\omega\over
{8\pi^2}}[\sin 2F(1-2g\gamma r)-4g\alpha r\sin^2F]\cr}
$$
\noindent {\hfill (A.4)}
\medskip

$$\eqalign{
\ddot\omega&=m_{\omega}^2\omega-2{\dot\omega\over r}-{\beta_{\omega}\over
{2\pi^2r^2}}\bigg\{\dot F[\sin^2 F+2g\alpha r\sin
2F(1-2g\gamma r)\cr
&+2g\gamma r\cos 2F+2g^2r^2(\alpha^2-\gamma^2)\cos 2F]+g[2
(\dot\alpha r+\alpha)(\sin^2F(1-2g\gamma r)\cr
&+g\alpha r\sin 2F)+(\dot\gamma r+\gamma)(\sin 2F(1-2g\gamma
r)-4g\alpha r\sin^2F)]\bigg\}\cr}
\eqno(A.5)
$$
\medskip

$$
\ddot\xi=-2{\dot\xi\over r}+\xi\bigg[(\dot F+2g\beta)^2+2({\sin F
\over
r}+2g(\alpha\cos F-\gamma\sin F))^2\bigg]
+4\lambda\xi(\xi^2-\Gamma^2)
\eqno(A.6)
$$
\medskip

\noindent La fonction $\beta$ ob\'eit \`a la contrainte suivante:
$$\eqalign{
\beta={1\over {r^2(m_{\rho}^2+\Delta m^2)}}\bigg[-gu+4g\alpha
rt+2v(1-2g\gamma r)\bigg]\cr}
\eqno(A.7)
$$
\bigskip

\noindent{\bfmagc Appendice B}

\bigskip

Dans cet appendice on explicite les fonctions $\alpha$
et $\beta$ qui apparaissent dans le texte (\'equation (6.14)). Pour
les calculer il faut d' abord s\'eparer les termes
quadratiques ($\upsilon_2$) des termes quartiques ($\upsilon_4$)
en la matrice $C$ dans les
expressions du potentiel, et ensuite prendre diff\'erentes projections
de cette derni\`ere. On a bien pris le soin d' \'ecrire ces formules sous
leur forme la plus g\'en\'erale, car elles sont ind\'ependantes du
mod\`ele consid\'er\'e.

I) Termes quadratiques

$$\eqalign{
\alpha_1=&\upsilon_2(i\tau_1)\cr
\alpha_2=&\upsilon_2(I)-\upsilon_2(i\tau_1)\cr
\alpha_3=&\upsilon_2(i\tau_3)-\upsilon_2(i\tau_1)\cr}\eqno(B.1)
$$
\medskip

II) Termes quartiques

$$\eqalign{
\beta_1=&\upsilon_4(i\tau_1)\cr
\beta_2=&4\upsilon_4((I+i\tau_1)/\sqrt{2})-3\upsilon_4(i\tau_1)-
\upsilon_4(I)\cr
\beta_3=&-{1\over
4}\big[\upsilon_4(i\tau_3)-25\upsilon_4(i(2\tau_1+i\tau_3)/\sqrt{5})\big]-6
\upsilon_4(i\tau_1)\cr
\beta_4=&2\big[\upsilon_4(I)+\upsilon_4(i\tau_1)\big]-4\upsilon_4((I+i\tau_1)/
\sqrt{2})\cr
\beta_5=&4\upsilon_4((I+i\tau_3)/\sqrt{2})+7\upsilon_4(i\tau_1)
-4\upsilon_4((I+i\tau_1)/\sqrt{2})-{3\over 4}\upsilon_4(i\tau_3)\cr
&-{25\over 4}\upsilon_4(i(2\tau_1+i\tau_3)/\sqrt{5})\cr
\beta_6=&5\upsilon_4(i\tau_1)+{5\over 4}\upsilon_4(i\tau_3)-
{25\over 4}\upsilon_4(i(2\tau_1+i\tau_3)/\sqrt{5})\cr}\eqno(B.2)
$$
\medskip

\bigskip

\noindent{\bfmagc Appendice C}

\bigskip

Dans cet appendice nous allons montrer qu'\`a la limite des grandes masses
pour les m\'esons $\rho,\omega,A_1,\epsilon$ le Lagrangien (3.1) tend
vers l' expression donn\'ee par l' \'equation (3.15). Nous allons
n\'egliger dans ce qui suit le terme de Wess-Zumino.

\noindent Tout d' abord il est clair que quand la masse du m\'eson scalaire
tend vers
l' infini ($\lambda=\infty$) le champ scalaire $\epsilon$ dispara\^\i t
car \`a cette limite $\xi=\Gamma$.La solution de l' \'equation du
mouvement du champ $\omega_\mu$ quand les param\`etres
$m_\omega,\beta_\omega$ tendent vers l' infini (en gardant le rapport
$\beta_\omega/m_\omega$ fixe) est
$\displaystyle{\omega_\mu=-{\beta_\omega\over{m_\omega^2}}B_\mu}$, ce qui
implique que le Lagrangien du syst\`eme $\pi\omega$ tend vers l'
expression $\displaystyle{-{{\beta_\omega^2}\over{2m_\omega^2}}B_\mu B^\mu}$
(\'equation (3.15)).

\noindent En ce qui concerne les champs de jauge,
l' \'equation d' Euler-Lagrange pour le champ gauche s' \'ecrit:

$$\eqalign{
-{1\over 2}ig\Gamma^2 L_{\mu}+{1\over 2}(m_{\rho}^2+{\Delta m^2\over
2}+g^2\Gamma^2)X_{\mu}-{1\over 2}(g^2\Gamma^2+{\Delta m^2\over
2})UY_{\mu}U^\dagger=\dnuup\bigg[{{\partial\Lgls}\over {\partial(\dnuup
X^{\mu})}}\bigg]\cr}\eqno(C.1)
$$
\medskip

\noindent L' \'equation du mouvement pour le champ droit est obtenue en
rempla\c cant dans (C.1) $X_{\mu}$ par $Y_{\mu}$ et $U$ par $U^\dagger$.
Faisons tendre les masses des m\'esons physiques $\rho$ et $A_1$ vers l'
infini. A cette fin il faut prendre les limites $m_\rho\to\infty$, $\Delta
m^2\to -\infty$
en respectant la relation $m_\rho^2+\Delta m^2=2g^2f_\pi^2$
pour que le d\'enominateur de l' \'equation (4.2) s' annule. Alors le
coefficient $\Gamma$, proportionnel \`a $m_{A_1}$,  devient tr\`es
grand. Nous pouvons alors n\'egliger les termes figurant dans le membre de
droite de l'
\'equation (C.1). La solution de cette \'equation \`a cette limite et
pour $g$ fini est:

$$\eqalign{
X^{\mu}\to{{ig\Gamma^2}\over m_A^2}\dmuup UU^\dagger={i\over
2g}L^{\mu}\cr
Y^{\mu}\to{{ig\Gamma^2}\over m_A^2}\dmuup U^\dagger U={i\over
2g}R^{\mu}\cr}\eqno(C.2)
$$
\medskip

\noindent donnant pour les tenseurs $X_{\mu\nu}$ et $Y_{\mu\nu}$ les
identit\'es suivantes:

$$\eqalign{
X^{\mu\nu}\to{i\over 2g}(1-{{g^2\Gamma^2}\over
m_A^2})[L^{\mu},L^{\nu}]={i\over 2g}{1\over 2}[L^{\mu},L^{\nu}]\cr
Y^{\mu\nu}\to{i\over 2g}(1-{{g^2\Gamma^2}\over
m_A^2})[R^{\mu},R^{\nu}]={i\over 2g}{1\over
2}[R^{\mu},R^{\nu}]\cr}\eqno(C.3)
$$
\medskip

\noindent En rempla\c cant les expressions (C.2) et (C.3) dans le Lagrangien
(3.2) nous obtenons le mod\`ele $sigma$ non-lin\'eaire plus un
 terme d' ordre quatre, antisym\'etrique par
rapport aux d\'eriv\'ees du champ du pion. La constante de Skyrme
correspondante est donn\'ee par $e=\sqrt{2}g$, ce qui donne le facteur
${1\over {64g^2}}$ figurant dans l' \'equation (3.15).
\eject
\rightline {}
\rightline {}
\rightline {}
{\rightline {}}
{\rightline {}}
\bigskip
\bigskip
\bigskip
\bigskip
\centerline {\bfmag Chapitre II}
\bigskip
\bigskip
\bigskip
\centerline {\bfmagd Sur le d\'eveloppement semi-classique}
\centerline {\bfmagd de la masse du soliton}
\eject
\bigskip
\bigskip
\bigskip

Les solutions solitoniques du type consid\'er\'e dans le chapitre
pr\'ec\'edent, \'etudi\'ees d' abord par Skyrme puis reprises par la
r\'ef. [17] et les travaux ult\'erieurs sont des solutions classiques
de th\'eories de champs
non-lin\'eaires. Leur quantification consiste \`a construire
des \'etats quantiques autour de ces solutions. Les observables physiques
se d\'eveloppent alors en puissances de $\hbar$ [10]. Pour le cas du
nucl\'eon, il a \'et\'e montr\'e [12] que le param\`etre de
d\'eveloppement semi-classique est en fait $\displaystyle{{\hbar\over{N_c}}}$,
de sorte que sa masse se met sous la forme:
$$\eqalign{
M_N= \ N_c \ \big[{\cal{M}}_0 \ + \ {\hbar\over{N_c}}{\cal{M}}_1 \ + \
{{\hbar^2}\over{N^2_c}}{\cal{M}}_2 \  + \ ...\big]\cr}\eqno(I.1)
$$
\medskip
\noindent La convergence de ce d\'eveloppement repose fortement
sur les valeurs des coefficients ${\cal{M}}_0$, ${\cal{M}}_1,{\cal{M}}_2$
qui sont d\'ependants du mod\`ele.

Le but de ce
chapitre est d' illustrer cette d\'ependance pour les
diff\'erents mod\`eles bas\'es sur les Lagrangiens de m\'esons. Nous
allons montrer que certains d' entre eux ont plus de chances que d'
autres de fournir un d\'eveloppement (I.1) bien d\'efini.
Nous allons, dans ce qui va suivre nous int\'eresser \`a la premi\`ere
correction quantique \`a la masse du soliton, le terme ${\cal{M}}_1$. Nous
allons \'etablir que le terme ${\cal{M}}_1$ est une
\'energie de Casimir [40], et donner son expression.
Cette \'energie poss\`ede une divergence ultraviolette qui peut \^etre
r\'egularis\'ee  dans le sch\'ema de la fonction z\'eta.
Nous  calculons num\'eriquement le rapport
$\displaystyle{{{{\cal{M}}_1}\over{N_c{\cal{M}}_0}}}$
pour le mod\`ele de Skyrme et une de ses g\'en\'eralisations possibles.
Les r\'esultats obtenus apportent quelques conclusions sur la
validit\'e du d\'eveloppement semi-classique pour la masse du nucl\'eon.
\bigskip
\leftline{\bfmagc 1. L' \'energie de Casimir du soliton}
\bigskip

En effectuant des fluctuations quantiques sur des degr\'es de libert\'e
collectifs autour de la solution du
mod\`ele de Skyrme, les auteurs de [17] ont obtenu une partie des
contributions \`a ${\cal{M}}_2$ (\'eq. (I.1)). Mais curieusement le terme
${\cal{M}}_1$ n' a re\c cu que tr\`es peu d' attention dans le pass\'e.
Ce terme  est certainement plus difficile
\`a \'evaluer car il met en jeu des fluctuations non-collectives. Afin
de trouver son expression, effectuons des fluctuations  autour de la solution
classique $U_0$ d\'efinies par les param\`etres $\vec\alpha$:
$$\eqalign{
U=U_0 \ \exp{(i\vec\tau.\vec\alpha)}\cr}\eqno(1.1)
$$
\medskip
\noindent et consid\'erons la fonctionnelle g\'en\'eratrice des
fonctions de Green en l' absence de sources:
$$\eqalign{
W_E={\cal{N}}\int {\cal{D}}[U] e^{-S_E(U)}\cr}\eqno(1.2)
$$
\medskip
\noindent $S_E$ \'etant l' action euclidienne associ\'ee au Lagrangien du
mod\`ele consid\'er\'e et ${\cal{N}}$ une constante de normalisation. Cette
action peut \^etre d\'evelopp\'ee autour de la configuration classique
$U_0$ qui est un point stationnaire de $S_E$:
$$\eqalign{
S_E(U)=S_E(U_0)+{1\over 2}{{\delta^2 S_E}\over{\delta
U_0^2}}(\delta
U)^2+...\cr}\eqno(1.3)
$$
\medskip
\noindent La deuxi\`eme variation de l' action est un
op\'erateur qui d\'epend de la solution classique:
$$\eqalign{
({{\delta^2 S_E}\over{\delta U_0^2}})_{ab}
=-\partial_\mu\partial_\mu \
\delta_{ab}+V_{ab}(U_0)\cr}\eqno(1.4)
$$
\medskip
\noindent Dans ce qui
va suivre, seule la contribution du mod\`ele sigma non-lin\'eaire
\`a l' op\'erateur potentiel $V_{ab}$ va \^etre prise en compte:
$\displaystyle{V_{ab}(U_0)=i \epsilon_{abc} \tr(\tau_c
U_0^\dagger\partial_\mu U_0) \partial_\mu}$. Nous allons justifier par la suite
cette approximation.

\noindent $W_E$ peut \^etre \'evalu\'ee dans l' approximation de
la phase stationnaire. Cette m\'ethode consiste \`a int\'egrer les
fluctuations de $S_E$ qui sont quadratiques en $\vec\alpha$:
$$\eqalign{
W_E={\cal{N}}e^{-S_E(U_0)} \int {\cal{D}}[\vec\alpha]
\exp{(-{1\over2}\alpha_a  \ ({{\delta^2 S_E}\over{\delta
U_0^2}})_{ab} \ \alpha_b)}\cr}\eqno(1.5)
$$
\medskip
\noindent L' int\'egrale Gaussienne peut \^etre effectu\'ee exactement [41]:
$$\eqalign{
W_E=\tilde {\cal{N}} e^{-S_E(U_0)} \big[
\det \  '{(-\nabla^2_\mu\delta_{ab}+V_{ab}(U_0))}\big]^{-1/2}\cr}\eqno(1.6)
$$
\medskip
\noindent Dans cette expression, le prime sur le d\'eterminant indique
que les valeurs propres nulles de l'
op\'erateur $\displaystyle{({{\delta^2 S_E}\over{\delta U_0^2}})_{ab}
}$ sont omises [44]. Pour extraire de cette formule
$\displaystyle{{\cal{M}}_1}$, il
suffit d' \'ecrire l' expression de $W_E$ sous la forme
$W_E=\exp{(-MT)}$ , o\`u M est la masse du soliton et $T$ est
un facteur de temps Euclidien. Apr\`es avoir effectu\'e la
soustraction de l' \'energie des fluctuations du vide, l' \'energie de Casimir
du soliton s' \'ecrit:
$$\eqalign{
{\cal{M}}_1={1\over {2T}}\bigg\{\tr '\log
{(-\nabla^2_\mu\delta_{ab}+V_{ab})}-\tr \log
{(-\nabla^2_\mu\delta_{ab})}\bigg\}
=-{1\over T}\big[S_{eff}(U_0)-S_{eff}(0)\big]\cr}\eqno(1.7)
$$
\medskip
\noindent Sous cette forme, on comprend pourquoi
$\displaystyle{{\cal{M}}_1}$ est
une \'energie de Casimir. Il s' agit de la diff\'erence entre l' \'energie des
fluctuations du vide avec et sans la pr\'esence de la solution
classique. En remettant les facteurs $\hbar$ \`a leur place, il n' est
pas difficile de voir qu' il s' agit de la premi\`ere correction quantique
\`a la masse. Mais il
 est aussi manifeste sur l' expression (1.7) que
cette \'energie poss\`ede une divergence ultraviolette \`a
laquelle il faudra pr\^eter attention. En effet, l' action effective $S_{eff}$
se d\'eveloppe en th\'eorie des perturbations comme une somme sur
les diagrammes \`a une boucle engendr\'es par le potentiel d'
interaction $V(U_0)$ [42]. Comme le Lagrangien de base est
non-renormalisable, on peut s' attendre \`a des probl\`emes dans le
calcul du $\tr \log$ de l' \'equation (1.7).

\noindent En fait nous n' aurons besoin que d'
un nombre fini de contretermes pour r\'egulariser l' \'energie de
Casimir. Par exemple si $V$ est obtenu en effectuant des
fluctuations autour de la solution classique du Lagrangien du mod\`ele
$\sigma$ non-lin\'eaire, alors toutes les fonctions de Green \`a une
boucle peuvent \^etre rendues finies par des contretermes
d' ordre quatre qui sont connus [43].

\noindent Nous allons maintenant simplifier l' \'equation (1.7) en
tenant compte du fait que le potentiel $V$ ne d\'epend pas du temps
Euclidien. R\'eecrivons la diff\'erence des $\tr\log$ sous la forme [45]:
$$\eqalign{
{\cal{M}}_1={1\over {2T}} \  \intif d\tau \ \tau^{-1}
\tr(e^{\tau\partial_0^2}) \
\tr(e^{-\Omega_0\tau}-e^{-\Omega\tau})\cr}\eqno(1.8)
$$
\medskip
\noindent avec les op\'erateurs tridimensionnels $\Omega=-\nabla_i^2 \
\delta_{ab} \ + \
V_{ab}$ \ \ et $\Omega_0=-\nabla_i^2 \ \delta_{ab}$. Sur la premi\`ere
trace de cette \'equation on reconna\^\i t la fonction de partition de la
particule libre \`a une dimension:
$\displaystyle{\tr(e^{\tau\partial_0^2})={T\over\sqrt{2\pi\tau}}}$. L'
int\'egration sur $\tau$ donne l' expression connue [42] de l' \'energie
de Casimir :
$$\eqalign{
{\cal{M}}_1=&{1\over 2}\tr(\Omega^{1/2}-\Omega_0^{1/2})\cr
=&{1\over 2}(\sum_n \omega_n \ - \sum_n \omega^0_n)\cr}\eqno(1.9)
$$
\medskip
\noindent o\`u $\omega_n^2$ et $(\omega_n^0)^2$ repr\'esentent les valeurs
propres des op\'erateurs $\Omega$ et $\Omega_0$ respectivement.
Notre estimation de $\displaystyle{{\cal{M}}_1}$ sera
bas\'ee sur cette expression.

\noindent Mis \`a part la divergence de la somme sur les
modes qui est une particularit\'e de la th\'eorie quantique des champs,
l' \'equation (1.9) est tout \`a fait naturelle. En effet,
rappelons nous du cas d' une particule non-relativiste \'evoluant
dans un puits de potentiel unidimensionnel $\tilde V$ poss\'edant
un minimum en $x_0$:
$$\eqalign{
\tilde V(x)=\tilde V(x_0)\ + \ {1\over 2}\omega^2 (x-x_0)^2 \ + \lambda
(x-x_0)^4\cr}\eqno(1.10)
$$
\medskip

\noindent Classiquement, la trajectoire dans l' espace des phases
minimisant l' \'energie est celle
pour laquelle la particule reste immobile ($\dot x=0$) en $x_0$.
 Son \'energie classique est $M_{cl}=\tilde V(x_0)$.
Si le couplage est faible (si $\lambda$ est petit et positif) le premier \'etat
quantique a une \'energie
$\displaystyle{M=\tilde V(x_0)+{1\over 2}\hbar \omega}$. En d' autres termes
$M$ est la somme de

a) l' \'energie d' une solution ind\'ependante du temps
des \'equations du mouvement classique et

b) d' une correction quantique
proportionnelle \`a $\omega$, la racine car\'ee de la d\'eriv\'ee seconde
du potentiel prise au point stationnaire $x=x_0$.

L' analogue du terme a) en th\'eorie des
champs est la masse classique du soliton (le terme $N_c{\cal{M}}_0$) et
l' analogue de la correction b) (avec
la d\'eriv\'ee seconde  remplac\'ee par une
d\'eriv\'ee fonctionnelle) est l' \'energie de Casimir du soliton, le
terme ${\cal{M}}_1$. La diff\'erence en th\'eorie quantique
des champs est que pour obtenir la premi\`ere correction quantique il faut
sommer sur une infinit\'e de modes et aussi soustraire l' \'energie de vide.
En particulier l' \'energie de Casimir peut \^etre n\'egative.

\noindent En pratique, le calcul de la trace dans  l' \'equation (1.9) se fait
plus
facilement en se pla\c cant dans un volume fini (mais grand) dans l' espace.
Le spectre des op\'erateurs $\Omega, \Omega_0$ devient alors discret. Le
passage
\`a la limite continue se fait en multipliant les valeurs propres $\omega_n$
associ\'ees au vecteur d' onde $k_n$ par la densit\'e des niveaux. On obtient
alors
l' expression [46]:
$$\eqalign{
{\cal{M}}_1={1\over{2\pi}}\intif dE \ E \ {{d\delta(E)}\over
{dE}}\cr}\eqno(1.11)
$$
\medskip
\noindent o\`u $\delta(E)$ est le d\'ephasage associ\'e au potentiel
$V_{ab}$ pour l' \'energie propre $E$. Comme $V$ est \`a sym\'etrie
sph\'erique, les fonctions
propres se d\'eveloppent sur une base d' harmoniques sph\'eriques
vectorielles indic\'ees par $j$, avec $j=\hat l+\hat
1$, o\`u $\hat l$ est le moment angulaire spatial. Alors le d\'ephasage s'
\'ecrit:
$$\eqalign{
\delta(E)=\sum^\infty_j(2j+1)\delta_j(E)\cr}\eqno(1.12)
$$
\medskip
\noindent Le calcul de l' \'energie de Casimir consiste donc \`a \'evaluer les
d\'ephasages associ\'es au potentiel $V$ de l' op\'erateur tridimensionnel
$\Omega=-\nabla^2\delta_{ab}+V_{ab}$. L' expression de $V$ en fonction
de la solution classique $F$ est:
$$\eqalign{
V_{ab}=2 \ \epsilon_{abc}\bigg\{-{{\sin 2F}\over {2r}}(\delta_{ci}-\hat
r_c \hat r_i)-{{dF}\over{dr}} \hat r_c \hat r_i+ {{\sin^2 F}\over r}
\epsilon_{cim} \hat r_m\bigg\} \nabla_i\cr}\eqno(1.13)
$$
\medskip
\noindent Nous allons nous concentrer maintenant au calcul des d\'ephasages
associ\'es \`a l'
op\'erateur $\Omega$ et \`a leur r\'egularisation.
\bigskip
\leftline{\bfmagc 2. Calcul des d\'ephasages.}
\leftline{\bfmagc \ \ \ Traitement des divergences ultraviolettes}
\bigskip

Les propri\'et\'es de transformation par des rotations d' espace de la
solution classique permettent d' \'ecrire les fonctions propres de l'
op\'erateur perturb\'e $\Omega$ comme une somme sur les ondes partielles $j$:
$$\eqalign{
\vec\alpha=\sum_{j=0}^\infty (u_j(r) \vec Y_{jj-1}+v_j(r) \vec
Y_{jj}+w_j(r) \vec Y_{jj+1})\cr}\eqno(2.1)
$$
\medskip
\noindent Les fonctions $\vec Y$ sont les harmoniques sph\'eriques
vectorielles. L' \'equation aux valeurs propres $\Omega \
\vec\alpha=E^2 \vec\alpha$
se d\'ecompose, pour chaque valeur de $j$,
 en trois \'equations  radiales sur les composantes
$u_j,v_j,w_j$:
$$\eqalign{
&\nabla^2_ru_j=({j(j-1)\over r^2}-E^2+(j-1){2\sin^2F\over r^2})u_j
+C_j(-(j+1){\sin2F\over r^2}v_j-2 {{dF}\over{dr}}\dot v_j)\cr
&\nabla^2_rv_j=({j(j+1)\over r^2}-E^2-{2\sin^2F\over r^2})v_j
+C_j(-(j-1){\sin2F\over r^2}u_j+2 {{dF}\over{dr}}\dot u_j)\cr
&\phantom{\nabla^2_rv_j=({j(j+1)\over r^2}-E^2-{2\sin^2F\over r^2})v_j+}
+S_j(-(j+2){\sin2F\over r^2}w_j-2 {{dF}\over{dr}}\dot w_j)\cr
&\nabla^2_rw_j=({(j+1)(j+2)\over r^2}-E^2-(j+2){2\sin^2F\over r^2})w_j
+S_j(-j    {\sin2F\over r^2}v_j+2 {{dF}\over{dr}}\dot v_j)\cr}\eqno(2.2)
$$
\medskip
\noindent avec $C_j=\sqrt{(j+1)/(2j+1)}$, \ $S_j=\sqrt{j/(2j+1)}$.

\noindent Le comportement asymptotique de la solution ($u_j,v_j,w_j$) de ces
\'equations aux valeurs propres donne le d\'ephasage d\^u au
potentiel $V$ dans l' \'equation de Schrodinger.
En effectuant la somme des trois d\'ephasages
associ\'es aux trois \'equations (2.2) on obtient le d\'ephasage total
$\delta_j(E)$
pour l' onde partielle j,  avec
le facteur de d\'eg\'en\'erescence $(2j+1)$.

\noindent Nous avons montr\'e dans la r\'ef. [47] que les fonctions $\delta_j$
pour
$j=1,2,3...$ se comportent \`a
haute \'energie comme $1/E$ ce qui veut dire que m\^eme avant
de sommer sur toutes les ondes partielles, l' int\'egrale en (1.11)
diverge logarithmiquement \`a grand $E$. L'
introduction d' un sch\'ema de r\'egularisation de fa\c con \`a ce que
l' int\'egrale (1.11) soit d\'efinie s' impose. Elle s' impose d' autant plus
que la
sommation sur les ondes partielles donne naissance \`a une divergence
encore plus violente de sorte que le comportement \`a grand
moment $p$ de la phase totale est:
$$\eqalign{
\delta(p)=\sum_{j=0}^\infty (2j+1) \delta_j(p)\approx a_1 p + {{a_2}\over p} +
...\cr}\eqno(2.3)
$$
\medskip
\noindent Les coefficients $a_1, a_2$ peuvent \^etre reli\'es
au d\'eveloppement du noyau de la chaleur de l' op\'erateur $\Omega$ [48].
Notre m\'ethode de r\'egularisation consiste
\`a calculer non pas la somme divergente $\displaystyle{\sum_n
{(\omega_n^2)}^{1\over 2}}$, mais plut\^ot la somme $\displaystyle{\sum_n
{(\omega_n^2)}^{{1\over 2}-s}}$ o\`u s est une variable complexe. Cette
fonction
est bien d\'efinie sur le plan complexe. En termes de d\'ephasages
$$\eqalign{
{1\over 2}\tr(\Omega^{{1\over 2}-s}-\tr\Omega_0^{{1\over
2}-s})={1\over{2\pi}}\intif dp\ (p^2+M^2)^{{1\over 2}-s} \
{{d\delta(p)}\over {dp}}\cr}\eqno(2.4)
$$
\medskip
\noindent o\`u l' on a introduit une masse $M$ comme r\'egulateur
infrarouge. L' \'energie de Casimir est \'egale \`a cette expression \`a
la limite o\`u $s \to 0$. Pour extraire la partie finie
de cette \'energie, il sera pratique d' utiliser une m\'ethode susceptible de
r\'egulariser aussi les fonctions de Green \`a une boucle qui sont \`a
l' origine du comportement (2.3). Ces
fonctions sont engendr\'ees par l' action effective
(1.7) qui est \'egale \`a $\displaystyle{{1\over
2}{{d\zeta_O}\over{ds}}(s=0)}$. Dans cette expression $\zeta_O$ est
la fonction z\'eta [49]
associ\'ee \`a l' op\'erateur $O=-\partial_0^2+\Omega+M^2$:
$$\eqalign{
\zeta_O(s)={{\mu^{2s}}\over{\Gamma(s)}}\tr\int_0^\infty d\tau
\tau^{s-1}\exp{(-\tau O)}\cr}\eqno(2.5)
$$
\noindent o\`u on a introduit une \'echelle $\mu$.
En fonction des d\'ephasages la fonction $\zeta$ s' \'ecrit:
$$\eqalign{
\zeta_O(s)={T\over{2\pi}}{{\Gamma(s-1/2)}\over{\Gamma(-1/2)\Gamma(s)}}\mu^{2s}\intif
\ dp\ (p^2+M^2)^{{1\over 2}-s} \ {{d\delta(p)}\over {dp}}\cr}\eqno(2.6)
$$
\medskip
\noindent L' astuce consiste maintenant \`a soustraire et rajouter
le comportement asymptotique (2.3) dans l' expression des d\'ephasages,
tout en introduisant un r\'egulateur infrarouge $M^2$:
$$\eqalign{
\delta(p) \ = \
(\delta(p)-a_1p-{{a_2}\over\sqrt{p^2+M^2}})+a_1p+{{a_2}\over\sqrt{p^2+M^2}}\cr}
$$
\noindent En remplacant  cette forme dans l' expression de la
fonction $\zeta$, l' \'energie de Casimir s' \'ecrit comme la somme
d' une int\'egrale finie sur les phases et d' un terme d\'ependant de l'
\'echelle $\mu$. Cette d\'ependance est explicitement calcul\'ee en
utilisant des fonctions b\'eta [45]. Nous obtenons l' expression:
$$\eqalign{
{\cal{M}}_1(\mu)=-{1\over {2T}}{{d\zeta_O}\over{ds}}(s=0)=&{1\over
{2\pi}}\bigg\{ \intif dp
\sqrt{p^2+M^2}\big[{{d\delta(p)}\over{dp}}-a_1+{{a_2}\over{p^2+M^2}}\big]\cr
&+{1\over 4}a_1 M^2\big(1+\log{{\mu^2}\over{M^2}}\big)-{1\over
2}a_2\big(2+\log{{\mu^2}\over{M^2}}\big)-M\delta(0)\bigg\}\cr}\eqno(2.7)
$$
\medskip
\noindent Cette derni\`ere int\'egrale sur les d\'ephasages est finie.
La divergence est implicitement contenue dans la
d\'ependance en $\mu$ de l' expression (2.7). On utilise ensuite l'
identit\'e [45] $\displaystyle{\intif dp
\big\{(p^2+M^2)^{-1/2}-(p^2+\mu^2)^{-1/2}\big\}={1\over 2} \log{
{{\mu^2}\over{M^2}}}}$. Nous int\'egrons par
parties la somme sur les d\'ephasages, pour finalement
prendre la limite $M\to 0$ et arriver \`a l' expression simple
et finie de ${\cal{M}}_1(\mu)$:
$$\eqalign{
{\cal{M}}_1(\mu)=-{1\over{2\pi}}\intif \  dp \
\big(\delta(p)-a_1p-{{a_2}\over\sqrt{p^2+\mu^2}}\big)\cr}\eqno(2.8)
$$
\medskip
\noindent A cette expression il faut ajouter maintenant les
contretermes. Leur r\^ole est pr\'ecis\'ement de faire dispara\^\i tre
la d\'ependance par rapport \`a l' \'echelle de l' \'energie de Casimir.
Les contretermes d' ordre chiral quatre r\'egularisant les
diagrammes \`a une boucle engendr\'es par l' op\'erateur $O$,
ont \'et\'e d\'etermin\'es dans la r\'ef. [43]. On
ajoute leur contribution \`a l' expression de la
correction quantique \`a la masse du soliton. Il vient,
$$\eqalign{
{\cal{M}}_1={\cal{M}}_1(\mu)&-\bigg\{{1\over{32\pi^2}}(\bar
l_1-1+\log{{m_\pi^2}\over{\mu^2}})-{3\over{4e^2}}\bigg\}\int d^3 x {1\over 3}
(\vec L_\mu.\vec L_\mu)^2\cr
&-\bigg\{{1\over{32\pi^2}}(\bar
l_2-1+\log{{m_\pi^2}\over{\mu^2}})+{3\over{8e^2}}\bigg\}\int d^3 x {2\over 3}
(\vec L_\mu.\vec L_\nu)^2\cr}\eqno(2.9)
$$
\medskip
\noindent avec $\vec L_\mu=(1/2i)\tr(\vec\tau U_0^\dagger \partial_\mu
U_0)$, $U_0$ \'etant la solution classique dans le secteur \`a charge
baryonique unit\'e. Les
param\`etres $\bar l_1,\bar l_2$ contiennent la partie finie et
physiquement observable des contretermes d' ordre chiral quatre.
Leurs valeurs num\'eriques ont \'et\'e d\'etermin\'ees dans la r\'ef.
[50]:
$$\eqalign{
\bar l_1=-0.97 \ \pm \ 1.22, \ \ \bar l_2=5.77 \ \pm \
0.72\cr}
$$
\noindent C' est l' expression (2.9) que nous avons utilis\'e
pour les r\'esultats num\'eriques. Il faut pr\'eciser ici que cette
expression de ${\cal{M}}_1$ n' est pas totalement ind\'ependante de l'
\'echelle $\mu$ parce que nous avons tronqu\'e  l' op\'erateur $V$
de fa\c con \`a ne garder que la contribution de la fluctuation autour
de ${\cal{L}}_2$. En effet, on peut montrer que le coefficient
asymptotique $a_2$ associ\'e au potentiel $V$ tronqu\'e, est \'egal \`a un
facteur
multiplicatif pr\`es \`a:
$$\eqalign{
a_2=T\int d^3x \bigg\{{1\over 3}(\vec L_\mu.\vec L_\mu)^2+{2\over
3}(\vec L_\mu.\vec L_\nu)^2-(\partial_\mu \vec
L_\mu)^2\bigg\}\cr}\eqno(2.10)
$$
\medskip
\noindent La d\'ependance en $\mu$ des deux premiers termes va \^etre
absorb\'ee par les contretermes d' ordre quatre de l' \'equation (2.9).
Par contre, pour faire dispara\^\i tre la d\'ependance en $\mu$
provenant du terme en $(\partial_\mu\vec L_\mu)^2$ il faudrait inclure des
contretermes d' ordre sup\'erieur \`a quatre. Il est clair qu' en raison de l'
\'equation du mouvement classique du
mod\`ele de Skyrme:
$$\eqalign{
\partial_\mu L_\mu^k={1\over{(ef_\pi)^2}}\partial_\mu\big[L_\mu^k(\vec
L_\nu.\vec L_\nu)-L_\nu^k(\vec L_\mu.\vec L_\nu)\big],\cr}\eqno(2.11)
$$
\medskip
\noindent la quantit\'e $(\partial_\mu \vec
L_\mu)^2$ est d' ordre chiral huit. L'
\'energie de Casimir sera  compl\`etement ind\'ependante de l' \'echelle $\mu$
si on y ajoute les contretermes correspondants .
Nous n' allons pas effectuer cette derni\`ere soustraction ici, la partie
finie (et donc physique) de ces contretermes n' \'etant pas connue \`a l'
heure actuelle. Nous justifierons par la suite cette approximation.

\noindent En l' absence du terme de Skyrme (sa contribution
\`a l' \'equation du mouvement figure dans le membre de droite de l'
\'equation (2.11)), la quantit\'e $\partial_\mu L_\mu$ est
identiquement nulle, et l' \'energie de Casimir peut-\^etre
rendue compl\`etement ind\'ependante de l' \'echelle.
Mais dans ce cas il n' y a pas de solution classique de type soliton.

Nous allons maintenant pr\'esenter les r\'esultats pour la premi\`ere
correction \`a la masse du soliton. Nous calculerons ${\cal{M}}_1$ dans
deux cas: celui du mod\`ele de Skyrme et celui o\`u l' on ajoute \`a ce
dernier un terme d' ordre six.
\bigskip
\leftline{\bfmagc 3. R\'esultats}
\bigskip

Les phases de l' op\'erateur $\Omega$ ont \'et\'e calcul\'ees dans la r\'ef.
[47]. Elles ont \'et\'e calcul\'ees pour les solutions classiques

\noindent I) du mod\`ele de Skyrme:
$\displaystyle{{\cal{L}}_{SK}={{f_\pi^2}\over
4}\tr(\partial_\mu U\partial_\mu
U^\dagger)+{1\over{32e^2}}\tr([\partial_\mu UU^\dagger,\partial_\nu
UU^\dagger]^2)}$

\noindent II) du mod\`ele o\`u l' on inclut le terme d' ordre six d\'ej\`a
rencontr\'e lors du
chapitre pr\'ec\'edent:
$\displaystyle{{\cal{L}}_{2+4+6}={\cal{L}}_{SK}-{b\over{2f_\pi^2}}
\big[{{\epsilon_{\mu\nu\alpha\beta}}\over{24\pi^2}}\tr(\partial^\nu
UU^\dagger\partial^\alpha UU^\dagger\partial^\beta UU^\dagger)\big]^2}$

Nous avons trouv\'e que les ondes partielles (\`a l'
\'exception de l' onde $j=1$) ont un d\'ephasage nul \`a $p=0$,
et que pour $j$ grand les phases s' \lq\lq aplatissent" \`a petit $p$. Le
fait le plus important pour l' \'energie de Casimir est que
le  d\'ephasage pour l' onde $j=1$ soit \'egal \`a $2\pi$ pour $p=0$.
Ce comportement n' est pas d\^u \`a la pr\'esence de vrais \'etats li\'es de l'
op\'erateur
$\Omega$, mais \`a la pr\'esence de deux \lq\lq modes z\'ero" i.e.
deux fonctions propres \`a \'energie nulle. Leur existence traduit l'
invariance de l' \'energie ${\cal{M}}_0[U_0]$
par des transformations continues de la solution classique $U_0$. Il
existent quatre telles op\'erations continues et deux d' entre elles
donnent naissance \`a des modes propres normalisables. Chacun contribue
d' un facteur $\pi$ au d\'ephasage \`a l' origine de sorte que l' on
ait: $\displaystyle{\delta_{j=1}(p=0)=2\pi}$. Ce r\'esultat est fortuit
car \`a priori l' op\'erateur tronqu\'e $V_{ab}$ n' a pas
les m\^emes modes z\'ero que l' op\'erateur complet. Cette circonstance
justifie partiellement notre approximation de n\'egliger les
contributions \`a l' op\'erateur de fluctuation $V$, provenant du terme
de Skyrme.

\noindent En effectuant la somme sur $j$ des d\'ephasages $\delta_j$
(\'equation (1.12)), nous obtenons la
fonction de phase totale $\delta(p)$. Nous avons v\'erifi\'e que pour des
valeurs de $p$ grandes (en pratique $p>5\mu$) le comportement de la  solution
num\'erique $\delta(p)$ est conforme aux pr\'edictions analytiques
(\'equation (2.3)). L' int\'egrand de l' \'equation
(2.8) est montr\'e sur la figure (2.1).

\noindent Observons que les courbes partent d' une valeur proche de
$6\pi$ et qu' elles deviennent pratiquement nulles vers $p\approx
0.5-0.7$ GeV. Ceci veut dire que ce sont les fluctuations \`a basse
\'energie qui dominent l' \'energie de Casimir. Ce r\'esultat est tr\`es
important car il justifie en quelque sorte la validit\'e d' un Lagrangien
effectif tronqu\'e \`a un ordre chiral donn\'e.  Il est clair que
l' \'energie de Casimir du soliton est contr\^ol\'ee par
les propri\'et\'es d' invariance de la solution classique et
par son extension spatiale.

Est-ce que les d\'ephasages vont changer si l' on inclut dans l' op\'erateur
$V_{ab}$ les fluctuations provenant du terme de Skyrme et du terme d'
ordre six qui ont \'et\'e n\'eglig\'ees
dans notre travail? La r\'eponse est qu' ils vont diff\'erer
sensiblement mais seulement \`a haute \'energie. Les phases de l'
\'op\'erateur complet du mod\`ele de Skyrme ont \'et\'e calcul\'ees il y
a assez longtemps par les auteurs de la r\'ef. [51]. A premi\`ere vue
ces phases ne ressemblent en rien aux n\^otres. En particulier, elles
divergent lin\'eairement en $p$. Ceci veut dire que la phase totale
de l' op\'erateur complet poss\`ede  une divergence cubique: $\delta(p)\approx
a_0 p^3+...$, mais d' apr\`es ce que nous venons de voir, elle n'
intervient pas dans l' \'energie de
Casimir, car elle est essentiellement soustraite de l'
expression \`a int\'egrer (2.8). C' est le comportement \`a
basse \'energie qui domine ${\cal{M}}_1$ et on s' aper\c coit que dans la
r\'egion des petits $p$ les phases trouv\'ees dans [51] sont
qualitativement proches des n\^otres. On \'esp\`ere aussi que la
valeur \`a l' origine de la phase totale $\delta(0)=6\pi-{{a_2}\over
\mu}$ ne va pas \^etre tr\`es modifi\'ee pour le cas de l' op\'erateur
complet $V_{ab}$. Les
r\'esultats de la r\'ef. [52] indiquent que dans ce cas le coefficient $a_2$
pour le mod\`ele de Skyrme ne serait pas tr\`es diff\'erent du n\^otre,
de sorte que la relation $\delta(0)\approx 6\pi$ ait un caract\`ere
g\'en\'eral.

Pour le mod\`ele de Skyrme nous avons  trouv\'e pour la premi\`ere
correction quantique \`a la masse du soliton:
$({\cal{M}}_1)_{SK}=-1.169+(-0.070)$ GeV \hfill (3.1)
\medskip
\noindent Entre parenth\`eses figure la contribution du contreterme.
Comme nous l' avions annonc\'e, c' est l' int\'egrale sur les d\'ephasages
qui domine le r\'esultat, la contribution du contreterme est de $\approx
\ 6$ \%. On
s' attend \`a ce que les contretermes (inconnus) d' ordre sup\'erieur
\`a quatre soient encore plus petits. M\^eme si le
r\'esultat (3.1) n' est pas compl\'etement ind\'ependant de l'
\'echelle $\mu$ choisie, nous pensons qu' il constitue une bonne approximation
de l'
\'energie de Casimir du soliton pour des \'echelles typiques des masses
des hadrons. Nous avons \'etudi\'e la variation du r\'esultat en fonction de l'
\'echelle, et nous avons
trouv\'e que si au lieu de $\mu=m_\rho$ on prend $\mu=2m_\rho$ par exemple,
${\cal{M}}_1$ change \`a peu pr\`es de $10$ \% ce qui semble
raisonnablement petit.

Pour ce qui est de la valeur de ${\cal{M}}_1$ remarquons d' abord que
son signe est n\'egatif. Ce signe est en accord avec les r\'efs.
[53-54]. Par contre la valeur de ${\cal{M}}_1$ est
quatre \`a cinq fois plus grande que celle trouv\'ee dans ces travaux.
La raison physique de ce d\'esaccord est tr\`es simple; les calculs [53-54]
sont bas\'es sur une approximation de Born des
d\'ephasages ignorant l' existence d' \'etats propres non-perturbatifs comme
les modes
z\'ero. Or ces \'etats, nous l' avons vu, dominent l' \'energie de Casimir par
leur
contribution au d\'ephasage $\delta_1$, et ne peuvent \^etre n\'eglig\'es.
De fa\c con tr\`es g\'en\'erale et \`a cause du th\'eor\`eme de Levinson, ces
modes z\'ero \lq\lq forcent" la fonction $\delta(p)$ d' \^etre \'egale \`a
$6\pi$ \`a l' origine
 $p=0$. Remarquons ici que la premi\`ere correction quantique \`a la
masse des solitons dans les th\'eories en $1+1$ dimensions est donn\'ee par une
simple
somme sur les \'etats li\'es de l' op\'erateur de fluctuation [55].

Le r\'esultat (3.1) nous am\`ene \`a
\'emettre des doutes sur la validit\'e de l' approximation
semi-classique pour le calcul de la masse du nucl\'eon dans le mod\`ele de
Skyrme.
Le rapport de la premi\`ere correction quantique sur la masse
classique (qui est \'egale \`a $1.23$ GeV) est:
{\bf $$\eqalign{
({{{\cal{M}}_1}\over{N_c{\cal{M}}_0}})_{SK} \ = \ -1\cr}
$$}
\medskip
\noindent Cette valeur nous fait penser que l'
approximation semi-classique ne constitue pas un tr\`es bon cadre pour traiter
le soliton
du mod\`ele de Skyrme car ce dernier est sujet \`a des effets quantiques
tr\`es importants. Ces derniers ne peuvent pas vraiment \^etre interpr\'et\'es
comme des \lq\lq corrections" devant la masse classique. Le \lq\lq skyrmion"
est
donc un soliton \`a couplage fort, ce qui n' est pas adapt\'e \`a
l' approche des Lagrangiens effectifs o\`u l' on cherche \`a construire
une approximation \`a couplage faible.

Nos espoirs pour la description semi-classique du nucl\'eon ne sont pas
vains moyennant quelques consid\'erations ph\'enom\'enologiques du m\^eme type
que
celles qui conduisent \`a g\'en\'eraliser le mod\`ele de Skyrme en
incluant les m\'esons de basse \'energie. Pour \'etudier leurs effets,
prenons pour simplifier le Lagrangien local
${\cal{L}}_{2+4+6}$ qui correspond
\`a la limite des grandes masses du mod\`ele unifi\'e propos\'e lors du
chapitre I.

\noindent Calculant l' \'energie de Casimir avec la solution classique de
ce nouveau Lagrangien (les phases sont trac\'ees sur la figure (2.1))
 nous trouvons qu' elle diminue notablement:
$({\cal{M}}_1)_{2+4+6}=-0.795+(-0.035)$ GeV \hfill (3.2)

\noindent La contribution du contreterme est aussi diminu\'ee par rapport
au mod\`ele de Skyrme. La masse classique, $N_c{\cal{M}}_0$ est plus
grande ($1.59$ GeV) dans ce mod\`ele et le rapport de la premi\`ere correction
quantique sur le terme dominant de la masse du nucl\'eon est:
{\bf $$\eqalign{
({{{\cal{M}}_1}\over{N_c{\cal{M}}_0}})_{II} \ = \ -0.52\cr}
$$}
\medskip
\noindent La diminution de l' \'energie de Casimir du soliton dans le
mod\`ele ${\cal{L}}_{2+4+6}$ peut \^etre expliqu\'ee physiquement par une
augmentation
de la taille du soliton dans ce dernier par rapport au soliton du
mod\`ele de Skyrme. On pourrait \`a la limite \lq\lq deviner" ce r\'esultat
simplement en observant la courbe des phases (figure (2.1)). Sur cette courbe
il est clair que l'  inclusion du terme d' ordre six a comme effet de
r\'eduire la r\'egion pour laquelle la phase est importante, et par
cons\'equent l' effet Casimir sur la masse du soliton. Cet effet a \'et\'e
aussi
observ\'e dans [56] dans une \'etude des effets quantiques
vibrationnels \`a la masse. Le fait que les effets quantiques soient plus
petits pour un
objet ayant une extension spatiale plus importante n' est qu' une simple
manifestation du principe d' incertitude de Heisenberg.

\noindent On trouve donc que contrairement au mod\`ele de Skyrme, une
simple g\'en\'eralisation \`a un Lagrangien effectif qui contient les effets
des interactions du pion avec  un m\'eson $\omega$ infiniment massif (ou
tout simplement l' inclusion d' un terme d' ordre sup\'erieur dans le
d\'eveloppement chiral)
conduit \`a un secteur baryonique consistent avec le d\'eveloppement
semi-classique. Le profil classique dans ce dernier mod\`ele  conduit
\`a une hi\'erarchie bien d\'efinie parmi les diff\'erentes contributions
\`a la masse du nucl\'eon:
$$\eqalign{
N_c {\cal{M}}_0 \ > \ {\cal{M}}_1 \ >
{1\over{N_c}}{\cal{M}}_2\cr}\eqno(3.3)
$$
\medskip
\noindent Le profil de la solution classique d\'ecro\^\i t plus vite
en pr\'esence du terme d' ordre six, de sorte que l' on ait:
$$\eqalign{
-\int d^3 x {\cal{L}}_2 \ > \ -\int d^3 x {\cal{L}}_4 \ > \ -\int d^3 x
{\cal{L}}_6\cr}\eqno(3.4)
$$
\medskip
\noindent ce qui justifie encore plus notre approximation qui consiste
\`a n\'egliger les contretermes d' ordre six, huit etc. dans le
d\'eveloppement chiral.

\noindent Une derni\`ere remarque concernant la constante de couplage axiale
du nucl\'eon $g_A$. La valeur de cette derni\`ere  est grande
au niveau classique pour le mod\`ele ${\cal{L}}_{2+4+6}$ ($g_A=1.37$). M\^eme
si
nous n' avons pas calcul\'e les contributions des boucles de pions \`a
cette observable on peut penser, \`a cause des r\'esultats qualitatifs
de la r\'ef. [53], que ces corrections vont dans la bonne direction. Il faut
pr\'eciser ici qu' un calcul complet de ces corrections quantiques \`a
$g_A$ va probablement se compliquer par l' introduction des coordonn\'ees
collectives du nucl\'eon.

\eject

\leftline{\bfmagc 4. Conclusions}

\bigskip

Nous avons vu dans ce chapitre que le coefficient ${\cal{M}}_1$
constituant la premi\`ere correction quantique \`a la masse du
soliton-nucl\'eon d\'epend fortement de la structure sp\'ecifique du
Lagrangien effectif. En particulier, nous avons montr\'e la n\'ecessit\'e
d' inclure des termes d' ordre sup\'erieur \`a quatre dans
le d\'eveloppement chiral. Il est maintenant clair que le
d\'eveloppement en $\hbar/N_c$ a des chances de fournir une bonne
description du secteur baryonique, seulement quand le mod\`ele de Skyrme
est g\'en\'eralis\'e de fa\c con \`a inclure certains des effets du m\'eson
$\omega$.
Ces conclusions restent valables quand on envisage d' effectuer les calculs de
l' \'energie de Casimir avec l' op\'erateur $\Omega$ complet [57].

\noindent Il est
int\'eressant de remarquer qu' on arrive qualitativement \`a la m\^eme
conclusion qu' au chapitre I, o\`u nous proposons  un Lagrangien
effectif pour la description unifi\'ee des m\'esons et des baryons. Le
fait que le d\'eveloppement semi-classique des observables baryoniques
(\'equation (I.1)) est meilleur dans le mod\`ele
${\cal{L}}_{2+4+6}$ est en accord avec l' id\'ee d' inclure
dans le Lagrangien Effectif plus de degr\'es de libert\'e
que ceux qui sont pr\'esents dans le mod\`ele de Skyrme [58].

Le cadre th\'eorique pour calculer syst\'ematiquement les corrections
quantiques \`a la masse du nucl\'eon
dans le contexte de la th\'eorie de perturbation chirale a
\'et\'e r\'ecemment propos\'e [59].
Connaissant le r\^ole important que jouent les m\'esons vecteurs
$\rho,\omega$ et scalaires dans la physique de la diffusion $\pi\pi$ il
serait tr\`es interessant dans l' avenir d' aller plus loin et de calculer
l' \'energie de Casimir dans un mod\`ele o\`u les m\'esons scalaires et
vecteurs sont explicitement pr\'esents. L' issue d' un tel calcul
sera de la plus haute importance pour le d\'eveloppement
semi-classique des observables baryoniques et de son utilit\'e pratique.

\noindent Un aspect qui ne doit pas \^etre n\'eglig\'e dans les futures
\'etudes
des fluctuations quantiques autour des solutions classiques des
Lagrangiens effectifs est la d\'etermination des contretermes d' ordre
sup\'erieur. M\^eme si comme on a vu leur partie finie ne doit pas
affecter beaucoup la magnitude de la masse du nucl\'eon, ils
sont n\'ecessaires pour \'eliminer la d\'ependance par rapport \`a l'
\'echelle $\mu$ des observables physiques.

\noindent Pour ce qui est des  ordres sup\'erieurs \`a quatre
 dans le d\'eveloppement chiral, tr\`es peu d' auteurs ont entrepris une
\'etude
syst\'ematique pour justifier le choix des termes \`a y inclure. Ces
\'etudes sont compliqu\'ees par le manque d' informations
experimentales sur les constantes qui figurent devant ces
termes d' ordre sup\'erieur. On peut \'eventuellement envisager d'
introduire des contraintes pour sp\'ecifier la forme du Lagrangien
effectif aux ordres chiraux \'elev\'es, comme il a \'et\'e sugg\'er\'e
par quelques auteurs [60-61].

\noindent Terminons ce chapitre en faisant une derni\`ere
remarque sur les calculs de la r\'esonance Roper du nucl\'eon avec ces
mod\`eles de soliton topologique. D' apr\`es nos r\'esultats il ne
semble pas \'etonnant que le calcul
de cette observable dans le cadre du mod\`ele de Skyrme soit en
d\'esaccord avec la ph\'enom\'enologie. Une premi\`ere am\'elioration de la
pr\'ediction pour la masse de cette r\'esonance peut-\^etre envisag\'ee apr\`es
l' inclusion du terme d' ordre six.

\eject

\rightline {}
\rightline {}
\rightline {}
{\rightline {}}
{\rightline {}}
\bigskip
\bigskip
\bigskip
\bigskip
\centerline {\bfmag Chapitre III}
\bigskip
\bigskip
\bigskip
\centerline {\bfmagd Sur la stabilit\'e des solitons topologiques}
\eject

Le nucl\'eon \'etant une particule stable, cette
stabilit\'e doit \^etre retrouv\'ee dans les mod\`eles o\`u le
nucl\'eon est consid\'er\'e comme un soliton topologique.
Dans ce contexte, il faut donc s' assurer non seulement de l' existence de
la solution de type soliton, mais aussi de la positivit\'e du spectre des
fluctuations autour de cette derni\`ere.

Cette question de la stabilit\'e a \'et\'e d' abord pos\'ee par Skyrme
dans son travail original. Il a remarqu\'e que le mod\`ele $\sigma$
non-lin\'eaire ne poss\`ede pas de solution non-triviale
stable par rapport aux dilatations. Pour construire une solution de type
soliton il a propos\'e d' ajouter \`a ce mod\`ele un terme d' ordre
quatre par rapport au gradient du champ unitaire $U$. Plus tard il
a \'et\'e propos\'e de stabiliser le mod\`ele $\sigma$ non-lin\'eaire
par des termes couplant le champ du pion au m\'eson $\rho$, ce
dernier \'etant consid\'er\'e comme un champ de jauge chirale. La
solution classique du syst\`eme $\pi\rho$ trouv\'ee dans ce contexte [62]
est instable comme il a \'et\'e montr\'e plus tard [63]. Ce r\'esultat m\'erite
inspection car \`a la limite o\`u le champ du
$\rho$ devient tr\`es lourd, le syst\`eme
$\pi\rho$ consid\'er\'e dans ces travaux, tend vers le mod\`ele de
Skyrme qui bien s\^ur poss\`ede des solutions stables. L'
instabilit\'e du soliton quand le pion se couple au champ du $\rho$ est d'
autant plus \'etonnante, si on se rappelle que le syst\`eme $\pi\omega$
[23], du point de vue de la stabilit\'e, est parfaitement
coh\'erent [32] avec sa limite locale.
\bigskip
\leftline{\bfmagc 1. R\'ealisations non-lin\'eaires de la sym\'etrie
chirale}
\bigskip

Dans ce chapitre nous allons \'etudier en d\'etail le syst\`eme
$\pi\rho$ dans le but d' \'eclaircir ce probl\`eme d' instabilit\'e, et
de lever les ambiguit\'es qui existent dans la litt\'erature.
L' accent sera mis sur les diff\'erentes mani\`eres
possibles d' introduire le m\'eson $\rho$ dans les Lagrangiens
ph\'enom\'enologiques, quand la sym\'etrie chirale est
r\'ealis\'ee d' une fa\c con non-lin\'eaire [64].
La r\'ealisation non-lin\'eaire est d\'efinie en sp\'ecifiant
l' action de l' \'el\'ement $G$ de
$SU(2)\otimes SU(2)$ sur les \'el\'ements $u(\vec\pi)$  de l' espace
quotient $SU(2)\otimes SU(2)/SU(2)_V$:
$$\eqalign{
u(\vec\pi)\to^{G} \  g_L u(\vec\pi) h^\dagger (\vec\pi)=h(\vec\pi) u(\vec\pi)
g_R^\dagger\cr}\eqno(1.1)
$$
\noindent $u(\vec\pi)$ \'etant la racine carr\'ee du champ chiral:
$u=U^{1/2}=e^{\displaystyle{i\vec\tau.\vec\pi/2}}$. $g_L$ et $g_R$ sont
des matrices de $SU(2)_L$ et $SU(2)_R$ respectivement. Il est
clair sur l' \'equation (1.1) que cette loi de
transformation pour $u$ est non-lin\'eaire car la matrice de rotation $h$
doit d\'ependre du champ du pion pour satisfaire \`a l' \'egalit\'e (1.1). Avec
cette
d\'efinition le champ chiral $U$ se transforme
lin\'eairement: $U \to \  g_L U g_R^\dagger$.

Deux repr\'esentations diff\'erentes pour le
champ du m\'eson $\rho$ vont \^etre consid\'er\'ees ici.
L' une, conventionnelle [65], consiste \`a supposer que ce champ
vecteur ($\tilde V_\mu$) se transforme comme un
boson de jauge du groupe \lq\lq cach\'e"  $h(\pi)$:
$$\eqalign{
\tilde V_\mu\to  h(\pi) \tilde V_\mu h^\dagger(\pi)+{i\over g}
h(\pi)\partial_\mu
h(\pi)^\dagger\cr}\eqno(1.2)
$$
\noindent
\noindent Dans la section 2 nous allons revoir ce qui a \'et\'e fait
dans la litt\'erature en ce qui concerne l' instabilit\'e du soliton dans
cette approche. La section 3 sera consacr\'ee \`a l' \'etude d\'etaill\'ee de
la stabilit\'e du soliton dans le cas d'
une transformation homog\`ene pour les champs vecteurs:
$$\eqalign{
V_\mu\to  h(\pi) V_\mu h^\dagger(\pi) \cr}\eqno(1.3)
$$
\medskip

\bigskip

\leftline{\bfmagc 2. Le soliton du syst\`eme $\pi\rho$
dans une formulation \lq\lq Yang-Mills"}

\bigskip

\noindent Il n' est pas difficile de montrer que le Lagrangien minimal
respectant la sym\'etrie chirale dans la repr\'esentation d\'efinie par les
lois (1.1) et (1.2), s' \'ecrit comme:
$$\eqalign{
{\cal{L}}_{\pi\rho}^{YM}=
-{1\over 4}\tr (\tilde V_{\mu\nu}\tilde V^{\mu\nu}) +{1\over 2}
M_\rho^2\tr(\big[\tilde V_\mu-{i\over g}\Gamma_\mu\big]^2)+{{f_\pi^2}\over
4}\tr(u^\mu u_\mu)\cr}\eqno(2.1)
$$
\medskip
\noindent avec $\tilde V_{\mu\nu}=\partial_\mu \tilde V_\nu-\partial_\nu
\tilde V_\mu-ig
[\tilde V_\mu,\tilde V_\nu]$, $\Gamma_\mu={1\over 2}(u^\dagger\partial_\mu
u+u\partial_\mu
u^\dagger)$ et $u_\mu=i(u^\dagger\partial_\mu u-u\partial_\mu
u^\dagger)$.

Le Lagrangien $\displaystyle{{\cal{L}}_{\pi\rho}^{YM}}$ contient outre
les termes quadratiques, des termes cubiques et
quartiques par rapport aux champs vecteurs. Le couplage
$\pi\rho$ dans le secteur des m\'esons est en accord avec
la notion de dominance vectorielle [65], et de fa\c con
g\'en\'erale ce Lagrangien d\'ecrit raisonnablement bien les
interactions entre les m\'esons $\pi$ et $\rho$.

\noindent Passons maintenant au secteur baryonique et remarquons
qu' \`a la limite o\`u la masse du m\'eson
$\rho$ devient grande, le terme de Yang-Mills [66] $\tilde
V_{\mu\nu}^2$ donne naissance \`a un terme de Skyrme dans le
d\'eveloppement local. Par cette observation on pourrait penser que
(2.1) poss\`ede des solutions classiques
stables. Il n' en est rien. Pour le voir il faut d' abord \'ecrire
la masse du soliton en termes de profils sph\'eriques en adoptant l'
ansatz du h\'erisson pour le champ chiral et la configuration la
plus g\'en\'erale pour les composantes spatiales du champ $\tilde V_\mu$
classique:
$$\eqalign{
\tilde V_{i}=\tilde v_1(\tau_i-(\vec\tau.\hat r)
\hat r_i)+\tilde v_2(\vec\tau.\hat r)\hat r_i+
\tilde v_3(\vec\tau\times\hat r)_i\cr}\eqno(2.2)
$$
\medskip
\noindent Alors la masse statique du soliton, exprim\'ee en termes des
profils radiaux $F(r),\tilde v_i(r) \ (i=1,2,3)$ se met sous la forme
$$\eqalign{
M=&4\pi\intif dr r^2 \big\{{{f_\pi^2}\over 2}\big[\dot F^2+2{{\sin^2
F}\over {r^2}}\big]+M_\rho^2\big[2\tilde v_1^2+\tilde v_2^2+2(\tilde
v_3+{1\over{2gr}}(1-\cos F))^2\big]\cr
&+4({{\tilde v_3}\over r}+g(\tilde v_1^2+\tilde v_3^2))^2+2(\tilde{\dot
v}_3+{{\tilde v_3}\over r}+2g\tilde v_1\tilde v_2)^2+2(\tilde{\dot
v}_1+{{\tilde v_1-\tilde v_2}\over r}-2g\tilde v_2\tilde v_3)^2\big\}
\cr}\eqno(2.3)
$$
\medskip
\noindent Il existe des configurations sp\'eciales pour lesquelles $M$
poss\`ede
des solutions classiques. Par exemple, les auteurs de la r\'ef. [62] ont
montr\'e qu' il
existe une solution non-triviale des \'equations du mouvement ${F(r),\tilde
v_3(r), \tilde v_1=\tilde v_2=0}$ (avec $\tilde v_3={{-G(r)}\over{2r}}$ pour
retrouver leurs
notations). Cependant, il a \'et\'e ult\'erieurement
montr\'e par les auteurs de la r\'ef. [63], que cette solution n' est pas
stable
car elle peut \^etre rendue triviale ($F=\tilde v_3=0$) par une
d\'eformation continue des composantes $\tilde v_1,\tilde v_3$.
Cette d\'eformation est possible \`a \'effectuer dans l' approche de
Yang-Mills, car les
composantes $\tilde v_1,\tilde v_2$ se couplent \`a $\tilde v_3$.  Comme
on peut voir sur l' \'equation (2.3) ces couplages d\'estabilisants ont
lieu dans les termes cubiques et quartiques de la densit\'e
d' \'energie de Yang-Mills.   Ces instabilit\'es se manifestent aussi
dans le cas le plus g\'en\'eral qui consiste \`a
extr\'emiser la fonctionnelle (2.3) non seulement par rapport aux
fluctuations de $F,\tilde v_3$ mais aussi par rapport \`a celles
de $\tilde v_1,\tilde v_2$. Ceci a \'et\'e \'etudi\'e dans [67]
o\`u il est trouv\'e que la seule solution est celle qui correspond au vide
trivial $F=\tilde v_i=0 \ (i=1,2,3)$, donnant ainsi
$M=0$. Il n' existe pas de solution stable dans le secteur non-trivial
du Lagrangien (2.3).

Une autre propri\'et\'e des solitons du Lagrangien (2.3) doit
retenir notre attention ici. Supposons que l' on ajoute \`a
${\cal{L}}_{\pi\rho}^{YM}$ des termes stabilisateurs d' ordre
quatre en d\'eriv\'ees du champ $U$. Dans ce cas, il existe des solutions
classiques non-triviales pour les fonctions $F,
\tilde v_1,\tilde v_2,\tilde v_3$ [67], mais un nouveau probl\`eme
appara\^\i t: l' existence de deux solitons
d\'eg\'en\'er\'es dans le spectre classique. Ceci est
d\^u \`a l' invariance de la masse classique du
soliton par rapport \`a la transformation discr\`ete des
fonctions de profils :
$$\eqalign{
\tilde v_1 \ & \to -\tilde v_1\cr
\tilde v_2 \ & \to -\tilde v_2\cr
\tilde v_3 \ & \to \tilde v_3\cr}\eqno(2.4)
$$
\medskip
\noindent Les solutions $(F,\tilde v_1,\tilde v_2,\tilde v_3)$ et
$(F,-\tilde v_1,-\tilde v_2,\tilde v_3)$ forment un
doublet de solitons. Une fois
quantifi\'e, ce doublet de solitons donnera naissance \`a deux \'etats
baryoniques
quasi-d\'eg\'en\'er\'es. Ceci n' est \'evidemment pas observ\'e dans la
Nature.

\noindent Quelques mots sur ce doublet de baryons. Les auteurs de
[67] l' ont interpret\'e comme un \lq\lq doublet de parit\'e". Pour
notre part, nous nous gardons de l' appeller ainsi, car la transformation
(2.4) n' a rien d' une op\'eration de parit\'e telle qu' elle est d\'efinie en
th\'eorie des perturbations [68]. Pour le voir il suffit d' \'ecrire le
champ du vecteur de fa\c con \`a expliciter les indices de spin et d'
isospin dans la configuration (2.2):
$\tilde V_i=\tau_k\tilde V_{ki}$
$$\eqalign{
\tilde V_{ki}=\tilde v_1(\delta_{ki}-\hat r_k\hat r_i)+
\tilde v_2\hat r_k\hat r_i-\tilde v_3\epsilon_{kim}\hat
r_m\cr}\eqno(2.5)
$$
\noindent Alors la
transformation (2.4) pour les composantes du champ du $\rho$,
s' exprime de la  fa\c con compacte:
$$\eqalign{
\tilde V_{ki} \ \to -\tilde V_{ik}\cr}\eqno(2.6)
$$
\medskip
\noindent Cette transformation n' a rien d' habituel, car il s' agit d'
inverser le signe du champ
et \'echanger les indices de spin et d' isospin. Cette
sym\'etrie ne peut exister que dans le secteur topologique du
Lagrangien (2.1), o\`u pr\'ecis\'ement les indices de spin et d' isospin
des champs sont confondus \`a ceux de l' espace de configuration pour un
h\'erisson. C' est la raison pour laquelle nous
pensons qu' il ne faut pas interpr\'eter
cette transformation comme une transformation de parit\'e.

Pour r\'esumer donc, quand le m\'eson $\rho$ est suppos\'e se transformer
comme un boson de jauge du groupe cach\'e $h$ (\'equation (1.2)), le
syst\`eme $\pi\rho$ ne poss\`ede pas de solution
classique stable en l' absence de termes d'
ordre sup\'erieur. En pr\'esence de termes stabilisateurs,
il appara\^\i t un doublet non-physique d' \'etats baryoniques.

\bigskip

\leftline{\bfmagc 3. Transformation homog\`ene des champs vecteurs}

\bigskip

\noindent  Comme nous pensons qu' il n' y a aucune raison
viable de consid\'erer que le m\'eson $\rho$ est un boson
de jauge nous postulons une loi de
transformation homog\`ene dans ce qui suit:
$$\eqalign{
V_\mu\to  h(\pi) V_\mu h^\dagger(\pi) \cr}\eqno(3.1)
$$
\medskip

\noindent \noindent $h$ \'etant la matrice d\'efinie par l'
\'equation (1.1). Puisque $h$ est locale, les gradients doivent \^etre
g\'en\'eralis\'es pour se transformer eux aussi comme les champs
$V_\mu$. Pour cela il faut y ajouter un terme contenant la
quantit\'e $\Gamma_\mu$ d\'efinie  dans la section pr\'ec\'edente:
$$\eqalign{
\nabla_{\mu} V_\nu=&\partial_{\mu} V_\nu+ [\Gamma_\mu,V_\nu]\cr}
$$
\medskip
\noindent Le Lagrangien du
syst\`eme $\pi\rho$, \`a l' ordre le plus bas par rapport aux champs vecteurs,
et
invariant par les transformations (1.1) et (3.1), est donn\'e par:
$$\eqalign{
{\cal{L}}_V=-{1\over
4}\tr\bigg\{(V_{\mu\nu}+i{{g_V}\over{\sqrt{2}}}[u_\mu,u_\nu])^2\bigg\}+{{M_{\rho}^2}\over
2}\tr(V_\mu V^\mu)+{{f_\pi^2}\over 4}\tr(u^\mu u_\mu)\cr}\eqno(3.2)
$$
\medskip
\noindent o\`u $\displaystyle{V_{\mu\nu}=\nabla_\mu V_\nu-\nabla_\nu
V_\mu}$,  $g_V$ est la constante de couplage du pion au
m\'eson $\rho$. Comme il est montr\'e dans l' Appendice A, le
Lagrangien (3.2) est le
plus simple conduisant \`a un Hamiltonien qui soit born\'e
inf\'erieurement. Ce Lagrangien est quadratique par rapport
aux champs vecteurs.

\noindent Nous allons \'etudier la
stabilit\'e du secteur topologique dans le syst\`eme $\pi\rho$, mais nous
voulons aussi \'etablir une analogie entre le m\'ecanisme de stabilisation
dans le syst\`eme $\pi\rho$ et celui du syst\`eme $\pi\omega$. En fait,
on peut montrer que dans ce dernier mod\`ele,  la stabilit\'e est d\^ue \`a une
contrainte secondaire \`a laquelle le champ du $\omega$ ob\'eit. Pour mieux
illustrer
le r\^ole jou\'e par les contraintes dans le
m\'ecanisme de stabilisation du soliton pour le
syst\`eme $\pi\rho$,  nous avons d\'elib\'er\'ement
choisi de ne pas consid\'erer dans ce qui va suivre la th\'eorie (3.2),
dans laquelles les champs contraints $V_0$ ne peuvent pas se coupler au
courants statiques du pion. Nous allons plut\^ot \'etudier une
th\'eorie avec des contraintes, \'equivalente \`a (3.2),
l' \'equivalence \'etant comprise au sens des transformations
canoniques et sera montr\'ee dans l' Appendice A.
Dans cette th\'eorie le champ du $\rho$ est d\'efini en termes de champs
tenseurs antisym\'etriques $\WMN$ se transformant comme $\WMN\to h(\pi)\WMN
h^\dagger(\pi)$
par une rotation chirale, et la densit\'e Lagrangienne est donn\'ee par
[43]:
$$\eqalign{
\Lag=&-{1\over
2}\tr(\nabla^{\mu}\WMN\nabla_{\sigma}W^{\sigma\nu})+{{M_{\rho}^2}\over 4}
\tr(\WMN
W^{\mu\nu})\cr
&+{{f_\pi^2}\over 4}\tr(u_\mu u^\mu)
+i{{G_\rho}\over{2\sqrt{2}}}\tr(\WMN [u^\mu, u^\nu])\cr}\eqno(3.3)
$$
\medskip
\noindent avec $W_{\mu\nu}=\tau_k W_{\mu\nu}^k$ \  $k$  \ \'etant un indice d'
isospin.  La contribution de l' \'echange d' un $\rho$ \`a la
diffusion $\pi\pi$ avec les couplages du Lagrangien (3.3) a
 \'et\'e syst\'ematiquement \'etudi\'ee dans [69] et
[70] pour le cas de trois saveurs. Il a \'et\'e trouv\'e par les auteurs
de [70] que cette
contribution est identique \`a celle d\^ue au Lagrangien (2.1).
L' \'equivalence des deux approches dans les ordres dominants
de la diffusion $\pi\pi$ est mieux per\c cue \`a la limite des grandes
masses pour le m\'eson $\rho$. En effet, l' \'equation du mouvement du
champ $W_{\mu\nu}$, \`a cette limite,  nous dit que
$W_{\mu\nu}\to \displaystyle{{G_\rho}\over{i\sqrt{2}M_\rho^2}}
[u_\mu,u_\nu]$. En rempla\c cant
cette derni\`ere expression de $\WMN$ dans (3.3), on obtient:
$$\eqalign{
\Lag(M_\rho,G_\rho\to\infty) \ \to {{f_\pi^2}\over 4}
\tr(\partial_\mu U\partial^\mu
U^\dagger)+{{G_\rho^2}\over{8M_\rho^2}}\tr([\partial_\mu
UU^\dagger ,\partial_\nu U U^\dagger ]^2)\cr}
\eqno(3.4)
$$
\medskip
\noindent en utilisant la relation $u_\mu=iu^\dagger (\partial_\mu
UU^\dagger ) u$. Comme nous l' avons mentionn\'e dans la section
pr\'ec\'edente, la limite locale du Lagrangien
${\cal{L}}_{\pi\rho}^{YM}$ est aussi donn\'ee par l' \'equation
(3.4). La contribution du prochain terme
(d' ordre six) du
d\'eveloppement en puissances du champ du pion,  \`a la composante temporelle
du tenseur
impulsion-energie, est n\'egative, comme on peut le voir sur son expression:
$$\eqalign{
{\cal{L}}_{\pi\rho}^{(6)} \ \approx \tr(\nabla^\mu
[u_\mu,u_\nu]\nabla_\sigma [u^\sigma,u^\nu])
\cr}\eqno(3.5)
$$
\medskip
\noindent En vertu de cette relation
nous pouvons anticiper que l' energie de la solution classique de
l' \'equation (3.3), si cette solution existe,
sera inf\'erieure \`a celle du Skyrmion de l' \'equation (3.4).

\noindent Repr\'esenter des champs vecteurs par des tenseurs antisym\'etriques
suppose que l' on contraint les degr\'es de libert\'e redondants.
Il est donc naturel de se consacrer en premier \`a la construction du
Hamiltonien associ\'e au Lagrangien (3.3), afin d' \'eliminer les champs
ob\'eissant \`a des contraintes.

\bigskip

{\centerline{\bf a. Le Hamiltonien}}

\bigskip

\noindent Il n' est pas difficile de voir
que les composantes spatiales du tenseur
$W_{\mu\nu}$ ne sont pas des degr\'es de libert\'e physiques par
construction. En
effet, les moments canoniquement conjugu\'es du champ $W_{\mu\nu}$ que l'
on appelera $\pi_{\mu\nu}$ sont donn\'es par:
$$\eqalign{
\pi^{\mu\nu}={{\partial\Lag}\over{\partial[\partial_0 W_{\mu\nu}]}}=
-2\eta^{\mu 0}\nabla_\sigma W^{\sigma\nu}\cr}\eqno(3.6)
$$
\medskip
\noindent $\eta$ \'etant la m\'etrique de l' espace-temps.
Il est imm\'ediat de
d\'eduire que $\pi_{ij}=0$, et par cons\'equent les champs $W_{ij}$ ne se
propagent pas. Contrairement au cas des th\'eories de Yang-Mills
massives o\`u les champs contraints $\tilde V_0$ se couplent \`a des
d\'eriv\'ees
temporelles disparaissant \`a la limite statique,
ici les champs contraints $W_{ij}$ se couplent au courants statiques du pion.
Pour trouver la contrainte \`a laquelle les $W_{ij}$
ob\'eissent il faut calculer le Hamiltonien primaire. La terminologie
\lq\lq primaire" utilis\'ee dans le formalisme de Dirac-Bergmann
[71], sous-entend
que cet Hamiltonien contient tous les champs, m\^eme ceux qui sont contraints.
Il sera utile de param\'etriser de la fa\c con suivante le champ chiral:
$$\eqalign{
U=\exp{(i\vec\tau \hat F F)}\cr}\eqno(3.7)
$$
\medskip
\noindent Avec cette d\'efinition le Lagrangien du mod\`ele
$\sigma$ non-lin\'eaire s' \'ecrit sous la forme compacte:
$$\eqalign{
{\cal{L}}_{\sigma NL}=&{{f_\pi^2}\over 2} \ \partial_\mu \vec F
 \ {\cal{G}} \ \partial^\mu \vec F\cr
{\cal{G}}_{ab}=&\hat F_a\hat F_b+{{\sin^2 F}\over {F^2}}(\delta_{ab}-\hat
F_a\hat
F_b)\cr}\eqno(3.8)
$$
\medskip
\noindent L' isotenseur ${\cal{G}}$ est en fait la \lq\lq m\'etrique" du groupe
$SU(2)$ et son spectre est positif. Avec ces notations,
$\Gamma_\mu$ et les gradients du pion $u_\mu$ sont donn\'es par:
$$\eqalign{
\Gamma_\mu=&i{{\sin^2(F/2)}\over F}\tau_a\epsilon_{abc}\hat
F_b\partial_\mu F_c\cr
u_\mu=&-\tau_a\big[\hat F_a\hat F_b+{{\sin F}\over F}(\delta_{ab}-\hat F_a\hat
F_b)\big]\partial_\mu F_b\cr}\eqno(3.9)
$$
\medskip
\noindent On a \'egalement besoin de l' expression du moment conjug\'e
$\vec\phi$ du champ $\vec  F$. Il est de la forme:
$$\eqalign{
\vec\phi={{\partial\Lag}\over{\partial[\partial_0 \vec F]}}=f_\pi^2
{\cal{G}} \ \partial_0 \vec
F+{{4G_\rho}\over{\sqrt{2}}}{\cal{N}}^\dagger_i\vec W_{0i}
-{\cal{M}}^\dagger_{0i}\vec\pi_{0i}\cr}\eqno(3.10)
$$
\medskip
\noindent avec la d\'efinition suivante des op\'erateurs d' isospin
 ${\cal{N}}_i$ et ${\cal{M}}_{0i}$ :
$$\eqalign{
({\cal{N}}_i)_{ab}=&\epsilon_{pqa}\partial_i F_r{{\sin F}\over
F}\big[(1-{{\sin F}\over F})(\hat F_p\hat F_b \delta_{qr}+\hat F_q\hat F_r
\delta_{pb})+{{\sin F}\over F}\delta_{qr}\delta_{pb}\big]\cr
({\cal{M}}_{0i})_{ab}=&-{{2\sin^2(F/2)}\over F}\big[\delta_{ab}(\hat F.\vec
W_{0i})-\hat F_a (W_{0i})_b\big]\cr}\eqno(3.11)
$$
\medskip
\noindent On \'elimine les d\'eriv\'ees temporelles des champs, en inversant
les  \'equations (3.6) et (3.10), ce qui permet d'
\'ecrire le Hamiltonien  de la  th\'eorie  en termes de variables
canoniquement conjug\'ees: $\displaystyle{H_P=\int d^3 x
(\vec\pi_{\mu\nu}\partial_0\vec
 W^{\mu\nu}+\vec\phi\partial_0\vec F-\Lag)}$.  N\'egligeant
les termes de surface, on obtient apr\`es int\'egration par parties
l' expression:
$$\eqalign{
H_P=\int d^3 x \bigg\{ &{{\vec\pi_{0i}^2}\over
4}+{1\over{2f_\pi^2}}\big[\vec\phi
-{{4G_\rho}\over{\sqrt{2}}}\vec W_{0i} {\cal{N}}_i
+\vec\pi_{0i}{\cal{M}}_{0i}\big] \ {\cal{G}}^{-1} \ \big[\vec\phi
-{{4G_\rho}\over{\sqrt{2}}} {\cal{N}}^\dagger_i \vec W_{0i}
+{\cal{M}}^\dagger_{0i} \vec\pi_{0i}\big]\cr
&+{{f_\pi^2}\over 2}\partial_i\vec F \ {\cal{G}} \ \partial_i\vec
F+\vec A_{i0i}^2+M_\rho^2 \vec W_{0i}^2\cr
&-{{M^2_\rho}\over 2} \vec W_{ij}^2-{{iG_\rho}\over{\sqrt{2}}}
(W_{ij})_k[u_i,u_j]_k-\vec\pi_{0i}\vec A_{jji}\bigg\}\cr}\eqno(3.12)
$$
\medskip
\noindent avec les notations:
$$\eqalign{
(A_{\alpha\beta\gamma})_k=&{1\over 2}\tr(\tau_k\nabla_\alpha
W_{\beta\gamma})\cr
[u_i,u_j]_k=&{1\over 2}\tr(\tau_k[u_i,u_j])\cr}\eqno(3.13)
$$
\medskip
\noindent  D\'efinissons maintenant les crochets de Poisson
fondamentaux, liant les champs canoniquement
conjugu\'es:
$$\eqalign{
\bigg\{W^k_{\mu\nu}(x),\pi^l_{\rho\sigma}(y)\bigg\}_{[x_0=y_0]}={1\over
2}\delta^{kl}(\eta_{\mu\rho}\eta_{\nu\sigma}-\eta_{\mu\sigma}\eta_{\nu\rho})\delta^3
(\vec x-\vec y)\cr}\eqno(3.14)
$$
\medskip
\noindent o\`u l' on a antisymetris\'e les indices de spin dans l'
expression habituelle, $k$ et $l$ sont des indices d'
isospin. Remarquons que cette \'equation est incompatible
avec la relation $\vec\pi_{ij}=0$. Pour s' en sortir il suffit de traiter les
variables
$\vec\pi_{ij}$ de fa\c con un peu particuli\`ere. Imposons
la conservation dans le temps de la contrainte primaire. Les
variations des champs \'etant \'egales \`a leur crochets de Poisson avec
le Hamiltonien, cette conservation est assur\'ee par:
$$\eqalign{
\big\{\pi_{ij}(x),H_P\big\}=0\cr}\eqno(3.15)
$$
\medskip
\noindent En reportant l' expression de $H_P$ (eq. (3.12)) dans ce
crochet, et en utilisant les crochets fondamentaux d\'efinis plus haut,
on arrive \`a l' \'equation de la contrainte
secondaire apr\`es quelques int\'egrations par parties. Cette derni\`ere
peut-\^etre \'ecrite sous forme matricielle:
$$\eqalign{
W_{ij}=&{1\over {2M_\rho^2}}\bigg\{-i\sqrt{2}G_\rho[u_i
,u_j]+\nabla_i {\cal{\pi}}_{0j}-\nabla_j {\cal{\pi}}_{0i}
\bigg\}\cr}\eqno(3.16)
$$
\medskip
\noindent Cette formule  est l' \'equivalente de la loi de Gauss [72].

\noindent L' \'equation (3.16)
est formellement analogue \`a celle obtenue
par les auteurs de [32], pour le syst\`eme $\pi\omega$:

$$\eqalign{
\omega_0={1\over{m_\omega^2}}(-\beta_\omega B_0+\partial_i\Omega_i)\cr}
$$
\medskip

\noindent concernant la composante temporelle du champ du m\'eson $\omega$
dans le mod\`ele de la r\'ef. [23]. $\Omega_i$ sont les moments
conjug\'es des champs $\omega_i$ et $B_0$ est la densit\'e de charge
baryonique.

Nous sommes maintenant en position d' \'ecrire le
Hamiltonien physique (secondaire), en remplacant $W_{ij}$ (\'equation
(3.16)) dans
$H_P$. Nous obtenons:
$$\eqalign{
H_S=\int d^3 x \bigg\{ &{{\vec\pi_{0i}^2}\over
4}+{1\over{2f_\pi^2}}\big[\vec\phi
-{{4G_\rho}\over{\sqrt{2}}}\vec W_{0i} {\cal{N}}_i
+\vec\pi_{0i}{\cal{M}}_{0i}\big] \ {\cal{G}}^{-1} \ \big[\vec\phi
-{{4G_\rho}\over{\sqrt{2}}} {\cal{N}}^\dagger_i \vec W_{0i}
+{\cal{M}}^\dagger_{0i} \vec\pi_{0i}\big]\cr
&+{{f_\pi^2}\over 2}\partial_i\vec F \ {\cal{G}} \ \partial_i\vec
F+\vec A_{i0i}^2+M_\rho^2 \vec W_{0i}^2\cr
&+{1\over{2M_\rho^2}}\big[{1\over
2}(\nabla_i\pi_{0j}-\nabla_j\pi_{0i})_k-{{iG_\rho}\over{\sqrt{2}}}[u_i,u_j]_k\big]^2\bigg\}
\cr}\eqno(3.17)
$$
\medskip
\noindent En principe, il faut ajouter \`a cet Hamiltonien des
multiplicateurs de Lagrange, pour s' assurer que les
contraintes primaire et secondaire soient prises en compte dans la
dynamique. Ici nous simplifions le probl\`eme, en posant ces
multiplicateurs identiquement \'egaux \`a z\'ero,  mais ceci n' alt\'erera
pas la g\'en\'eralit\'e de nos conclusions.

Avant de passer \`a la construction des
solutions statiques remarquons qu' en vertu de l' \'equation de la contrainte
secondaire, toute tentative d' annuler le champ du $\rho$ par une d\'eformation
continue va se heurter \`a la r\'epulsion d\^ue au terme de Skyrme qui est
explicitement pr\'esent dans le
Hamiltonien secondaire. Ceci nous rappelle fortement le m\'ecanisme de
stabilisation du soliton dans le cas du syst\`eme $\pi\omega$ [23], o\`u
la r\'epulsion stabilisatrice
provient d' un terme d' ordre six dans le d\'eveloppement chiral.

Nous allons maintenant chercher les solutions classiques de l'
\'equation (3.17) dans le secteur de charge baryonique unit\'e.

\bigskip

{\centerline{\bf b. Solutions classiques}}

\bigskip

\noindent Consid\'erons  l' ansatz du h\'erisson,
d\'ej\`a rencontr\'e au cours des chapitres pr\'ec\'edents:
$$\eqalign{
&\vec F=\hat r F,\quad W_{0i}=f_\pi\big[w_1(\tau_i-(\vec\tau.\hat r)\hat
r_i)+w_2(\vec\tau.\hat r)\hat r_i
-w_3(\vec\tau\times\hat r)_i\big]\cr
&\vec \phi=\hat r {{f_\pi^2}\over r}\phi,\quad \pi_{0i}={{f_\pi}\over
r}\big[\pi_1(\tau_i-(\vec\tau.\hat r)
\hat r_i)+\pi_2(\vec\tau.\hat r)\hat r_i
-\pi_3(\vec\tau\times\hat r)_i\big]
\cr}\eqno(3.18)
$$
\medskip
\noindent La
forme (3.18) pour les champs vecteurs $W_{0i}$ et $\pi_{0i}$ est
la configuration la plus g\'en\'erale, compatible avec la sym\'etrie
sph\'erique. Les profils $F,\phi,w_1,w_2,w_3,\pi_1,\pi_2,\pi_3$ sont
des fonctions de la variable radiale r (et du temps \'eventuellement,
mais de toute fa\c con cette d\'ependance n' est qu' implicite dans le
formalisme Hamiltonien). Une fois les configurations (3.18) report\'ees
dans le Hamiltonien
(3.17), il est naturel de s\'eparer ce dernier en deux parties, car les champs
se
d\'ecouplent exactement:
$$\eqalign{
H=4\pi f_\pi^2\intif
dr\big[{\cal{H}}_Q(F,\phi,w_1,w_2,w_3,\pi_1,\pi_2)+{\cal{H}}_{\pi\rho}(F,\pi_3)
\big]
\cr}\eqno(3.19)
$$
\medskip
\noindent Le premier terme,  $\displaystyle{{\cal{H}}_Q}$,  est
positif et quadratique dans les champs
$\displaystyle{\phi,w_1,w_2,w_3,\pi_1,\pi_2}$:
$$\eqalign{
{\cal{H}}_Q=&{1\over{2}}(\phi+2g_\rho
w_3 \sin F)^2+{{\pi_2^2}\over 4}+{{\pi_1^2}\over 2}
+(r \dot w_2 +2w_2-2\cos F w_1)^2\cr
&+(M_\rho r)^2(w_2^2+2w_1^2+2w_3^2)
+{1\over{2M_\rho^2}}(\dot\pi_1-\cos F{{\pi_2}\over r})^2
\cr}\eqno(3.20)
$$
\medskip
\noindent $g_\rho$ est la constante de couplage sans dimension,
$g_\rho=2\sqrt{2}G_\rho / f_\pi$.

\noindent Cet Hamiltonien est nul
\`a la limite statique. Pour le prouver, consid\'erons  l' ensemble d'
\'equations $\displaystyle{\delta\big[\int_0^\infty
dr{\cal{H}}_Q\big]=0}$ pour des champs statiques. Tout d' abord, celles
concernant les
champs $\phi$ et $w_3$ donnent identiquement $\phi=w_3=0$. Pour les
autres champs il suffit de remplacer les
variables  $w_1,w_2,\pi_1,\pi_2$ par les variables $p_1$ et $p_2$:
$$\eqalign{
w_1=&{{\cos F}\over{M_\rho^2 r^2}}p_1 \ , \ w_2={{\dot
p_1}\over{M_\rho^2 r}}\cr
\pi_1=&{{\dot p_2}\over{M_\rho^2}} \ \ \  \ , \ \pi_2={{2\cos F}\over{M_\rho^2
r}}p_2\cr}\eqno(3.21)
$$
\medskip
\noindent Les \'equations de Hamilton se r\'eduisent \`a:
$$\eqalign{
\ddot p_i=&(M_\rho^2+{{2\cos^2 F}\over{r^2}})p_i \ \ \ \
i=1,2\cr}\eqno(3.22)
$$
\medskip
\noindent En multipliant cette \'equation par $p_i$ et en int\'egrant
par parties le membre de gauche on arrive \`a l' expression:
$$\eqalign{
\int^\infty_0 \ dr \ \big[\dot p_i^2 \ + \ p_i^2(M_\rho^2+{{2\cos^2
F}\over{r^2}})\big]=0\cr}\eqno(3.23)
$$
\medskip
\noindent La seule solution est \'evidemment la solution triviale
$p_i=\dot p_i=0$, par
cons\'equent $w_1=w_2=\pi_1=\pi_2=0$ ce qui compl\`ete la
d\'emonstration. Pour des solutions de type soliton, nous avons donc
montr\'e que:
$$\eqalign{
({\cal{H}}_Q)_{statique}=0\cr}\eqno(3.24)
$$
\medskip
\noindent Il est clair que la d\'eg\'en\'erescence au niveau de la
masse classique rencontr\'ee dans la section 2,
ne peux pas avoir lieu ici,   les champs $w_1,w_2,\pi_1,\pi_2$
sont nuls de fa\c con triviale. En plus, ils se
d\'ecouplent de la composante non-nulle \`a
la limite statique $\pi_3$.

En ce qui concerne la deuxi\`eme partie du Hamiltonien, nous trouvons l'
expression suivante pour ${\cal{H}}_{\pi\rho}$:
$$\eqalign{
{\cal{H}}_{\pi\rho}=&{1\over 2}\big[(r\dot F)^2+2\sin^2
F+\pi_3^2\big]
+{1\over{2M_{\rho}^2}}\big\{(g_\rho\sin F\dot F-\dot\pi_3)^2
+2(g_\rho{{\sin^2 F}\over {2r}}-\cos F{{\pi_3}\over
r})^2\big\}
\cr}\eqno(3.25)
$$
\medskip

\noindent Cherchons maintenant des solutions ayant une charge
topologique unit\'e. Pour cela, il faut imposer des conditions aux
limites $\displaystyle{F(0)=\pi}$ et $\displaystyle{F(\infty)=0}$ et
r\'esoudre les deux \'equations de Hamilton, qui vont rendre
$\displaystyle{{\cal{H}}_{\pi\rho}}$ stationnaire par rapport aux variations
arbitraires des champs  $\displaystyle{F}$ and $\displaystyle{\pi_3}$.
Ces \'equations sont de la forme:
$$\eqalign{
\dot p=&\sin 2F+g_\rho\pi_3\sin F+2D{{\pi_3\sin F}\over{M_\rho^2
r}}\cr}\eqno(3.26a)
$$
$$\eqalign{
\dot C=&2D{{\cos F}\over r}-M_\rho^2\pi_3
\cr}\eqno(3.26b)
$$
\noindent avec $C=g_\rho\sin F \dot F-\dot\pi_3$, \
$D=g_\rho\displaystyle{{{\sin^2 F}\over{2r}}-\cos F {{\pi_3}\over
r}}$ \  et $p=r^2\dot F$. Ces \'equations sont r\'esolues
num\'eriquement (voir figure (3.1)).

\noindent On v\'erifie tr\`es ais\'ement que la forme
asymptotique de la solution des \'equations (3.26a-b) est compatible avec
les conditions aux limites impos\'ees au champ chiral. La solution asymptotique
de
(3.26b) pour le champ $\pi_3$ est de la forme $(Cte)\displaystyle{e^{-M_\rho
r}}$.
La valeur de $g_\rho$ utilis\'ee ($g_\rho=2.1$) reproduit la
largeur de d\'esint\'egration $\rho\to\pi\pi$ au premier ordre des
perturbations.

\noindent Il est int\'eressant d' observer que notre solution classique pour
les
champs $F, \pi_3$ est qualitativement
semblable \`a celle obtenue par les auteurs de [62] pour les champs
correspondants $F, \tilde v_3$. La masse classique de la solution pr\'esente
\`a la figure (3.1) est de
$1.14$ GeV, l\'eg\'erement inf\'erieure \`a celle que l' on trouve dans
le mod\`ele de Skyrme correspondant ($e=\sqrt{2}M_\rho/f_\pi g_\rho$),
comme nous l' avions pr\'edit au d\'ebut de cette section. Cette \lq\lq
attraction" provenant
de l' interaction du champ chiral avec le m\'eson $\rho$ est aussi \`a
l' origine de la compression l\'eg\`ere que la solution subit par rapport au
mod\`ele de Skyrme. En effet, si l' on calcule le rayon
isoscalaire du soliton
$$\eqalign{
<r^2>_{I=0}={2\over \pi}\intif r^2dr(-\dot F\
sin^2 F)\cr}\eqno(3.27)
$$
\medskip
\noindent nous trouvons $<r^2>^{1/2}_{I=0}=0.35$ fm, contre $0.4$ fm dans
le mod\`ele de Skyrme.

\noindent Nous avons aussi v\'erifi\'e
qu' il existe une solution des \'equations (3.26a-b) pour toute valeur de
la constante de couplage $g_\rho$.

\bigskip

{\centerline{\bf c. Stabilit\'e}}

\bigskip

On va s' int\'eresser au signe de la seconde variation du
Hamiltonien (3.19) par rapport aux petites
fluctuations des champs autour de la solution classique. On consid\`ere des
fluctuations monopolaires [73], correspondant \`a un mode de
vibration scalaire:
$$\eqalign{
&F=F_0(r)+\delta F(r,t), \ \ \ \ \ \ \ \phi=\phi_0(r)+\delta\phi(r,t)\cr
&\pi_i=\pi^0_i(r)+\delta\pi_i(r,t), \ \ \ \  w_i=w_i^0(r)+\delta w_i(r,t) \ \ \
i=1,2,3\cr}\eqno(3.28)
$$
\medskip

Consid\'erons d' abord le terme ${\cal{H}}_Q$. En vertu des r\'esultats
analytiques
vus plus haut, la partie statique des champs
$\pi_1,\pi_2,w_1,w_2,w_3,\phi$ est nulle. Alors, leur
contribution \`a la variation du Hamiltonien se d\'ecouple
des fluctuations de $F$, et on a la relation simple
$$\eqalign{
\delta_{(2)}{\cal{H}}_Q(F,\phi,w_1,w_2,w_3,\pi_1,\pi_2)={\cal{H}}_Q(F_0,\delta\phi,\delta
w_1,\delta w_2,\delta w_3,\delta\pi_1,\delta\pi_2)\cr}\eqno(3.29)
$$
\medskip
\noindent Nous avons montr\'e analytiquement que ${\cal{H}}_Q$ ne peut avoir
qu' une contribution positive \`a la fluctuation de la masse. Il
en r\'esulte que les composantes $\pi_1,\pi_2,w_1,w_2$ ne
d\'estabilisent pas le soliton.

Int\'eressons nous maintenant au terme ${\cal{H}}_{\pi\rho}$, pour
d\'eterminer si des fluctuations arbitraires des champs $F$ et $\pi_3$ ne
d\'estabilisent pas  la solution classique. D' abord, effectuons
un test pr\'eliminaire pour d\'eterminer si cette solution
est un minimum local par rapport \`a une classe de transformations
d' \'echelle. Consid\'erons \`a cette fin l' int\'egrale $I=\intif
dr\big[{\cal{H}}_{\pi\rho}(F,\pi_3)\big]$. Cette int\'egrale est \'egale \`a la
masse du soliton \`a un facteur multiplicatif pr\`es, et elle peut s' exprimer
comme une somme $I=I_1+I_2+I_3+I_4+I_5$ avec
$$\eqalign{
&I_1={1\over 2}\intif dr \pi_3^2\cr
&I_2={1\over 2}\intif dr \big[(r\dot F)^2+2\sin^2 F\big]\cr
&I_3={1\over {2M_\rho^2}}\intif dr\big[(g_\rho \sin F \dot
F)^2+2(g_\rho{{\sin^2 F}\over r})^2\big]\cr
&I_4=-{1\over {2M_\rho^2}}\intif dr\big[2\dot\pi_3(g_\rho \sin F \dot
F)+4\cos F {{\pi_3}\over r}(g_\rho{{\sin^2 F}\over r})\big]\cr
&I_5={1\over {2M_\rho^2}}\intif dr\big[\dot\pi_3^2+2\cos^2 F {{\pi^2_3}\over
{r^2}}\big]\cr}\eqno(3.30)
$$
\noindent Effectuons maintenant la transformation suivante de la solution
classique:
$$\eqalign{
F_0(r) \ \to \ F_0(\lambda r) \ , \ \pi_3^0(r) \ \to \ \gamma \pi_3^0(\lambda
r)\cr}\eqno(3.31)
$$
\medskip
\noindent $\lambda$ et $\gamma$ \'etant deux param\`etres arbitraires.
Par cette transformation d' \'echelle globale, $I$ devient:
$$\eqalign{
I_{\lambda\gamma}={1\over\lambda}(\gamma^2 I_1+I_2)
+\lambda(I_3+\gamma I_4+\gamma^2 I_5)\cr}\eqno(3.32)
$$
\medskip
\noindent D\'eveloppons cette expression autour de la solution classique
($\lambda=1+\epsilon_\lambda,\gamma=1+\epsilon_\gamma$) avec
$\epsilon_\lambda,\epsilon_\gamma \ << \ 1$. Au premier ordre,
la stationnarit\'e de l'
Hamiltonien  fournit des relations entre les diff\'erentes
contributions \`a la masse. On trouve
$$\eqalign{
I_1+I_2=&I_3+I_4+I_5\cr
2(I_1+I_5)=&-I_4\cr}\eqno(3.33)
$$
\medskip
\noindent relations que l' on v\'erifie avec la solution
de la section pr\'ec\'edente. Par ailleurs ceci constitue un excellent
test de la m\'ethode de r\'esolution num\'erique.
\noindent La variation quadratique maintenant peut se mettre
sous la forme:
$$\eqalign{
\delta_{(2)}I_{\lambda\gamma}=
\pmatrix{\epsilon_\lambda&\epsilon_\gamma\cr}
\pmatrix{2(I_1+I_2)&-4I_1\cr-4I_1&2(I_1+I_5)\cr}
\pmatrix{\epsilon_\lambda\cr\epsilon_\gamma\cr}
}\eqno(3.34)
$$
\medskip
\noindent Parmi les valeurs propres de la matrice de cette forme
quadratique une est positive. La
positivit\'e de la deuxi\`eme valeur propre est moins simple \`a
\'etablir. On peut formuler la condition n\'ecessaire pour la stabilit\'e
de la solution classique, en d' autres termes le th\'eor\`eme de
Derrick g\'en\'eralis\'e pour le  syst\`eme $\pi\rho$, par l' in\'egalit\'e:
$$\eqalign{
(I_1+I_2)(I_1+I_5)-4I^2_1 \ > \ 0\cr}\eqno(3.35)
$$
\medskip
\noindent Nous n' avons malheureusement pas pu \'etablir analytiquement que
la solution de la section pr\'ec\'edente satisfait \`a cette in\'egalit\'e.
Malgr\'e tout le membre de gauche de (3.35) peut se calculer
num\'eriquement et en effet, nous avons trouv\'e qu' il est positif
pour toute solution classique (en fait pour toute valeur de la constante
$g_\rho$).
Le soliton est donc un minimum local par rapport \`a la transformation (3.31).

\noindent Le fait que la
solution soit un minimum local pour des transformations d' \'echelle, ne
constitue pas une condition suffisante pour sa stabilit\'e. Pour
d\'eterminer une telle condition, il
faut consid\'erer des fluctuations plus g\'en\'erales, de la forme
$F=F_0+\delta F(r,t)$
$\pi_3=\pi^0_3+\delta\pi_3(r,t)$.

\noindent La variation de l' Hamiltonien au second ordre par rapport \`a
ces fluctuations est:
$$\eqalign{
\delta{\cal{H}}_{\pi\rho}=2\pi f_\pi^2\intif \psi^\dagger{\cal{M}}\psi dr\cr}
\eqno(3.36)
$$
\medskip
\noindent o\`u $\psi$ est un vecteur \`a deux composantes $\displaystyle{
\psi^\dagger=(\delta F \,   \delta \pi_3)}$ et ${\cal{M}}$ est un
op\'erateur hermitique, fonction de la variable $r$. Ses \'el\'ements de
matrice sont:

$$\eqalign{
{\cal{M}}_{(\delta F,\delta F)}=&-{d\over
{dr}}(r^2+{{g_\rho^2}\over {M_\rho^2}}\sin^2F_0){d\over {dr}}
+2\cos 2F_0\cr
&-{{g_\rho^2}\over {M_\rho^2}}\big[
\sin 2F_0 \ddot F_0+ \cos 2F_0 \dot F_0^2-{{\sin^2 F_0}\over
{r^2}}(3-4\sin^2 F_0)\big]\cr
&+{1\over{M_\rho^2 r^2}}\big[g_\rho \cos F_0(\ddot \pi_3^0 r^2+\pi_3^0
(9\sin^2 F_0-2))-2(\pi_3^0)^2\cos 2F_0\big]\cr
{\cal{M}}_{(\delta \pi_3,\delta \pi_3)}=&{1\over
{M_\rho^2}}\big[-{{d^2}\over{dr^2}}+(M_\rho^2+2{{\cos^2 F_0}\over
{r^2}})\big]\cr
{\cal{M}}_{(\delta F,\delta \pi_3)}=&{1\over {M_\rho^2}}\big[
g_\rho\sin F_0 {{d^2}\over {dr^2}}
+{1\over{r^2}}(-2\sin 2F_0\pi_3^0+g_\rho(3\sin^2 F_0-2)\sin F_0)\big]\cr
{\cal{M}}_{(\delta \pi_3,\delta F)}=&({\cal{M}}_{(\delta F,\delta
\pi_3)})^\dagger
\cr}\eqno(3.37)
$$

\noindent Il est int\'eressant de noter que sur les deux premi\`eres lignes de
cette
\'equation, et parce que les
termes en $\pi_3^0$ dans (3.37) aussi bien que le potentiel sont d'
ordre $1/M_\rho^4$ ou plus,
on retrouve l' op\'erateur de fluctuation du Skyrmion. Cet
op\'erateur a un spectre positif comme il est montr\'e dans la r\'ef.
[74].

Pour montrer finalement la stabilit\'e il faut montrer que le spectre de
${\cal{M}}$ se situe du c\^ot\'e des fr\'equences positives. C' est ce
qui a \'et\'e fait dans [72] en utilisant deux m\'ethodes
diff\'erentes. La premi\`ere consiste \`a diagonaliser une approximation
discr\'etis\'ee de ${\cal{M}}$, dans laquelle nous n' avons trouv\'e que
des valeurs propres positives et la deuxi\`eme consiste \`a r\'esoudre
les \'equations aux valeurs propres pour des \'energies n\'egatives, pour
laquelle nous avons trouv\'e
que le determinant de Jost [75] associ\'e \`a ${\cal{M}}$ ne change pas de
signe.

En d\'efinitive,  nous avons
montr\'e que le m\'eson $\rho$ peut stabiliser le soliton
d' une  mani\`ere tr\`es similaire \`a celle que l' on rencontre quand on
introduit le m\'eson $\omega$.

\bigskip

\leftline{\bfmagc 4. Conclusions}

\bigskip

\noindent Il a \'et\'e montr\'e par les auteurs de la r\'ef. [70], qu'  avec
le Lagrangien donn\'e par l' \'equation (3.2) (ou (3.3)),
les couplages effectifs induits dans le secteur des pions par
l' \'echange d' un $\rho$ sont \'equivalents aux couplages produits
par des Lagrangiens du m\^eme type que celui de l' \'equation (2.1).
Cependant, nous avons not\'e une diff\'erence cruciale entre les
secteurs baryoniques correspondants:

\noindent Quand la transformation du champ du
m\'eson $\rho$ par une rotation chirale \lq\lq cach\'ee",
est suppos\'ee \^etre homog\`ene,
les solitons topologiques du syst\`eme $\pi\rho$
sont stables par rapport aux fluctuations des champs m\'esoniques, alors
que dans les approches \lq\lq non-minimales", o\`u le m\'eson
$\rho$ est coupl\'e au pion comme un boson de jauge de ce m\^eme groupe
cach\'e,  le soliton est instable.  De plus, dans la premi\`ere approche
il n' y a pas de doublet de baryons.

Il est donc clair que les ambiguit\'es du syst\`eme $\pi\rho$
existant dans la litt\'erature pr\'ec\'edant la r\'ef\'erence [72] sont
d\^ues au choix particulier de r\'ealisation de la sym\'etrie chirale
par les champs vecteurs dans ces travaux.

Il n' est pas difficile en fait de montrer [70] qu' il existe une
transformation non-lin\'eaire entre les champs $\tilde V_\mu$ et
$V_\mu$, par le biais de laquelle on peut trouver
une relation entre les Lagrangiens (2.1) et (3.2):
$$\eqalign{
\tilde V_\mu=&V_\mu+{i\over g}\Gamma_\mu, \ \ \ g={1\over{\sqrt{8} g_V}}\cr
{\cal{L}}_{\pi\rho}^{YM}(\tilde
V_\mu)=&{\cal{L}}_V(V_\mu)+{1\over{32g_V^2}}\tr([V_\mu,V_\nu]^2)+
{i\over{\sqrt{32} g_V}}\tr\big\{(V_{\mu\nu}+i
{{g_V}\over{\sqrt{2}}}[u_\mu,u_\nu])[V^\mu,V^\nu]\big\}\cr}\eqno(4.1)
$$
\medskip

\noindent Nous pensons
que c' est justement le dernier terme (proportionnel \`a ${1\over{g_V}}$)
dans le membre de droite de
cette \'equation qui est responsable des instabilit\'es dans le secteur
baryonique du Lagrangien ${\cal{L}}_{\pi\rho}^{YM}$.

\noindent D' un autre c\^ot\'e, et dans le secteur des
m\'esons, ces termes cubiques et quartiques en champs vecteurs
(\'equation (4.1)) contribuent \`a des
processus mettant en jeu plusieurs m\'esons $\rho$ (trois ou quatre),
et du fait de la sym\'etrie de jauge des relations ad-hoc existent
entre ces processus. La question de savoir si ces couplages, et surtout
si ces relations, sont r\'ealistes du point de vue exp\'erimental reste
ouverte.

\noindent Il faut noter ici que les lois (1.2) et (1.3) ont
la bonne propri\'et\'e de former un groupe, et il est probable qu'
elles soient les seules \`a la poss\'eder [76]. Par exemple,
la transformation inhomog\`ene $\hat V_\mu \ \to
\ h(\pi)\hat V_\mu h(\pi)^\dagger + \alpha \partial_\nu h(\pi) \hat
V_\mu \partial^\nu h^\dagger (\pi)$ ne satisfait pas \`a l'
associativit\'e.

\noindent Finissons ce chapitre en disant qu' il sera int\'eressant de
clarifier la structure topologique de la solution classique quand le
Lagrangien contient non seulement le champ $U$ mais aussi des champs
axiaux etc.. On sait que la structure topologique de la solution
joue un r\^ole non-trivial dans la stabilit\'e des solitons pour les
th\'eories en $1+1$ dimensions [8],[9]. Mais la situation se complique beaucoup
dans les th\'eories \`a $3+1$ dimensions [77] et plus
particuli\`erement pour les  mod\`eles o\`u le champ chiral se
couple \`a des m\'esons. On peut se poser la
question de savoir si les propri\'et\'es topologiques
non-triviales des champs de jauge [78] sont vraiment \`a l' origine de l'
instabilit\'e des solitons des Lagrangiens tels que (2.1).

\eject

\leftline{\bfmagc Appendice A.}

\bigskip

Dans cet appendice nous allons \'etudier une description des m\'esons vecteurs
en termes de champs vecteurs $V_{\mu}$, se transformant comme $V_\mu\to
h(\pi)V_\mu h^\dagger(\pi)$
par une rotation chirale non-lin\'eaire. Notre but est d' abord de
montrer que le Lagrangien invariant chiral, qui  satisfait aux conditions
suivantes:

\noindent a) \^etre d' ordre le plus bas en champs vecteurs

\noindent b) conduire  \`a un Hamiltonien born\'e inf\'erieurement

\noindent est donn\'e par le
Lagrangien de l' \'equation (3.2). L' \'equivalence formelle de
la formulation en termes de vecteurs \`a celle en termes de tenseurs
antisym\'etriques sera ensuite montr\'ee.

\noindent Consid\'erons d' abord
un Lagrangien qui satisfait seulement \`a la condition a):
$$\eqalign{
{\cal{L}}_V'=&-{1\over 4}\tr(V_{\mu\nu}V^{\mu\nu})+{{M_{\rho}^2}\over
2}\tr(V_\mu V^\mu)\cr
&+{{f_\pi^2}\over 4}\tr(u_\mu u^\mu)
-i{{g_V}\over{2\sqrt{2}}}\tr(V_{\mu\nu} [u^\mu, u^\nu])\cr}\eqno(A.1)
$$
\medskip

\noindent avec $\displaystyle{V_{\mu\nu}=\nabla_\mu V_\nu-\nabla_\nu
V_\mu}$. Ce Lagrangien a \'et\'e \'etudi\'e par les auteurs de la r\'ef.
[70] dans un autre contexte. Il n' est pas difficile de s' apercevoir que la
limite locale (m\'eson $\rho$ tr\`es lourd) de ce Lagrangien, contient
le terme \`a deux d\'eriv\'ees, un terme \`a six d\'eriv\'ees etc...:l' ordre
quatre est absent de ${\cal{L}}_V'$. En fait, il a
\'et\'e montr\'e dans [70] qu' il est n\'ecessaire d' ajouter
\`a ${\cal{L}}_V'$ un terme local (en fait, le terme de Skyrme en
$SU(2)$) pour que l' amplitude de diffusion $\pi\pi$ satisfasse \`a l'
unitarit\'e. Nous verrons
qu' en demandant au Hamiltonien de satisfaire \`a la
condition b), on peut arriver \`a la m\^eme conclusion que les auteurs de [70],
\`a
savoir qu' il faut ajouter un terme local d' ordre quatre \`a
${\cal{L}}_V'$.

\noindent Nous allons consid\'erer le Hamiltonien associ\'e au Lagrangien
(A.1).
Pour ceci faire, il faut d' abord calculer les moments conjugu\'es
$\bar\pi_\mu$ (la barre est pour \'eviter la confusion avec les
$\pi_{0i}$ de la section 3) et $\vec\phi$ de
$V_\mu$ et $\vec F$ respectivement. Ceux du champ vecteur sont donn\'es
par:
$$\eqalign{
\bar\pi_\mu=&-(2V_{0\mu}+i\sqrt{2} g_V[u_0,u_\mu])\cr}\eqno(A.2)
$$
\medskip
\noindent En vertu de  la relation qui en r\'esulte pour les composantes
spatiales $\bar\pi_i$, le moment du champ
chiral se met sous la forme:
$$\eqalign{
\vec\phi =& \ {\cal{A}} \ \partial_0\vec F \ + \ {\cal{B}}_i^\dagger\bar{\vec
\pi}_i\cr
{\cal{A}}={\cal{A}}^\dagger=&f_\pi^2 \ {\cal{G}} -{1\over 2}
({{4g_V}\over{\sqrt{2}}}{\cal{N}}^\dagger_i)
({{4g_V}\over{\sqrt{2}}}{\cal{N}}_i)\cr
{\cal{B}}_i=&{1\over 2}\big[{{4g_V}\over{\sqrt{2}}}{\cal{N}}_i \ - \
2{\cal{M}}_i\big]\cr}\eqno(A.3)
$$
\medskip
\noindent avec les matrices ${\cal{G}}, {\cal{N}}_i$ d\'efinies
par les \'equations (3.8,3.11). Pour avoir l' expression de
${\cal{M}}_i$ il suffit de remplacer $W_{0i}$ par $V_i$ dans l' expression
de la matrice ${\cal{M}}_{0i}$ d\'efinie par les m\^emes \'equations.

La contrainte primaire se lit sur l' \'equation (A.2): $\bar{\pi}_0=0$. Ce
qui veut dire que $V_0$ n' est pas un degr\'e de libert\'e de la
th\'eorie.  Pour l' \'eliminer, il faut i) inverser les \'equations
(A.2) et (A.3) pour obtenir le Hamiltonien
primaire et ii) imposer la conservation dans le temps de la contrainte
primaire en annulant son crochet de Poisson avec cet
Hamiltonien. Ces op\'erations conduisent \`a la relation secondaire,
$$\eqalign{
V_0 \ = \ {1\over{2M_\rho^2}} \ \nabla_i\bar{\pi}_i\cr}\eqno(A.4)
$$
\medskip
\noindent En utilisant cette relation, on arrive au Hamiltonien
exprim\'e en termes de vrais degr\'es de libert\'e  $\vec
F,\vec\phi, \vec V_i, \bar{\vec\pi}_i$ :
$$\eqalign{
H_S=\int d^3 x \bigg\{ &{{\bar{\vec\pi}_{i}^2}\over 4}+{1\over 2}\big[\vec\phi
-\bar{\vec\pi}_i {\cal{B}}_i\big] \ {\cal{A}}^{-1} \ \big[\vec\phi
-{\cal{B}}_i^\dagger\bar{\vec\pi}_i\big]
+{{f_\pi^2}\over 2}\partial_i\vec F \ {\cal{G}} \ \partial_i\vec F
+{{(\nabla_i\bar\pi_i)_k^2}\over{4M_\rho^2}}\cr
&+M_\rho^2\vec V_i^2
+{{\vec V_{ij}^2}\over 2}
+{{ig_V}\over{\sqrt{2}}}(V_{ij})_k[u_i,u_j]_k\bigg\}
\cr}\eqno(A.5)
$$
\medskip
\noindent Observons maintenant la partie d\'ependante de $\vec\phi$ de cet
Hamiltonien.
La contribution de ces termes n' est pas manifestement
positive, car  il n' est pas exclu que la matrice $\cal{A}$  ait des valeurs
propres
n\'egatives quand la constante de couplage $g_V$ est
non-nulle. Pour se rendre compte du probl\`eme, pla\c cons-nous dans la
configuration du h\'erisson pour le champ du pion $\vec F=F(r)
\hat r$. Il est alors
imm\'ediat d' inverser cette matrice:
$$\eqalign{
{\cal{A}}^{-1}_{ab}={1\over{f_\pi^2}}\bigg[\displaystyle{{1\over{(1-{{4g_V^2}\over{f_\pi^2}}{{2\sin^2F}\over
{r^2}})}}} \ \hat r_a \hat
r_b+{{F^2}\over{\sin^2F}}\displaystyle{{1\over{(1-{{4g_V^2}\over{f_\pi^2}}
[{{\sin^2F}\over
{r^2}}+\dot F^2])}}} \ (\delta_{ab}-\hat r_a \hat r_b)\bigg]\cr}\eqno(A.6)
$$
\medskip
\noindent Calculons maintenant la contribution \`a ${\cal{H}}_Q$ d\^ue
aux termes d\'ependant du moment du pion. Pour une
configuration sph\'erique du champ vecteur:
$$\eqalign{
\bar\pi_{i}={{f_\pi}\over
r}\big[\bar\pi_1(\tau_i-(\vec\tau.\hat r)
\hat r_i)+\bar\pi_2(\vec\tau.\hat r)\hat r_i
-\bar\pi_3(\vec\tau\times\hat r)_i\big]\cr}\eqno(A.7)
$$
\medskip
\noindent nous arrivons, en vertu des \'equations (A.5), (A.6) et (A.7),
\`a la forme suivante :
$$\eqalign{
{\cal{H}}_Q={1\over 2}
{
{(\phi+2\sqrt{2}g_V\bar\pi_3\sin F)^2}
\over
{1-8g_V^2\sin^2 F/ f_\pi^2 r^2} }
+\  ... \cr}\eqno(A.8)
$$
\noindent Cet Hamiltonien a un p\^ole, et n' est pas born\'e
inf\'erieurement si $g_V$ et $F$ sont non-nuls.
Ceci n' est \'evidemment pas acceptable. Il est important de
pr\'eciser ici, que m\^eme si nous avons fait usage d' un ansatz particulier
pour
compl\'eter la d\'emonstration, la conclusion sur la validit\'e de
la th\'eorie (A.1) est g\'en\'erale. Notamment, nous n' avons pas v\'erifi\'e
si
cette configuration satisfait aux \'equations du mouvement pour des
valeurs de la fonction $F$ diff\'erentes de z\'ero. M\^eme si ce
n' est pas le cas au niveau classique, ces configurations
non-perturbatives d\'estabilisent la th\'eorie
quantique.

\noindent Nous avions anticip\'e sur l' origine de cette pathologie en
observant que la
matrice $\cal{A}$ est susceptible d' acqu\'erir une contribution n\'egative du
couplage
pion-$\rho$. La fa\c con la plus simple de \lq\lq r\'egulariser" cette matrice,
et par cons\'equent le Hamiltonien, est d' ajouter un terme local \`a
${\cal{L}}_V'$ compensant cette contribution :
$$\eqalign{
{\cal{L}}_V={\cal{L}}_V'+{{g_V^2}\over 8}\tr([u^\mu, u^\nu][u_\mu,
u_\nu])\cr}\eqno(A.9)
$$
\medskip
\noindent Ce terme n' est rien d' autre que le terme de
Skyrme; le nouveau Hamiltonien est maintenant manifestement positif:
$$\eqalign{
H=\int d^3 x \bigg\{ &{{\bar{\vec\pi}_{i}^2}\over 4}+{1\over
{2f_\pi^2}}\big[\vec\phi
-\bar{\vec\pi}_i {\cal{B}}_i\big] \ {\cal{G}}^{-1} \ \big[\vec\phi
-{\cal{B}}_i^\dagger\bar{\vec\pi}_i\big]
+{{f_\pi^2}\over 2}\partial_i\vec F \ {\cal{G}} \ \partial_i\vec F
+{{(\nabla_i\bar\pi_i)_k^2}\over{4M_\rho^2}}\cr
&+M_\rho^2\vec V_i^2+{1\over 2}\big[(\vec
V_{ij})_k+{{ig_V}\over{\sqrt{2}}}[u_i,u_j]_k\big]^2\bigg\}
\cr}\eqno(A.10)
$$
\noindent Nous concluons donc que le
Lagrangien satisfaisant aux conditions a) et b) \'enonc\'ees au
d\'ebut de cet appendice est donn\'e par l' expression (3.2).
Remarquons que le seul fait d' imposer au
Hamiltonien d' \^etre positif, d\'etermine la valeur minimale de
la constante de Skyrme: $\displaystyle{e=(2g_V)^{-1}}$.

\noindent Il n' est pas difficile de montrer maintenant que le Hamiltonien
(A.10) est exactement \'egal au Hamiltonien en termes de tenseurs
antisym\'etriques (3.17):

\medskip

$H\big[\vec F,\vec\phi,V_i,\bar\pi_i\big]=H_S \big[\vec
F,\vec\phi,W_{0i},\pi_{0i}\big]$ .

\medskip

\noindent Il existe une {transformation canonique entre les
deux ensembles de champs $\displaystyle{\big\{\vec
F,\vec\phi,V_i,\bar\pi_i\big\}}$
et $\displaystyle{\big\{\vec F,\vec\phi,W_{0i},\pi_{0i}\big\}}$ en
effectuant le remplacement $g_V \to G_\rho/M_\rho$:
$$\eqalign{
V_i \ \to \ -{{\pi_{0i}}\over{2M_\rho}}\cr
\bar\pi_i \ \to \ 2M_\rho W_{0i}\cr}\eqno(A.11)
$$
\medskip

\medskip
\noindent Nous avons montr\'e l' \'equivalence de la formulation en
termes de vecteurs (3.2) et de celle en
termes de tenseurs antisym\'etriques (3.3). Il est donc
\'evident que les conclusions faites sur la stabilit\'e des solutions
classiques
du Lagrangien (3.3) s' appliquent sans aucune
ambiguit\'e et telles quelles au cas des
solitons topologiques
du Lagrangien (3.2).

\eject

\bigskip

\centerline{\bfmag Conclusions G\'en\'erales}

\bigskip

\bigskip

Dans cette revue nous nous sommes concentr\'e \`a la description
des nucl\'eons en tant que solitons topologiques d' une th\'eorie effective de
m\'esons.
Le but de cette \'etude est de construire une
th\'eorie unifi\'ee des m\'esons et des baryons, comme il a \'et\'e
sugg\'er\'e en [21]. Cette th\'eorie est sens\'ee mod\'eliser la
Chromodynamique Quantique (QCD) dans son r\'egime non-perturbatif,
aux \'echelles des distances des baryons ($1$ fm). Dans notre
approche, bas\'ee en partie sur une vieille id\'ee de Skyrme [2],
les m\'esons sont consid\'er\'es comme des champs
\'el\'ementaires et les baryons comme leurs excitations de
type soliton topologique. Les interactions (fortes) entre les m\'esons
sont d\'ecrites par un Lagrangien effectif qui doit respecter les
propri\'et\'es de la th\'eorie sous jacente,
la QCD. Ce Lagrangien doit ainsi respecter la
sym\'etrie chirale, la brisure de l' invariance d' \'echelle de la QCD
etc. Mais ces propri\'et\'es g\'en\'erales
ne sont pas assez contraignantes pour d\'eterminer la forme sp\'ecifique du
Lagrangien effectif. Ce degr\'e d' arbitraire dans la d\'etermination du
Lagrangien effectif peut \^etre r\'eduit si l' on tient compte de la
ph\'enom\'enologie bien connue du secteur des m\'esons.
C' est ce qui a \'et\'e propos\'e dans la r\'ef. [24]. Notre travail a
consist\'e d' abord \`a \'etudier une extension du mod\`ele de la r\'ef. [24]
dans
le but d' y inclure des degr\'es
de libert\'e scalaires. Ceci nous a conduit dans le chapitre I
\`a introduire les champs vecteurs comme
des champs de jauge non-ab\'eliens dans le mod\`ele $\sigma$
lin\'eaire [30]. Cette formulation nous am\`ene naturellement \`a un Lagrangien
effectif contenant les m\'esons les plus l\'egers
$(\pi,\rho,\omega,A_1,\epsilon)$. Les param\`etres du mod\`ele sont
d\'etermin\'es en ajustant les observables du secteur m\'esonique. Nous
construisons alors les solutions de type soliton dans le secteur \`a charge
baryonique unit\'e, et nous \'etudions les propri\'et\'es statiques des
baryons. Un autre test important et s\'ev\`ere pour le mod\`ele
consid\'er\'e  est fourni par les propri\'et\'es de l' interaction entre
les baryons. Nous avons \'etudi\'e les  interactions
statiques entre ces solitons. En projetant ces interactions sur les
diff\'erents canaux de spin et isospin [30] nous avons obtenu les
composantes centrale, spin-spin et tenseur de l' interaction
nucl\'eon-nucl\'eon.

\noindent Les r\'esultats montrent que m\^eme dans le cas o\`u l'
approximation du produit est adopt\'ee pour le syst\`eme \`a deux
solitons, des attractives ayant la bonne port\'ee
apparaissent dans le canal central de l' interaction $NN$. Ceci est
\`a opposer \`a tous les calculs pr\'ec\'edents \'effectu\'es \`a partir
du mod\`ele de Skyrme, qui n' aboutissent qu' \`a des forces r\'epulsives,
ou attractives mais \`a longue port\'ee. Nous pensons \`a cet \'egard
que la pr\'esence simultan\'ee des degr\'es de libert\'e m\'esoniques du pion,
et des r\'esonances de basse \'energie est d' une importance cruciale.

Ensuite nous avons effectu\'e une analyse semi-classique du
mod\`ele de Skyrme, en \'evaluant la premi\`ere correction
quantique \`a la masse du soliton [47]. Dans le chapitre II,
nous avons montr\'e que cette correction est tr\`es grande (elle est du
m\^eme ordre de grandeur que la masse
classique) dans le cas du mod\`ele de Skyrme original. Nos r\'esultats
sugg\`erent que ce mod\`ele ne peut pas \^etre consid\'er\'e comme
une th\'eorie effective r\'ealiste. Nous avons ensuite montr\'e que la
situation est nettement am\'elior\'ee si l' on envisage d' inclure dans le
Lagrangien effectif des termes d' ordre sup\'erieur, par exemple un terme d'
ordre six dans les d\'eriv\'ees du champ du pion. Ces r\'esultats fournissent
des motivations th\'eoriques tr\`es fortes pour g\'en\'eraliser le mod\`ele
de Skyrme afin de mod\'eliser QCD \`a grand $N_c$.

Un autre probl\`eme qui requiert attention est celui de la
stabilit\'e des solitons, car les nucl\'eons
sont stables par l' interaction forte.
Dans le chapitre III nous avons \'etudi\'e le r\^ole jou\'e par
les m\'esons vecteurs dans la stabilit\'e des solitons topologiques,
dans le cadre du syst\`eme $\pi\rho$. Aujourd' hui on sait
que l' introduction du m\'eson $\rho$ comme un
champ de Yang-Mills massif se couplant au pion, d\'estabilise
le soliton, et que la sym\'etrie de jauge est \`a l'
origine de l' existence d' un doublet quasi-d\'eg\'en\'er\'e dans le
spectre baryonique. Ces ambiguit\'es n' existent pas dans le cas
du m\'eson $\omega$. Nous avons montr\'e qu' elles peuvent
dispara\^\i tre aussi dans le cas du m\'eson $\rho$, si ce dernier est
suppos\'e se transformer d' une fa\c con homog\`ene par une rotation du groupe
chiral non-lin\'eaire [72]. Ainsi les m\'esons $\rho$ et $\omega$ jouent un
r\^ole similaire dans le m\'ecanisme de stabilisation du soliton. Ceci
est satisfaisant parce qu' \`a la limite o\`u ces m\'esons deviennent tr\`es
lourds (limite locale) la contribution du m\'eson $\rho$ tend vers le
terme de Skyrme (ordre quatre) et celle du $\omega$ tend vers un
terme d' ordre six. Ces deux termes \`a leur tour stabilisent le soliton
s\'epar\'ement.

Un certain nombre de questions en rapport avec les \'etudes
pr\'esent\'ees ici restent ouvertes:

\noindent D' abord, l' extraction du potentiel
nucl\'eon-nucl\'eon dans le mod\`ele que nous avons
propos\'e pour la description unifi\'ee des m\'esons et des baryons,
peut \^etre faite par des m\'ethodes num\'eriques de minimisation exacte
dans le secteur \`a deux solitons. Le r\'esultat d' un tel calcul
pourrait nous donner une indication quasi-d\'efinitive sur la validit\'e
de ce mod\`ele pour la description d' un grand \'eventail de
ph\'enom\`enes de la physique des hadrons.

\noindent Les observables des baryons
dans les th\'eories effectives  pr\'ec\'edentes s' expriment comme un
d\'eveloppement en puissances de $\hbar/N_c$. Il est important
de savoir si ce d\'eveloppement est perturbatif et convergent.
L' \'etude expos\'ee dans le chapitre II apporte les premiers
\'el\'ements de r\'eponse \`a cette question. Il est donc naturel
maintenant non seulement de calculer les premi\`eres corrections quantiques
\`a d' autres observables telle la constante de couplage axiale du nucl\'eon,
mais aussi et surtout d' \'evaluer ces corrections pour les mod\`eles plus
r\'ealistes o\`u le champ chiral se couple \`a des champs vecteurs, scalaires
etc... .

\noindent Notre \'etude sur la stabilit\'e des solitons montre le vrai
r\^ole que le m\'eson $\rho$ joue dans les interactions fortes en stabilisant
le nucl\'eon-soliton topologique. Cette \'etude ouvre un d\'ebat sur les
diff\'erentes fa\c cons de r\'ealiser la sym\'etrie chirale par les m\'esons
$\rho$ et $A_1$. Dans cette optique, nous avons r\'ecemment propos\'e un
lagrangien
minimal $\pi\rho A_1$ [79] g\'en\'eralisant ainsi notre th\'eorie
du chapitre III au cas o\`u le m\'eson vecteur-axial est aussi pr\'esent.

\eject

\centerline{\bfmag Remerciements}

\noindent L' auteur aimerait remercier le Professeur Robert Vinh Mau pour
son soutien vigoureux et pour avoir dirig\'e une partie de
ses travaux. Les membres de la Division de Physique Th\'eorique de
l' Institut de Physique Nucl\'eaire \`a Orsay sont remerci\'es pour leur
accueil durant la p\'eriode
1990/93. Cet article de revue a \'et\'e compl\'et\'e gr\^ace au soutien
financier
du {\it Engineering and Physical Sciences Research Council}, Angleterre.

\eject

\medskip
\medskip
\medskip
\medskip
\centerline{\bfmag R\'ef\'erences}
\bigskip
\medskip
\item{[1]}K. G. Wilson, Phys. Rev. D10 (1974) 2445;
J. Kogut et L. Susskind, Phys. Rev. D11 (1975) 395; A. M. Polyakov, Phys. Lett.
B59 (1975) 85
\medskip
\item{[2]}E. Witten, Nucl. Phys. B160 (1979) 57
\medskip
\item{[3]}T.H.R. Skyrme, Proc. Roy. Soc. A260 (1961) 127
\medskip
\item{[4]}J. S. Russel, Proc. of the British Association for the Advancement of
Science (1845) 311
\medskip
\item{[5]}D. J. Korteweg et G. de Vries, Phil. Mag. 39 (1895) 422
\medskip
\item{[6]}N. J. Zabusky et M. D. Kruskal, Phys. Rev. Lett. 15 (1965) 240
\medskip
\item{[7]}C. Rogers et W. F. Shadwick, {\it B\ae cklund Tranformations and
Their Applications}, Academic Press 1982
\medskip
\item{[8]}J. Rubinstein, J. Math. Phys. 11 (1970) 258
\medskip
\item{[9]}D. Finkelstein et C. Misner, Ann. of Phys. (N.Y.) 6 (1959)
230; D. Finkelstein, J. Math. Phys. 7 (1966) 1218
\medskip
\item{[10]}R. Rajamaran, Phys. Rep. 21 (1975) 227
\medskip
\item{[11]}R. Dashen, B. Hasslacher et A. Neveu, Phys. Rev. D10 (1975)
4114; {\it ibid} D10 (1975) 4130; {\it ibid} D10 (1975) 4138;
J. Goldstone et R. Jackiw, Phys. Rev. D11 (1975) 1486
\medskip
\item{[12]}S. Coleman, Phys. Rev. D11 (1975) 2088
\medskip
\item{[13]}G. t' Hooft, Nucl. Phys. B72 (1974) 461; B75 (1974) 461
\medskip
\item{[14]}E. Witten, Nucl. Phys. B223 (1983) 422; {\it ibid.} (1983) 433
\medskip
\item{[15]}J. Goldstone et F. Wilczek, Phys. Rev. Lett. 47 (1981) 986
\medskip
\item{[16]}G. H. Derrick, J. Math. Phys. Vol.5  No9 (1964) 1252
\medskip
\item{[17]}G.S. Adkins, C.R. Nappi et E. Witten, Nucl. Phys. B228 (1983) 552
\medskip
\item{[18]}R. Vinh Mau, M.Lacombe, B.Loiseau, W.N. Cottingham et P. Lisboa,
Phys. Lett. B150 (1985) 259
\medskip
\item{[19]}A. Jackson, A.D. Jackson et V. Pasquier, Nucl. Phys. A432 (1985) 567
\medskip
\item{[20]}T.S. Walhout et J. Wambach Phys. Rev. Lett. 67 (1991) 314
\medskip
\item{[21]}E. Witten, dans {\it  Solitons in Nuclear and elementary Particle
Physics}, eds. A. Chodos, E. Hadjimichael et C. Tze (World Scientific,
Singapore,1984)
\medskip
\item{[22]}K. Iketani, Kyushu Univ. preprint, 84-HE-2 (1984)
\medskip
\item{[23]}G. Adkins et C. Nappi, Phys. Lett. B137 (1984) 552
\medskip
\item{[24]}M. Lacombe, B. Loiseau, R. Vinh Mau et W.N. Cottingham, Phys. Rev.
Lett. 57 (1986) 170; Phys. Rev. D38 (1988) 1491
\medskip
\item{[25]}M. Chemtob, Nucl. Phys. A 466 (1987) 509

M. Mashaal, T. N. Pham et T. N. Truong, Phys. Rev. Lett. 56 (1986) 436;

Phys. Rev. D34 (1986) 3484
\medskip
\item{[26]}M. Lacombe, B. Loiseau, R. Vinh Mau et W.N. Cottingham, Phys. Rev.
D40 (1989) 3012
\medskip
\item{[27]}M. Lacombe, B. Loiseau, R. Vinh Mau et W.N. Cottingham, Phys. Lett.
B161 (1985)
31; M. Lacombe, B. Loiseau, R. Vinh Mau et  W.N. Cottingham, Phys. Lett. B169
(1985) 121
\medskip
\item{[28]}J. Schechter, Phys. Rev.  D34 (1986) 868
\medskip
\item{[29]}H. Yabu, B. Schwesinger et G. Holzwarth, Phys. Lett. B224 (1989) 25
\medskip
\item{[30]}D. Kalafatis et R. Vinh Mau, Phys. Lett. B283 (1992) 13; Phys. Rev.
D46 (1992) 3903
\medskip
\item{[31]}S. L. Adler, Phys. Rev. 177 (1969) 2426;
J. S. Bell et R. Jackiw, Nuovo Cim. 60 (1969) 147; W. A. Bardeen, Phys. Rev.
184 (1969) 1848
\medskip
\item{[32]}H. Forkel, A.D. Jackson et C. Weiss, Nucl. Phys. A526 (1991) 453
\medskip
\item{[33]}O. Kaymakcalan, S. Rajeev et J. Schechter, Phys. Rev. D30 (1984)
594;
H. Gomm, O. Kaymakcalan et J. Schechter, {\it ibid} 30 (1984) 2345
\medskip
\item{[34]}B. Schwesinger, H. Weigel, G.Holzwarth et  A.Hayashi, Phys. Rep. 173
(1989) 173
\medskip
\item{[35]}O. Dumbrajs et al., Nucl. Phys.  B216 (1983) 277
\medskip
\item{[36]}A. Hosaka, M. Oka et R. Amado, Nucl. Phys. A530 (1991) 507
\medskip
\item{[37]}M. Lacombe et al, Phys. Rev. C21 (1980) 861
\medskip
\item{[38]}R. Vinh Mau dans {\it Mesons in Nuclei}, eds M. Rho et D. Wilkinson
(North
Holland, Amsterdam, 1979)
\medskip
\item{[39]}V. Thorsson et I. Zahed, Phys. Rev. D45 (1992) 965.
\medskip
\item{[40]}H. B. G. Casimir, Proc. Kon. Ned. Akad. Wetenschap., ser. B, 51
(1948) 793
\medskip
\item{[41]}R. P. Feynman et A. R. Hibbs, {\it Quantum Mechanics and Path
Integrals}, eds. McGraw-Hill (1965)
\medskip
\item{[42]}C. Itzykson et J.-B. Zuber, {\it Quantum Field Theory}, eds.
McGraw-Hill (1980), Chap. 9
\medskip
\item{[43]}J. Gasser et H. Leutwyler, Ann. Phys. (N.Y.) 158 (1984) 142
\medskip
\item{[44]}S. Coleman, {\it Aspects of Symmetry}, Eds. Cambridge
University Press (1985)
\medskip
\item{[45]}M. Abramowitz et I. Stegun, {\it Hanbook of Mathematical
Functions}, Eds. Dover (1964)
\medskip
\item{[46]}J. Schwinger, Phys. Rev. 94 (1951) 1362
\medskip
\item{[47]}B. Moussallam et D. Kalafatis, Phys. Lett. B272 (1991) 196
\medskip
\item{[48]}P. Wilkey, {\it Index Theorems and the Heat Equation}, Eds.
Publish or Perish, Berkeley (1975)
\medskip
\item{[49]}P. Ramond, {\it Field Theory: A modern Primer}, Eds.
Addison-Wesley Reading (1981)
\medskip
\item{[50]}C. Riggenbach, J. Gasser, J. F. Donoghue et B. R. Holstein,
Phys. Rev. D43 (1991) 127
\medskip
\item{[51]}H. Walliser et G. Eckart, Nucl. Phys. A429 (1984) 251; M. P.
Mattis et M. Karliner, Phys. Rev. D31 (1985) 2833
\medskip
\item{[52]}S. C. Generalis et G. Williams, Nucl. Phys. A484 (1988) 620
\medskip
\item{[53]}I. Zahed, A. Wirzba et U.-G. Meissner, Phys. Rev. D33 (1986) 830
\medskip
\item{[54]}A. Dobado et J. Terron, Phys. Lett. B247 (1990) 581
\medskip
\item{[55]}K. Cahill, A. Comtet et R. J. Glauber, Phys. Lett. B64 (1976) 283
\medskip
\item{[56]}M. Chemtob, Nucl. Phys. A473 (1987) 613
\medskip
\item{[57]}B. Moussallam, Orsay pr\'epublication IPNO/TH 92-94, paru
dans "Baryons as Skyrme Solitons", World Scientific eds. (1993)
\medskip
\item{[58]}D. Kalafatis, contribution \`a \lq\lq QCD Vacuum Structure", p. 241,
eds
World Scientific (1993)
\medskip
\item{[59]}B. Moussallam, Annals of Physics (N. Y.) 225 (1993) 264
\medskip
\item{[60]}C. Bernard et R. Kerner, Lett. Math. Phys. 18 (1989) 193
\medskip
\item{[61]}L. Marleau, Phys. Rev. D43 (1991) 885
\medskip
\item{[62]}Y. Igarashi, M. Johmura, A. Kobayashi, H. Otsu, T. Sato et S.
Sawada, Nucl. Phys. B259 (1985) 721
\medskip
\item{[63]}Z. F. Ezawa et T. Yanagida, Phys. Rev. D33 (1986) 247; J. Kunz
et D. Masak, Phys. Lett. B179 (1986) 146
\medskip
\item{[64]}S. Coleman, J. Wess et B. Zumino, Phys. Rev. 177 (1969) 2239
; C. G. Callan, S. Coleman, J. Wess et B. Zumino, {\it ibid} p. 2247; S.
Weinberg, Phys. Rev. 166 (1968) 1568
\medskip
\item{[65]}M. Bando, T. Kugo et K. Yamawaki, Phys. Rep. 164 (1988) 217
\medskip
\item{[66]}C. N. Yang and R. L. Mills, Phys. Rev. 96 (1954) 191
\medskip
\item{[67]}A. Kobayashi, H. Otsu, T. Sato et S. Sawada, Nagoya Univ.
preprint, DPNU 91-50, IPC-91-04 ; K. Yang, S. Sawada et A. Kobayashi, Prog.
Theor. Phys. 87 (1992) 457
\medskip
\item{[68]}J. D. Bjorken et S. D. Drell, {\it Relativistic Quantum
Fields}, Mc-Graw Hill 1965
\medskip
\item{[69]}G. Ecker, J. Gasser, A. Pich et E. de Rafael, Nucl. Phys.
B321 (1989) 311
\medskip
\item{[70]}G. Ecker, J. Gasser, H. Leutwyler, A. Pich et E. de Rafael
, Phys. Lett.  B223 (1989) 425
\medskip
\item{[71]}K. Sundermayer, {\it Constrained Dynamics} dans {\it  Lecture Notes
in Physics}, Springer-Verlag 1982
\medskip
\item{[72]}A. Abada, D. Kalafatis et B. Moussallam, Phys. Lett. B300
(1993) 256
\medskip
\item{[73]}Y. Brihaye, C. Semay et J. Kunz, Phys. Rev. D44 (1991) 250
\medskip
\item{[74]}I. Zahed, U. G. Meissner et U. B. Kaulfuss, Nucl. Phys. A426 (1984)
525
\medskip
\item{[75]}R. G. Newton, {\it Scattering theory of waves and
particles}, Springer-Verlag eds., p. 346
\medskip
\item{[76]}R. Kerner; communication priv\'ee
\medskip
\item{[77]}E. B. Bogomol'nyi, Sov. J. Nucl. Phys. Vol 24 (1976) 449
\medskip
\item{[78]}M. Chaichian et N. F. Nelipa, {\it Introduction to Gauge Field
Theories}, Springer-Verlag Ed., p. 296
\medskip
\item{[79]}D. Kalafatis, Phys. Lett. B313 (1993) 115; D. Kalafatis et Michael
C. Birse ,
pr\'epublication Universit\'e de Manchester MC/TH 94/10.

\eject

{\rightline{}}
\centerline {\bfmagd Sommaire}

\bigskip

\noindent {\it Abstract} {\hfill} p. 2 \ \

\medskip

\noindent {\bf Introduction} {\hfill} p. 3 \ \

\medskip

\noindent {\bf I. Un mod\`ele des interactions fortes \`a basses
\'energies} {\hfill} p. 8 \ \

\noindent \ \ \ {\it 1. Les solitons du mod\`ele de Skyrme} {\hfill} p. 10 \ \

\noindent \ \ \ {\it 2. G\'en\'eralisations du mod\`ele de Skyrme} {\hfill} p.
14 \

\noindent \ \ \ {\it 3. Un mod\`ele unifiant les m\'esons et les baryons}
{\hfill} p. 16 \

\noindent \ \ \ {\it 4. Le secteur des m\'esons} {\hfill} p. 23 \

\noindent \ \ \ {\it 5. Le secteur baryonique} {\hfill} p. 26 \

\noindent \ \ \ {\it 6. L'interaction nucl\'eon-nucl\'eon} {\hfill} p. 28 \

\noindent \ \ \ {\it 7. Discussion des r\'esultats} {\hfill} p. 36 \

\noindent \ \ \ {\it Appendice A} {\hfill} p. 42 \

\noindent \ \ \ {\it Appendice B} {\hfill} p. 43 \

\noindent \ \ \ {\it Appendice C} {\hfill} p. 44 \

\bigskip

\noindent {\bf II. Sur le d\'eveloppement semi-classique
de la masse du soliton} {\hfill} p. 46 \

\noindent \ \ \ {\it 1. L' \'energie de Casimir du soliton} {\hfill} p. 47 \

\noindent \ \ \ {\it 2. Calcul des d\'ephasages. Traitement des
divergences ultraviolettes} {\hfill} p. 52 \

\noindent \ \ \ {\it 3. R\'esultats} {\hfill} p. 56 \

\noindent \ \ \ {\it 4. Conclusion} {\hfill} p. 61 \

\bigskip

\noindent {\bf III. Sur la stabilit\'e des solitons topologiques} {\hfill} p.
63 \

\noindent \ \ \ {\it 1. R\'ealisations non-lin\'eaires de la sym\'etrie
chirale} {\hfill} p. 64 \

\noindent \ \ \ {\it 2. Le soliton du syst\`eme $\pi\rho$ dans une
formulation \lq\lq Yang-Mills"} {\hfill} p. 65 \

\noindent \ \ \ {\it 3. Transformation homog\`ene des champs
vecteurs} {\hfill} p. 68 \

\noindent \ \ \ \ \ \ {a. le Hamiltonien} {\hfill} p. 70 \

\noindent \ \ \ \ \ \ {b. Solutions classiques} {\hfill} p. 74 \

\noindent \ \ \ \ \ \ {c. Stabilit\'e} {\hfill} p. 77 \

\noindent \ \ \ {\it 4. Conclusions} {\hfill} p. 80 \

\noindent \ \ \ {\it Appendice A} {\hfill} p. 83 \

\bigskip

\noindent {\bf Conclusions G\'en\'erales} {\hfill} p. 87

\bigskip

\noindent {\bf Remerciements} {\hfill} p. 90

\bigskip

\noindent {\bf R\'ef\'erences} {\hfill} p. 91

\bigskip

\noindent {\bf Figures} {\hfill} p. 97

\bye